\documentclass[superscriptaddress, twocolumn,showkeys,
amssymb,amsmath,nobibnotes,aps,prd,
nofootinbib]{revtex4-2}
\pdfoutput=1
\usepackage{graphicx,subfigure,bm,color,psfrag,hyperref}
\usepackage[export]{adjustbox}
\usepackage{amsfonts}
\usepackage{lipsum}
\usepackage{dcolumn}
\usepackage{mathtools}
\usepackage{verbatim}
\usepackage[normalem]{ulem}
\usepackage[dvipsnames]{xcolor}
\hypersetup{colorlinks,linkcolor={blue},citecolor={red},urlcolor={greenW}}  
\definecolor{greenW}{rgb}{0.0, 0.55, 0.1} 
\begin{document}

\title{Phantom cosmology with arbitrary potential: New accelerating scaling attractors}

\author{Sudip Halder}
\email{sudip.rs@presiuniv.ac.in}
\affiliation{Department of Mathematics, Presidency University, 86/1 College Street, Kolkata 700073, India}

\author{Supriya Pan}
\email{supriya.maths@presiuniv.ac.in}
\affiliation{Department of Mathematics, Presidency University, 86/1 College Street, Kolkata 700073, India}
\affiliation{Institute of Systems Science, 
Durban University of Technology, Durban 4000, Republic of South Africa}

\author{Paulo M.\ S\'a}
\email{pmsa@ualg.pt}
\affiliation{Departamento de F\'\i sica, Faculdade de Ci\^encias e Tecnologia, Universidade do Algarve, Campus de Gambelas, 8005-139 Faro, Portugal}
\affiliation{Centro de Investiga\c{c}\~ao e Desenvolvimento em Matem\'atica e Aplica\c{c}\~oes (CIDMA), Universidade de Aveiro, Campus de Santiago, 3810-193 Aveiro, Portugal}

\author{Tapan Saha}
\email{tapan.maths@presiuniv.ac.in}
\affiliation{Department of Mathematics, Presidency University, 86/1 College Street, Kolkata 700073, India}

\begin{abstract}
In this article, we investigate the existence of accelerating scaling solutions in coupled phantom cosmology without assuming any specific potential for the phantom scalar field. The coupling between phantom dark energy and dark matter is motivated by the warm inflationary paradigm, with the dissipation coefficient assumed to be either constant or variable. The evolution equations are written in the form of autonomous systems, whose stability is studied using methods of qualitative analysis of dynamical systems. For this analysis, the only requirement imposed on the otherwise arbitrary phantom potential is that a particular dynamical variable, defined in terms of the potential and its derivative, must be invertible. For such a generic potential, we show that accelerated scaling solutions do exist, for both constant and variable dissipation coefficients. Although there is a limitation to these scaling solutions -- specifically, the current stage of accelerated expansion is not preceded by a long enough matter-dominated era -- our results show that the existence of a direct coupling between phantom dark energy and dark matter yields great potential for addressing the cosmic coincidence problem.
\end{abstract}

\maketitle
\section{Introduction \label{Sec-1}}

The accelerating expansion of the Universe during its late evolution was a major discovery in cosmology and astrophysics~\cite{SupernovaSearchTeam:1998fmf, SupernovaCosmologyProject:1998vns}, which caused a significant impact on the scientific community.
What drives this accelerating expansion has remained mysterious till today. 
Usually, two well-known approaches are considered to explain this phenomenon. The first involves introducing a hypothetical dark energy (DE) fluid into the gravitational equations described by Einstein's General Relativity (GR)~\cite{Copeland:2006wr, Bamba:2012cp}. The second approach modifies GR in various ways, widely known as modified gravity theories~\cite{Nojiri:2006ri,DeFelice:2010aj,Clifton:2011jh, Cai:2015emx, Nojiri:2017ncd, Bahamonde:2021gfp}.

The simplest explanation for this accelerated expansion is based on GR and incorporates a positive cosmological constant $\Lambda$ that acts as dark energy.
This cosmological model, known as $\Lambda$-Cold Dark Matter (or $\Lambda$CDM), has been quite successful in explaining various phenomena in the Universe.
However, it also faces several problems, notably the cosmological constant problem~\cite{Weinberg:1988cp} and the cosmic coincidence problem, often referred to as the “why now” problem~\cite{Zlatev:1998tr}.
The $\sim\! 5 \sigma$ tension on the Hubble constant between Planck (within the $\Lambda$CDM paradigm)~\cite{Planck:2018vyg} and SH0ES (Supernovae and $H_0$ for the Equation of State of dark energy)~\cite{Riess:2021jrx} is another challenge that has increased the debate on the revision of the $\Lambda$CDM model.
Additionally, from a theoretical point of view, the independent conservation of DE and dark matter within the $\Lambda$CDM framework imposes a constraint on the dynamics of the dark components.
These issues collectively suggest that a revision of the $\Lambda$CDM model is welcome until the true cosmological model is found that can answer all the existing cosmic puzzles.

The simplest way to extend the $\Lambda$CDM model is to replace the cosmological constant $\Lambda$ with a time-varying candidate for DE. This includes a variety of models, such as scalar-field DE models (e.g., quintessence, phantom, tachyon, and k-essence),  as well as models with a redshift-dependent equation of state for DE, among others~\cite{Copeland:2006wr, Bamba:2012cp}.

In the present article, we particularly focus on the phantom scalar-field model~\cite{Caldwell:1999ew,Caldwell:2003vq,Elizalde:2004mq}, with the underlying gravitational field solely described by GR.
Since the total energy of the phantom field is unbounded from below, this model is appropriate for describing only the late-time evolution of the Universe.

We consider two distinct scenarios. The first is the uncoupled phantom scenario, where the phantom scalar field coexists with dark matter (DM) without any interaction between them. The second scenario involves a coupled phantom scalar field that interacts with DM in a non-gravitational manner, with the coupling function inspired by warm inflation.

An interacting cosmological scenario (see, for instance, Refs.~\cite{Amendola:1999er,Cai:2004dk,Huey:2004qv,Curbelo:2005dh,Barrow:2006hia,delCampo:2008sr,Valiviita:2008iv,delCampo:2008jx,Leon:2009dt,Clemson:2011an,Li:2013bya,Yang:2014gza,Yang:2014okp,Salvatelli:2014zta,Duniya:2015nva,Nunes:2016dlj,Yang:2016evp,Sharov:2017iue,DiValentino:2017iww,Mifsud:2017fsy,Yang:2017iew,Yang:2017zjs,Linton:2017ged,Yang:2018qec,Martinelli:2019dau,Paliathanasis:2019hbi,DiValentino:2020leo,Pan:2020bur,Yang:2020zuk,Pan:2020zza,Sa:2020a,Pan:2020mst,Sa:2020b,DiValentino:2019ffd,DiValentino:2019jae,DiValentino:2020kpf,Paliathanasis:2020sfe,Gao:2021xnk,Lucca:2021eqy,Sa:2021,Gomez:2020sfz,Nunes:2022bhn,Chatzidakis:2022mpf,Sa:2022,Teixeira:2023zjt,Zhai:2023yny,Paliathanasis:2023moe,Gomez:2022okq,Li:2023gtu,Giare:2024smz,Giare:2024ytc,Li:2024qso,Paliathanasis:2024abl,Li:2025owk,Leon:2025sfd,Tsedrik:2025cwc,Liu:2025pxy,Feng:2025mlo,Zhai:2025hfi,Silva:2025hxw,Li:2025ula,vanderWesthuizen:2025iam,Paliathanasis:2025xxm,Yan:2025iga,Wang:2025znm,vanderWesthuizen:2025vcb,vanderWesthuizen:2025mnw,Li:2025muv,Zhang:2025dwu}) is a generalized cosmological framework offering many interesting outcomes.
From a rational point of view, there is no objection to considering interactive scenarios, as long as they are not ruled out by observations.  
The aftermath of a direct interaction between the dark sectors is far-reaching ---  such an interaction can alleviate the cosmic coincidence problem~\cite{Cai:2004dk,Huey:2004qv,delCampo:2008jx} and the Hubble constant tension~\cite{Kumar:2017dnp,DiValentino:2017iww,Yang:2018euj,Pan:2020bur}.

We investigated both uncoupled and coupled phantom scenarios by applying the powerful techniques of dynamical systems analysis (for a review on the application of such techniques to cosmology, see Ref.~\cite{Bahamonde:2017ize}).

One of the main foci of the present article is to search for late-time accelerating scaling attractors in phantom scalar field models, since, according to the existing literature in this direction, only a few articles have found such a possibility~\cite{Guo:2004xx,Chen:2008ft,Halder:2024aan,Paliathanasis:2024jxo}.
Our work presents a novel contribution to the literature of phantom cosmologies by allowing the potential $V(\phi)$ for the phantom scalar field to be quite general, both for the uncoupled and coupled scenarios.
The only requirement on this potential is that a specific dynamical variable $\lambda(\phi)$, which is defined in terms of $V(\phi)$ and its derivative $\partial V/\partial\phi$ (see Eq.~(\ref{lambda})), must be invertible.
For such a rather generic potential, our findings indicate that in the uncoupled phantom model, no accelerating scaling attractors can exist.
In contrast, such solutions are present in the coupled phantom scenario. 
This is an interesting result, suggesting that a direct, non-gravitational interaction between the dark components of the Universe might help to explain recent observational evidence. 

The paper is organized as follows.
In section~\ref{Sec-2} we present the key gravitational equations of the uncoupled/coupled phantom DE cosmological model.
Section~\ref{Sec-3} presents the dynamical system analysis for the uncoupled case, while Section~\ref {Sec-4} presents this analysis for the coupled case, considering both constant and variable dissipation coefficients in the interaction term between DE and DM.
Finally, in section~\ref{Sec-conclusions}, we summarize the main findings of the article.

\section{The phantom cosmological model}
\label{Sec-2}

We assume the gravitational sector of the Universe to be described by GR, and the matter sector, which is minimally coupled to gravity, to include a pressureless DM fluid and a phantom scalar field acting as DE.
For simplicity, radiation and ordinary baryonic matter are neglected.

Furthermore, we assume a flat Friedmann-Lema\^{i}tre-Robertson-Walker (FLRW) metric, 
\begin{equation}
 ds^2 = -dt^2 + a^2(t) d\Sigma^2, 
\end{equation}
where $a(t)$ is the scale factor and $d\Sigma^2$ is the metric of the three-dimensional Euclidean space. 
The evolution equations for the phantom cosmological model are then given by
\begin{gather}
H^2 = \frac{\kappa^2}{3} \left( -\frac{\dot{\phi}^2}{2} + V(\phi) 
+\rho_\texttt{DM} \right),   \label{friedmann}
\\
\dot{H} = -\frac{\kappa^2}{2}  \left( -\dot{\phi}^2 + \rho_\texttt{DM} \right),
\label{dotH}\\
\ddot{\phi} + 3H\dot{\phi} -\frac{\partial V}{\partial\phi}
= -\frac{Q}{\dot{\phi}},   \label{ddotphi1}
\\
\dot{\rho}_\texttt{DM}+3H\rho_\texttt{DM} = -Q.   \label{dotrho}
\end{gather}

In the above equations, $\phi$ is the phantom DE scalar field with potential $V(\phi)$, $\rho_\texttt{DM}$ is the energy density of a pressureless non-relativistic DM fluid, $H\equiv\dot{a}/a$ is the Hubble parameter, and $Q$ is the interaction term between DE and DM. 
An overdot denotes a derivative with respect to cosmic time $t$, and we use the notation $\kappa \equiv \sqrt{8\pi G} = \sqrt{8\pi/m_\texttt{P}}$, where $G$ is the gravitational constant and $m_\texttt{P}$ is the Planck mass.

The cosmological solutions of the uncoupled model ($Q=0$) and the coupled model ($Q=\Gamma\dot{\phi}^2$, with constant $\Gamma$) were investigated in detail in Ref.~\cite{Halder:2024gag} for the case of the simple exponential potential
\begin{equation}
    V(\phi)=V_0 e^{-\lambda \kappa \phi},
    \label{expV}
\end{equation}
where $V_0$ and $\lambda$ are positive constants.
In the present article, this investigation will be extended by considering an arbitrary potential $V$, as well as an interaction term with a variable dissipation coefficient $\Gamma$.

\section{Uncoupled phantom dark energy}
\label{Sec-3}

We first consider the uncoupled case, for which $Q=0$.
For the exponential potential (\ref{expV}), the evolution equations (\ref{friedmann})--(\ref{dotrho}) can be written as a two-dimensional dynamical system \cite{Halder:2024gag} by introducing the dimensionless variables 
\begin{equation}
     x=\frac{\kappa\dot{\phi}}{\sqrt6 H}
     \quad \mbox{and} \quad
     y=\frac{\kappa\sqrt{V}}{\sqrt3 H},
     \label{x-and-y}
\end{equation}
as well as a new “time” variable $N$ ($N = \ln(a/a_0)$ and $a_0$ is the present-day value of the scale factor), defined as
\begin{equation}
    \frac{dN}{dt}=H.
    \label{N}
\end{equation}

This is not the case for an arbitrary potential $V(\phi)$, where an extra variable is needed to close the dynamical system.
Let us denote this new variable by $\lambda$ and define it as ~\cite{Steinhardt:1999nw,delaMacorra:1999ff,Ng:2001hs}
\begin{equation}
    \lambda=-\frac{V_\phi}{\kappa V},
    \label{lambda}
\end{equation}
where a subscript $\phi$ denotes a derivative with respect to this variable. 

The evolution equations (\ref{friedmann})--(\ref{dotrho}) are now written as a three-dimensional dynamical system,
\begin{subequations} \label{DS-1}
\begin{align}
x_N &= - \frac{\sqrt{6}}{2} \lambda y^2 
	- \frac{3}{2} x (1+x^2+y^2),  \label{DS-1a}
\\
y_N &= \left[ -\frac{\sqrt{6}}{2} \lambda x
    + \frac{3}{2} (1-x^2 -y^2) \right] y, \label{DS-1b}
\\
\lambda_N &= -\sqrt{6} f(\lambda) x, \label{DS-1c}
\end{align}
\end{subequations}
where a subscript $N$ denotes a derivative with respect to $N$ and the function $f$ is given by
\begin{equation}
    f(\lambda)=\lambda^2 \left[ \frac{V V_{\phi\phi}}{(V_\phi)^2} -1 \right]. \label{f-lambda}
\end{equation}

In the above dynamical system, we have assumed $f$ to be a function of $\lambda$ only. This is only possible if $\lambda(\phi)$ is invertible, and we can use $\phi(\lambda)$ to eliminate any dependence of $\phi$ in the expression $V V_{\phi\phi}/(V_\phi)^2$ (see Ref.~\cite{Bahamonde:2017ize} for a detailed discussion).
Therefore, in what follows, we will restrict our analysis to the class of potentials that satisfy the above requirement, i.e., those for which $\lambda(\phi)=-V_\phi/(\kappa V)$ is an invertible function.\footnote{If $\lambda(\phi)$ is not invertible, the introduction of the extra variable $\lambda$ does not suffice to close the dynamical system and, consequently, one or more additional dynamical variables are needed. }

From Eqs.~(\ref{friedmann}) and (\ref{dotH}), it follows that the DM density parameter $\Omega_\texttt{DM}$ and the deceleration parameter $q$ are given, in terms of the variables $x$ and $y$, as
\begin{gather}
\Omega_\texttt{DM}\equiv \frac{\kappa^2\rho_\texttt{DM}}{3H^2}
    = 1+x^2-y^2, \label{omega-dm}
\\
q \equiv -\frac{\dot{H}}{H^2}-1
= \frac12 (1-3x^2-3y^2). \label{q}
\end{gather}
    
Furthermore, 
defining the energy density and pressure of the phantom scalar field $\phi$ as
\begin{equation}
     \rho_\phi=-\frac{\dot{\phi}^2}{2}+V(\phi) \quad \mbox{and}
     \quad p_\phi=-\frac{\dot{\phi}^2}{2}-V(\phi),
     \label{rho-p}
 \end{equation}
 respectively, 
the DE density parameter $\Omega_\phi$, the DE equation-of-state parameter $w_\phi$, and the total equation-of-state parameter $w_\texttt{tot}$ are given by
\begin{gather}
    \Omega_\phi\equiv \frac{\kappa^2\rho_\phi}{3H^2}
    = -x^2+y^2, \label{omega-phi}
\\
    w_\phi \equiv\frac{p_\phi}{\rho_\phi}
    =\frac{x^2+y^2}{x^2-y^2}, \label{w-phi}
\\
    w_\texttt{tot} \equiv \frac{p_\phi+p_\texttt{DM}}{\rho_\phi+\rho_\texttt{DM}}
    =-x^2-y^2. \label{w-tot}
\end{gather}

Because of the negative sign in the kinetic term of the phantom scalar field (see Eq.~(\ref{rho-p})), the DE density parameter $\Omega_\phi$ becomes negative for $x^2>y^2$ and the phantom equation-of-state parameter $w_\phi$ diverges for $x^2=y^2$. To avoid these unphysical situations, we restrict the dynamical-system analysis to the region $x^2<y^2$.

Inspection of the dynamical system (\ref{DS-1}) reveals that the plane $y=0$ is an invariant manifold. Moreover, from the evolution equation for the DM density parameter,
\begin{equation}
    \Omega_{\texttt{DM},N}=-3 \Omega_\texttt{DM} (x^2+y^2),
\end{equation}
it follows that the surfaces $y=\pm\sqrt{1+x^2}$ ($\Omega_\texttt{DM}=0$) are also invariant manifolds.
Keeping in mind that $\Omega_{\texttt{DM}}$ must be non-negative and restricting the analysis to non-contracting cosmologies, the phase space is confined to the region $\pm x< y\leq\sqrt {1+x^2}$ ($x\in\mathbb{R}$), where we have also taken into account the comment made in the previous paragraph.
The allowed values of $\lambda$, as well as the existence of invariant manifolds on this variable, cannot be specified at this point, since they depend on the specific form of the potential $V (\phi)$.
For now, we will allow $\lambda$ to assume any real value.

Let us now find the critical points of the dynamical system (\ref{DS-1}). From Eq.~(\ref{DS-1c}) it immediately follows that their existence requires either $x=0$ or $f(\lambda)=0$.

For $x=0$, Eqs.~(\ref{DS-1a}) and (\ref{DS-1b}) reveal the existence of a one-dimensional set of non-isolated critical points $A_1(0,0, \lambda)$ (critical line) and an isolated critical point $A_2(0,1,0)$.\footnote{The dynamical systems studied in this article allow for the existence of sets of non-isolated critical points, notably one-, two-, and three-dimensional sets, called critical lines, critical surfaces, and critical volumes, respectively. In what follows, for the sake of conciseness, we may use the expression “critical point” to refer to such sets of non-isolated critical points.}
We must emphasize that the existence of these critical points does not depend on the specific form of the potential $V(\phi)$.

If $f(\lambda)$ has some non-complex zeros, collectively denoted by $\overline{\lambda}$, then there exists a third critical point, namely, $A_3(-\overline{\lambda}/\sqrt6,\sqrt{1+\overline{\lambda}^2/6}, \overline{\lambda})$.

\begingroup
\setlength{\tabcolsep}{6pt}
\renewcommand{\arraystretch}{1.4}
\begin{table*}[t]
\centering
\begin{tabular}{c c c c c c c c}
\hline\hline
Critical points & Existence & $\Omega_\phi$ & $\Omega_\texttt{DM}$ & $w_\phi$ &  $w_\texttt{tot}$ & $q$ & Acceleration
\\
\hline
$A_{1}\left(0,0,\lambda\right)$ & Always & $0$ & $1$ & Indeterm. & $0$ & $\frac12$ & No
\\
$A_{2}\left(0,1,0\right)$ & Always  & $1$  & $0$ & $-1$ & $-1$ & $-1$ & Yes
\\
$A_{3}\left(-\frac{\overline{\lambda}}{\sqrt6}, \sqrt{1+\frac{\overline{\lambda}^2}{6}}, \overline{\lambda}\right)$ & $f(\overline{\lambda})=0$ & $1$ & $0$  & $-1-\frac{\overline{\lambda}^2}{3}$ & $-1-\frac{\overline{\lambda}^2}{3}$ & $-1-\frac{\overline{\lambda}^2}{2}$ & Yes
\\
\hline\hline
\end{tabular}
\caption{Phenomenological properties of the critical points of the dynamical system (\ref{DS-1}) for the uncoupled phantom DE cosmological model with an arbitrary potential $V(\phi)$. The critical points $A_1$ and $A_2$ exist for any $V$.
The critical point $A_3$ only exists if the equation $f(\lambda)=0$ has one or more non-complex roots, collectively denoted by $\overline{\lambda}$.
All critical points correspond to solutions dominated either by DM  ($\Omega_\texttt{DM}$=1) or DE ($\Omega_\phi$=1).}
	\label{table_uncoupled}
\end{table*}
\endgroup

Since all these critical points correspond to solutions that are completely dominated by either DM or DE (see Table~\ref{table_uncoupled}), we must conclude that in the uncoupled phantom cosmological model, described by the dynamical system (\ref{DS-1}), the final state of the Universe's evolution cannot be represented by a scaling solution whatever the potential $V(\phi)$ we choose (belonging, of course, to the class of potentials we are considering, i.e., potentials for which $\lambda(\phi)=-V_\phi/(\kappa V)$ is an invertible function).

\section{Coupled phantom dark energy}
\label{Sec-4}

Let us now move on to the coupled case.
Expressed in terms of the dimensionless variables $x$, $y$, and $\lambda$, introduced in the previous section, the evolution equations (\ref{friedmann})-(\ref{dotrho}) read
\begin{subequations} \label{DS-2}
\begin{align}
x_N &= - \frac{\sqrt{6}}{2} \lambda y^2 
	- \frac{3}{2} x (1+x^2+y^2)-\frac{\kappa^2 Q}{6 H^3 x},  \label{DS-2a}
\\
y_N &= \left[ -\frac{\sqrt{6}}{2} \lambda x
    + \frac{3}{2} (1-x^2 -y^2) \right] y, \label{DS-2b}
\\
\lambda_N &= -\sqrt{6} f(\lambda) x, \label{DS-2c}
\end{align}
\end{subequations}
where $f(\lambda)$ is defined in Eq.~(\ref{f-lambda}).
Depending on the functional form of $Q$, it may be necessary to introduce an extra dimensionless variable to render the system autonomous.

We assume the interaction term $Q$ to be of the dissipative type \cite{Sa:2023coi}
\begin{equation}
	Q=\Gamma \dot{\phi}^2,
	\label{interaction}
\end{equation}
where $\Gamma$ is the dissipation coefficient. It has the dimension of mass, implying that $Q$ has the dimension of (mass)$^5$, which is consistent with the dimensions of the terms $\dot{\rho}_\texttt{DM}$ ($\dot{\rho}_\phi$) and $3H\rho_\texttt{DM}$ ($3H\rho_\phi$) in the conservation equations, both with dimension (mass)$^5$.

This interaction term is inspired by warm inflation \cite{Berera:1995ie}.
According to this paradigm, during the inflationary period, energy is continuously transferred from the inflaton field $\psi$ to a radiation bath; as a result, the energy density of radiation $\rho_R$ is not diluted during inflation, and a smooth transition to a radiation-dominated era is ensured without a separate post-inflationary reheating phase.
The energy transfer between the inflation field and radiation is mediated by the dissipation term $Q=\Gamma \dot{\psi}^2$, where the dissipation coefficient $\Gamma$ is, in general, a function of the inflaton field and the temperature of the radiation bath (for recent reviews on warm inflation, see Refs.~\cite{Berera:2023liv,Kamali:2023}).

This kind of dissipation process could also be present at later stages of the Universe’s evolution.
In particular, it could mediate a direct, non-gravitational interaction between DE and DM.
This is a reasonable assumption. Indeed, at present, we lack a fundamental theory describing the interaction between the dark components of the Universe, implying that any choice of the putative interaction term is phenomenological. One could consider this interaction term to depend on non-local quantities like, for instance, the Universe’s expansion rate $H$, or one could consider it to depend on local dissipative effects as in the warm inflationary scenario.
The latter is our choice in this article. Therefore, we will consider the interaction term between DE and DM to be of the form~(\ref{interaction}), with a dissipation coefficient $\Gamma$ either a constant or a function of $\rho_\texttt{DM}$.

\subsection{Constant dissipation coefficient}

We start by considering the dissipation coefficient to be a constant with the dimension of mass,
\begin{equation}
 \Gamma=\mbox{constant}.
\end{equation}

For such choice of $\Gamma$, the evolution equations (\ref{friedmann})--(\ref{dotrho}) cannot be written as a three-dimensional dynamical system, as in the uncoupled case (see Eqs.~(\ref{DS-1})), since the interaction term $Q$ between DE and DM cannot be expressed as a function of the variables $x$, $y$, and $\lambda$, defined in Eqs.~(\ref{x-and-y}) and (\ref{lambda}); an additional variable $z$ is required to close the dynamical system (\ref{DS-2}), which therefore becomes four-dimensional.

We choose this additional variable to be \cite{Sa:2023coi}
\begin{equation}
	z=\frac{H_\ast}{H \Omega_\texttt{DM}+H_\ast},
	\label{z-variable}
\end{equation}
where $H_\ast$ is a positive constant with the dimension of mass.
Note that this choice of $z$ compactifies the phase space in the $z$ direction, which becomes comprised between $z=0$ (for $H=+\infty$) and $z=1$ (for $H=0$).
However, with such $z$, the interaction term appearing in the evolution equation for $x$ diverges as $z \rightarrow 1$. This divergence is removed by choosing a new “time” variable $\tau$, given by 
\begin{equation}
	\frac{d\tau}{dt}=\frac{H}{1-z}.
	\label{tau}
\end{equation}
It is worth mentioning that this choice of the dimensionless variable $z$ also guarantees that the hypersurface $\Omega_{\texttt{DM}}=1+x^2-y^2=0$ is an invariant manifold of the four-dimensional dynamical system, meaning that no phase-space trajectory can cross this hypersurface and enter the unphysical region of negative values of $\Omega_{\texttt{DM}}$.

In the variables $x$, $y$, $z$, $\lambda$, and $\tau$, the evolution equations (\ref{friedmann})--(\ref{dotrho}) for the coupled phantom DE cosmological model with a constant dissipation coefficient give rise to the four-dimensional dynamical system
\begin{subequations} \label{DS-3}
\begin{align}
x_\tau & = \left[ - \frac{\sqrt{6}}{2} \lambda y^2 - \frac{3}{2} x (1+x^2+y^2) \right] (1-z) \nonumber 
\\ 
& \hspace{3mm} - \alpha x (1+x^2 - y^2) z, \label{DS-3a}
\\
y_\tau &= \left[- \frac{\sqrt{6}}{2} \lambda x + \frac32 (1-x^2 -y^2) \right] y (1-z), \label{DS-3b}
\\
z_\tau &= \left[\frac{3}{2} (1 + x^2 +y^2) (1-z) + 2 \alpha x^2 z \right] z (1-z), \label{DS-3c}
\\
\lambda_\tau & = -\sqrt6 f(\lambda) x(1-z), \label{DS-3d}
\end{align}
\end{subequations}
where $\alpha \equiv \Gamma/H_{*}$ is the dimensionless coupling parameter, taken to be nonzero, and $f(\lambda)$ is defined in Eq.~(\ref{f-lambda}).
Here, again, as in the uncoupled case, we have assumed that $\lambda(\phi)=-V_\phi/(\kappa V)$ is invertible and $f$ is a function of $\lambda$ only.
If the potential $V$ is given by Eq.~(\ref{expV}), this dynamical system reduces to the one studied in Ref.~\cite{Halder:2024gag}.

The DM and DE density parameters $\Omega_\texttt{DM}$ and $\Omega_\phi$, the phantom and the total equation-of-state parameters $w_\phi$ and $w_\texttt{tot}$, and the deceleration parameter $q$ do not depend on the variable $z$ and, therefore, are given by the same expressions as in the uncoupled case (see Eq.~(\ref{omega-dm}), (\ref{q}), (\ref{omega-phi})--(\ref{w-tot})).

The hyperplanes $y=0$, $z=0$, and $z=1$ are invariant manifolds of the dynamical system (\ref{DS-3}). From the evolution equation for the DM density parameter,
\begin{equation}
    \Omega_{\texttt{DM},\tau}= - \Omega_\texttt{DM} \left[ 3(x^2+y^2)(1-z)+ 2\alpha x^2 z \right],
    \label{OmegaDM-coupled constant}
\end{equation}
it follows that the hypersurfaces $y=\pm\sqrt{1+x^2}$ ($\Omega_\texttt{DM}=0$) are also invariant manifolds. 
Taking into account that, by definition, $\Omega_{\texttt{DM}}\geq0$ and restricting ourselves to non-contracting cosmologies, the phase space is confined to the region $0\leq y \leq \sqrt{1+x^2}$, $0\leq z \leq 1$  ($x\in\mathbb{R}$).
Again, as in the uncoupled case, the phantom equation-of-state parameter $w_\phi$ diverges for $x^2=y^2$ and the DE density parameter $\Omega_\phi$ becomes negative for $x^2>y^2$. To avoid these unphysical situations, we further restrict the phase space to the region $\pm x< y \leq \sqrt{1+x^2}$, $0\leq z \leq 1$  ($x\in\mathbb{R}$).
The allowed values of $\lambda$, as well as the existence of invariant manifolds on this variable, cannot be specified at this point, since they depend on the specific form of the potential $V (\phi)$.
For now, we will allow $\lambda$ to assume any real value.

Note that the hyperplanes $y = \pm x$ are not invariant manifolds, meaning that phase-space trajectories can cross them.
These trajectories should only be assigned physical meaning once they cross the hyperplanes $y = x$ or $y = -x$ and enter the region where $\Omega_\phi > 0$.
Consequently, the phantom DE cosmological model should be viewed as a phenomenological model that describes the late-time evolution of the Universe (for further details, see Ref.~\cite{Halder:2024gag}).

\begingroup
\setlength{\tabcolsep}{3.5pt}
\renewcommand{\arraystretch}{1.4}
\begin{table*}[t]
\centering
\begin{tabular}{c c c c c c c c}
\hline\hline
Critical points & Existence & $\Omega_\phi$ & $\Omega_\texttt{DM}$ & $w_\phi$ &  $w_\texttt{tot}$ & $q$ & Acceleration
\\
\hline
$C_{1}\left(0,0,0,\lambda\right)$ & Always & $0$ & $1$ & Indeterm. & $0$ & $\frac12$ & No
\\
$C_{2}\left(0,1,0,0\right)$ & Always  & $1$  & $0$ & $-1$ & $-1$ & $-1$ & Yes
\\
$C_{3}\left(0,y,1,\lambda\right)$ & $0\leq y \leq 1$  & $y^2$  & $1-y^2$ & $-1$ & $-y^2$ & $\frac{1-3y^2}{2}$ &  $\frac{1}{\sqrt3}<y\leq1$
\\
$C_{4}\left(x,\sqrt{1+x^2},1,\lambda\right)$ & Always  & $1$  & $0$ & $-1-2x^2$ & $-1-2x^2$ & $-1-3x^2$ &  Yes
\\
$C_{5}\left(-\frac{\overline{\lambda}}{\sqrt6}, \sqrt{1+\frac{\overline{\lambda}^2}{6}},0, \overline{\lambda}\right)$ & $f(\overline{\lambda})=0$ & $1$ & $0$  & $-1-\frac{\overline{\lambda}^2}{3}$ & $-1-\frac{\overline{\lambda}^2}{3}$ & $-1-\frac{\overline{\lambda}^2}{2}$ & Yes
\\
$C_{6}\left(-\frac{\overline{\lambda}}{\sqrt6}, \sqrt{1+\frac{\overline{\lambda}^2}{6}},\frac{3(\overline{\lambda}^2+6)}{3(\overline{\lambda}^2+6)-2\alpha\overline{\lambda}^2}, \overline{\lambda} \right)$ & $f(\overline{\lambda})=0$, $\alpha<0$ & $1$ & $0$  & $-1-\frac{\overline{\lambda}^2}{3}$ & $-1-\frac{\overline{\lambda}^2}{3}$ & $-1-\frac{\overline{\lambda}^2}{2}$ & Yes
\\
$C_{7}\left(x,\sqrt{1-x^2},\frac{3}{3-2\alpha x^2},0 \right)$ & 
    \begin{tabular}{c}
    $\overline{\lambda}=0$, $\alpha<0$, \\ $|x|< 1/\sqrt2$ 
    \end{tabular}
& $1-2x^2$ & $2x^2$  & $-\frac{1}{1-2x^2}$ & $-1$ & $-1$ & Yes
\\
\hline\hline
\end{tabular}
\caption{Phenomenological properties of the critical points of the dynamical system (\ref{DS-3}) for the coupled phantom DE cosmological model with an arbitrary potential $V$ and a constant dissipation coefficient $\Gamma$. The critical points $C_1$, $C_2$, $C_3$, and $C_4$ exist for any $V$. The critical points $C_5$, $C_6$, and $C_7$ only exist if the equation $f(\lambda)=0$ has one or more non-complex roots, collectively denoted by $\overline{\lambda}$. For $C_7$, this root is zero. The critical points $C_3$ (for $0<y<1$) and $C_7$ (for $0<|x|<1/\sqrt2$) correspond to scaling solutions.}
	\label{table_coupled_constant}
\end{table*}
\endgroup

From Eq.~(\ref{DS-3d}) it follows that the existence of critical points of the dynamical system requires $x=0$, $z=1$, or $f(\lambda)=0$.

For $x=0$, we obtain a one-dimensional set of non-isolated critical points $C_1$ (critical line), an isolated critical point $C_2$, and a two-dimensional set of non-isolated critical points  $C_3$ (critical surface), while $z=1$ yields another critical surface $C_4$.
We shall reiterate here that the existence of these critical points\footnote{Recall that, for simplicity, when convenient, we are using the expression “critical point” to designate also a higher-dimensional set of non-isolated critical points.} does not depend on the specific form of the potential $V(\phi)$.
Note also that two of these points, $C_3$ and $C_4$, did not exist in the uncoupled case; they appear here due to the presence of the direct coupling between DE and DM.

If equation $f (\lambda)=0$ has non-complex roots, collectively denoted by $\overline{\lambda}$, then there are two additional isolated critical points $C_5$ and $C_6$, and an additional critical line $C_7$.
Note that the latter corresponds to a vanishing root, $\overline{\lambda}=0$.
Here, again, the presence of the interaction term is responsible for the appearance of two new critical points, $C_6$ and $C_7$.

The phenomenological properties of the critical points $C_1$ to $C_7$ are shown in Table~\ref{table_coupled_constant}.

Contrary to the uncoupled case, the coupled phantom cosmological model with a constant dissipation coefficient has critical points which correspond to scaling solutions for which the ratio $\Omega_\texttt{DM}/\Omega_\phi$ is finite and nonzero, namely, the critical points $C_3$ (for $0<y<1$) and $C_7$ (for $0<|x|<1/\sqrt2$).
To determine whether these points represent the final state of the Universe's evolution, and, consequently, whether they can address the cosmic coincidence problem, we need to investigate their stability. Specifically, we should examine whether these critical points are attractors and for what values of the parameter $\alpha$ this is the case.

Let us start with the critical surface $C_3$.
We consider a point $C_3(0,y_c,1,\lambda_c)$ on this surface, where $0\leq y_c \leq 1$ and $\lambda_c\in\mathbb{R}$.
At this point, the Jacobian matrix of the dynamical system (\ref{DS-3}) has the eigenvalues $\Lambda_1=\alpha(y_c^2-1)$ and $\Lambda_{2,3,4}=0$.
Since more than two eigenvalues are zero, the linear theory does not suffice to assess the stability of this critical point, and consequently, we have to resort to alternative methods.\footnote{If only two eigenvalues were zero (those whose corresponding eigenvectors are tangent to the surface $C_3$), this surface would be normally hyperbolic, and linear theory would suffice to study its stability. In this scenario, the signs of the two nonzero eigenvalues would determine whether $C_3$ is an attractor, saddle, or repellor in the remaining directions.}
In \ref{Sec-appendix1}, using the center manifold theory, we establish that this critical point is not an attractor for $0\leq y_c<1$, $\lambda_c\in\mathbb{R}$, and any value of $\alpha$.
For $y_c=1$ and $\lambda_c\in\mathbb{R}$, on the contrary, this critical point is indeed an attractor for $\alpha>0$, but, in this case, $\Omega_\phi=1$ and $\Omega_\texttt{DM}=0$ (see Table~\ref{table_coupled_constant}); therefore, it does not correspond to a scaling solution, but rather to a final state entirely dominated by phantom DE.

We now turn to the critical line $C_7$.
Let us consider a point $C_7(x_c,\sqrt{1-x_c^2},3/(3-2\alpha x_c^2),0)$ on this line, where $0<|x_c|< 1/\sqrt2$.
Taking into account that $f(\overline{\lambda}=0)=0$, the Jacobian matrix of the dynamical system (\ref{DS-3}) has the eigenvalues
\begin{gather}
    \Lambda_1=0, \quad \Lambda_{2,3}=\frac{3\alpha x_c^2\left( 1\pm \sqrt{1-4x_c^2} \right)}{3-2\alpha x_c^2}, \nonumber 
\\ 
    \mbox{and} \quad 
    \Lambda_4=\frac{2\sqrt6\alpha x_c^3 f'(0)}{3-2\alpha x_c^2},
    \label{eigenvalues-C7}
\end{gather}
where a prime denotes a derivative with respect to $\lambda$.
Since only one of the eigenvalues is zero, the critical line $C_7$ is normally hyperbolic, which means that linear theory is sufficient to decide its stability simply by analyzing the signs of the three remaining eigenvalues.
Taking into account that $C_7$ exists only for negative values of the parameter $\alpha$, we immediately conclude that the real parts of $\Lambda_2$ and $\Lambda_3$ are always negative and $\Lambda_4$ is negative if the condition $x_c f'(0)>0$ is satisfied.
Consequently, the critical line $C_7$ is an attractor for $\alpha<0$, $x_c f'(0)>0$, and $0<|x_c|< 1/\sqrt2$.
Furthermore, because $q=-1$ (see Table~\ref{table_coupled_constant}), this attractor corresponds to an accelerated scaling solution.

Note that $\alpha<0$ means that the dissipation coefficient $\Gamma$ is negative, which in turn implies that the interaction term $Q$ is also negative.
From Eq.~(\ref{dotrho}), it then follows that energy is continuously transferred from the phantom DE scalar field to the DM fluid, which explains why asymptotically $\Omega_\texttt{DM}=0$ does not vanish.

\begin{figure}[t]
\centering
\includegraphics[width=0.45\textwidth]{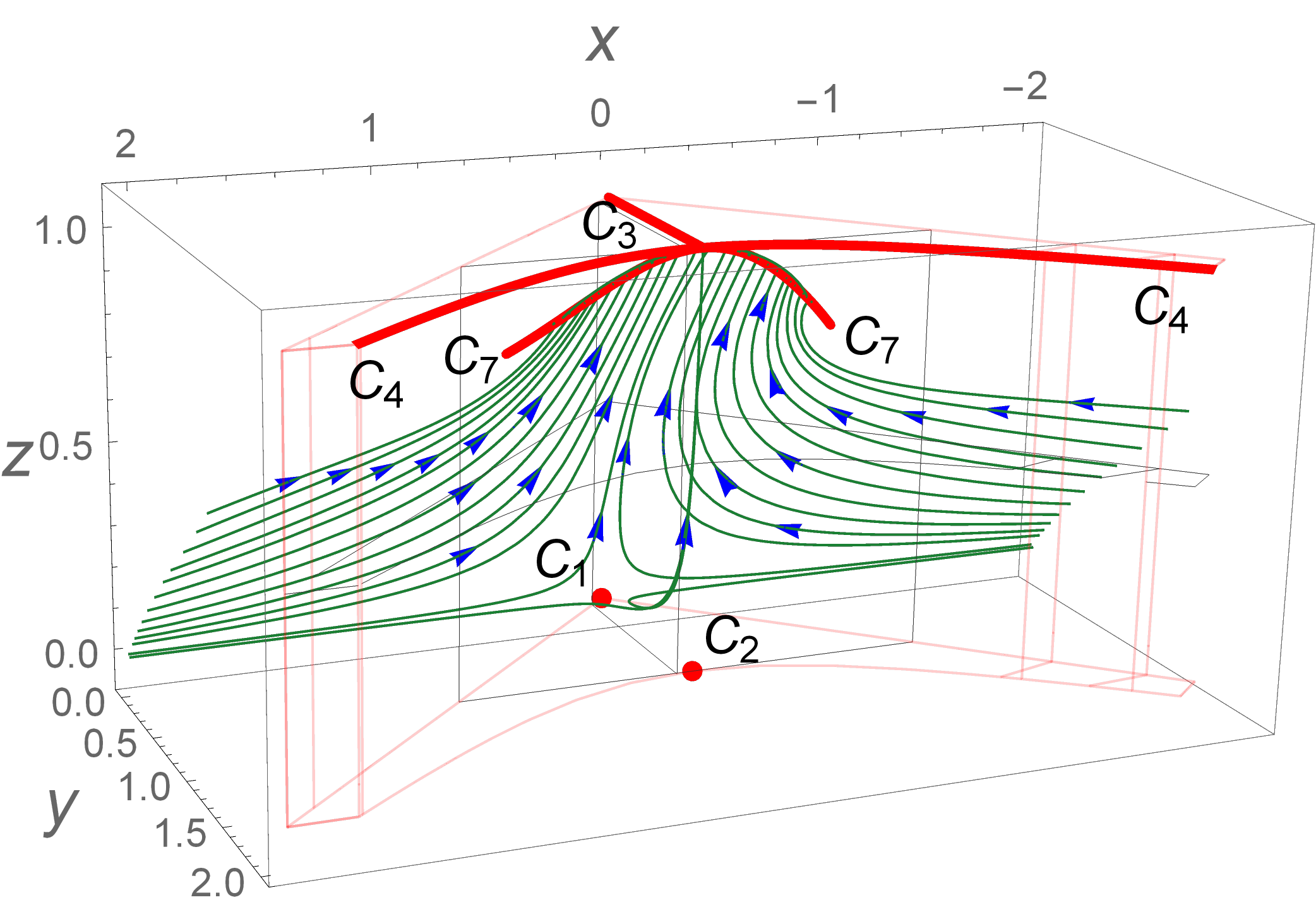}
\caption{Phase portrait of the three-dimensional invariant manifold $\lambda=0$ for $\alpha=-1$. This manifold fully contains the critical line $C_7$, which is an attractor for any value of $x$ ($|x| < 1/\sqrt2$).}
    \label{fig-3D_phase_space}
\end{figure}

If the function $f(\lambda)$ is such that it admits a zero root, then the hyperplane $\lambda=0$ is an invariant manifold, as can readily be seen from the dynamical system (\ref{DS-3}).
Furthermore, this hyperplane fully contains the critical line $C_7$. 
This circumstance allows us to visualize in a three-dimensional plot the trajectories\footnote{These trajectories are numerically computed from the three-dimensional dynamical system (\ref{DS-3a})--(\ref{DS-3c}) with $\lambda=0$, whose Jacobian has the eigenvalues $\Lambda_1$, $\Lambda_2$, and $\Lambda_3$ given by Eq.~(\ref{eigenvalues-C7}). For $\alpha <0$, the real parts of $\Lambda_{2}$ and $\Lambda_{3}$ are both negative, implying that $C_7$ is an attractor in the invariant manifold $\lambda=0$ for any value of $x$ ($|x| < 1/\sqrt2$).} that, coming from infinity, converge to this critical line (see Fig.~\ref{fig-3D_phase_space}).

Cosmological observations reveal that the present era of accelerated expansion was preceded by an era of matter-domination, long enough to allow the formation of large-scale structure. 
In our coupled phantom DE model, this period of matter domination occurs in the vicinity of $C_1$ (see Table~\ref{table_coupled_constant}).
Note that the trajectories that pass close to $C_1$ converge to points in $C_7$ for which $x\approx0$, implying that these solutions are asymptotically dominated by phantom dark energy (see Table~\ref{table_coupled_constant}).
On the other hand, trajectories that end at $C_7$ in points with $0\ll|x|<1/\sqrt2$, which correspond to scaling solutions suitable to solve the cosmic coincidence problem, do not pass near $C_1$, implying that the final stage of evolution is not preceded by a matter-dominated era. 

Combining these results with the ones obtained for the critical surface $C_3$, we conclude that the coupled phantom cosmological model with a constant dissipation coefficient $\Gamma$ and an arbitrary potential $V(\phi)$ admits accelerated scaling attractors.
However, these solutions have a limitation: the final stage of evolution is not preceded by a sufficiently long matter-dominated era.

The phase space associated with the dynamical system 
(\ref{DS-3}) is not bounded in the $x$, $y$, and $\lambda$ directions. Consequently, the existence of critical points at infinity cannot be ruled out. The asymptotic structure of the phase space and the stability of these putative points are investigated in~\ref{Sec-appendix3}. We find out that there are two critical points at infinity, both exhibiting unstable behavior. Consequently, the interaction term $Q$, given by Eq.~(\ref{interaction}), with a constant dissipation coefficient effectively avoids Big-Rip singularities.

\subsection{Variable dissipation coefficient}

We now consider the dissipation coefficient $\Gamma$ in the interaction term (\ref{interaction}) to depend on the DM energy density $\rho_\texttt{DM}$.
Such an interaction term between DE and DM was considered in recent years in the context of a steep-potential quintessence inflationary model ~\cite{Lima:2019,Das:2023}.
More specifically, aiming to unify early and late stages of the Universe's evolution through dissipative effects, the energy exchange between the quintessential scalar field and dark matter at late times was considered to be mediated by an interaction term with $\Gamma=M^2/\rho_\texttt{DM}^{1/4}$.
Here, we generalize this dissipation coefficient, choosing it to be of the form  
\begin{equation}
 \Gamma=\frac{M^{4n+1}}{\rho_\texttt{DM}^n},
 \label{Gamma-variable}
\end{equation}
where $M$ is a constant with the dimension of mass and $n>0$ ($n=0$ corresponds to the case of a constant dissipation coefficient analyzed in the previous subsection).

As in the case of constant $\Gamma$, the interaction term $Q$ between DE and DM cannot be expressed as a function of the variables $x$, $y$, and $\lambda$, defined in Eqs.~(\ref{x-and-y}) and (\ref{lambda}). An additional variable $z$ is needed to close the dynamical system  (\ref{DS-2}), which we choose to be given again by Eq.~(\ref{z-variable}).
Recall that such a choice, which compactifies the phase space in the $z$ direction, leads to a divergent interaction term in the evolution equation for $x$ (see discussion following Eq.~(\ref{z-variable})).
This divergence is removed by choosing a new “time” variable $\sigma$,
given by
\begin{equation}
\frac{d\sigma}{dt} = \frac{H}{(1-z)^{2n+1}}.
\end{equation}

In the variables $x$, $y$, $z$, $\lambda$, and $\sigma$, the evolution equations (\ref{friedmann})--(\ref{dotrho}) for the coupled phantom DE cosmological model with a variable dissipation coefficient give rise to the four-dimensional dynamical system
\begin{subequations} \label{DS-5}
\begin{align}
x_\sigma & = \left[ - \frac{\sqrt{6}}{2} \lambda y^2 - \frac{3}{2} x (1+x^2+y^2) \right] (1-z)^{2n+1} \nonumber  
\\ 
& \hspace{3mm} - \beta x (1+x^2 - y^2)^{n+1} z^{2n+1},  \label{DS-5a}
\\
y_\sigma &= \left[- \frac{\sqrt{6}}{2} \lambda x + \frac32 (1-x^2 -y^2) \right] y (1-z)^{2n+1}, \label{DS-5b}
\\
z_\sigma &= \Bigg[\frac{3}{2} (1 + x^2 +y^2) (1-z)^{2n+1} \nonumber 
\\ 
& \hspace{3mm} + 2 \beta x^2 (1+x^2 -y^2)^{n} z^{2n+1} \Bigg] z (1-z), \label{DS-5c}
\\
\lambda_\sigma & = -\sqrt6 f(\lambda) x (1-z)^{2n+1}, \label{DS-5d}
\end{align}
\end{subequations}
where $\beta \equiv \kappa^{2n} M^{4n+1} /(3^n H_{\ast}^{2n+1})$ is a dimensionless coupling parameter, taken to be nonzero, and we have assumed again that $\lambda(\phi)=-V_\phi/(\kappa V)$ is invertible and $f$, defined in Eq.~(\ref{f-lambda}), is a function of $\lambda$ only.
For $n=0$, the above dynamical system reduces to the one studied in the previous subsection (see Eq.~(\ref{DS-3})).

The parameters $\Omega_\texttt{DM}$, $\Omega_\phi$, $w_\phi$, $w_\texttt{tot}$, and $q$ do not depend on the variable $z$ and, therefore, are given by the same expressions as in the uncoupled model and the coupled model with constant dissipation term  (see Eq.~(\ref{omega-dm}), (\ref{q}), (\ref{omega-phi})--(\ref{w-tot})).

From the dynamical system (\ref{DS-5}) and the evolution equation for the DM density parameter,
\begin{equation}
    \Omega_{\texttt{DM},\sigma}= - \Omega_\texttt{DM} \left[ 3(x^2+y^2)(1-z)^{2n+1} +2\beta x^2  \Omega_\texttt{DM}^{n} z^{2n+1} \right],
\end{equation}
it follows that the hyperplanes $y=0$, $z=0$, and $z=1$ and the hypersurfaces $y=\pm\sqrt{1+x^2}$ ($\Omega_\texttt{DM}=0$) are invariant manifolds. 
Taking into account that $\Omega_{\texttt{DM}}\geq0$ and restricting ourselves to non-contracting cosmologies, the phase space is confined to the region $\pm x< y \leq \sqrt{1+x^2}$, $0\leq z \leq 1$  ($x\in\mathbb{R}$), where we have also excluded the regions $x^2\geq y^2$, in which $w_\phi$ diverges and $\Omega_\phi$ becomes negative (see discussion after Eqs.~(\ref{w-tot}) and (\ref{OmegaDM-coupled constant})). 
Since we are assuming the potential $V(\phi)$ to be arbitrary, the allowed values of $\lambda$ and the existence of invariant manifolds in this variable cannot be specified at this point. Therefore, we will allow $\lambda$ to take any real value.

According to Eq.~(\ref{DS-5d}), the existence of critical points of the dynamical system requires $x=0$, $z=1$, or $f(\lambda)=0$.

For $x=0$ and $z=1$, we obtain a critical line $E_1$, a critical point $E_2$, and two critical surfaces $E_3$ and $E_4$. Note that these critical points, which do not depend on the specific choice of the potential $V(\phi)$, are identical to those obtained within the coupled model with a constant dissipation coefficient $\Gamma$ (see Tables~\ref{table_coupled_constant} and \ref{table_coupled_variable}).
Therefore, changing from a constant to a variable $\Gamma$ does not affect the structure of this part of the phase space.

If equation $f (\lambda)=0$ has non-complex roots, collectively denoted by $\overline{\lambda}$, then there are one additional critical point $E_5$ and one additional critical line $E_6$, the latter corresponding to a vanishing root, $\overline{\lambda}=0$.
These points are identical to points $C_5$ and $C_7$ of the coupled model with a constant dissipation coefficient $\Gamma$ (see Tables~\ref{table_coupled_constant} and {\ref{table_coupled_variable}}). The critical point $C_6$ has no correspondence here.

The phenomenological properties of the critical points $E_1$ to $E_6$ are shown in Table~\ref{table_coupled_variable}.

\begingroup
\setlength{\tabcolsep}{5pt}
\renewcommand{\arraystretch}{1.4}
\begin{table*}[t]
\centering
\resizebox{1.0\textwidth}{!}{
\begin{tabular}{c c c c c c c c}
\hline\hline
Critical points & Existence & $\Omega_\phi$ & $\Omega_\texttt{DM}$ & $w_\phi$ &  $w_\texttt{tot}$ & $q$ & Acceleration
\\
\hline
$E_{1}\left(0,0,0,\lambda\right)$ & Always & $0$ & $1$ & Indeterm. & $0$ & $\frac12$ & No
\\
$E_{2}\left(0,1,0,0\right)$ & Always  & $1$  & $0$ & $-1$ & $-1$ & $-1$ & Yes
\\
$E_{3}\left(0,y,1,\lambda\right)$ & $0\leq y \leq 1$  & $y^2$  & $1-y^2$ & $-1$ & $-y^2$ & $\frac{1-3y^2}{2}$ &  $\frac{1}{\sqrt3}<y\leq1$
\\
$E_{4}\left(x,\sqrt{1+x^2},1,\lambda\right)$ & Always  & $1$  & $0$ & $-1-2x^2$ & $-1-2x^2$ & $-1-3x^2$ &  Yes
\\
$E_{5}\left(-\frac{\overline{\lambda}}{\sqrt6}, \sqrt{1+\frac{\overline{\lambda}^2}{6}},0, \overline{\lambda}\right)$ & $f(\overline{\lambda})=0$ & $1$ & $0$  & $-1-\frac{\overline{\lambda}^2}{3}$ & $-1-\frac{\overline{\lambda}^2}{3}$ & $-1-\frac{\overline{\lambda}^2}{2}$ & Yes
\\
$E_{6}\left(x,\sqrt{1-x^2},\frac{1}{1+\left(-\frac{2^{n+1}}{3} \beta x^{2n+2}\right)^{\frac{1}{2n+1}}},0 \right)$ & 
    \begin{tabular}{c}
    $\overline{\lambda}=0$, $\beta<0$, \\ $|x|< 1/\sqrt2$ 
    \end{tabular}
& $1-2x^2$ & $2x^2$  & $-\frac{1}{1-2x^2}$ & $-1$ & $-1$ & Yes
\\
\hline\hline
\end{tabular}
}
\caption{Phenomenological properties of the critical points of the dynamical system (\ref{DS-5}) for the coupled phantom DE cosmological model with an arbitrary potential $V$ and a variable dissipation coefficient $\Gamma$. The critical points $E_1$, $E_2$, $E_3$, and $E_4$ exist for any $V$. The critical points $E_5$ and $E_6$ only exist if the equation $f(\lambda)=0$ has one or more non-complex roots, collectively denoted by $\overline{\lambda}$. For $E_6$, this root is zero. The critical points $E_3$ (for $0<y<1$) and $E_6$ (for $0<|x|<1/\sqrt2$) correspond to scaling solutions.}
	\label{table_coupled_variable}
\end{table*}
\endgroup

Let us focus our attention on the critical surface $E_3$ and the critical line $E_6$, which correspond to scaling solutions for $0<y<1$ and $0<|x|<1/\sqrt2$, respectively.
To verify whether they are phase-space attractors and consequently correspond to a final state in the evolution of the Universe, we must study their stability.
For simplicity, this study is performed here for the case $n=1$, but it can be extended to other values of $n$.

Linear theory is not enough to assess the stability of $E_3$, since the Jacobian matrix has more than two zero eigenvalues.
Therefore, we resort to the center manifold theory to establish that this critical surface is an attractor only along the line $ y=1$, $\lambda\in\mathbb{R}$ (see \ref{Sec-appendix2}), implying that the corresponding final state is entirely dominated by phantom dark energy.

Regarding the critical line $E_6$, its stability properties can be established using linear theory alone  (see \ref{Sec-appendix2}), since only one of the eigenvalues of the Jacobian matrix is zero (the one whose corresponding eigenvector is tangent to the line $E_6$).
Two other eigenvalues have negative real parts, while the fourth is negative if $x_c f'(0)>0$, where $x_c$ is the $x$-coordinate of a specific point on the line $E_6$ and the prime denotes the derivative with respect to $\lambda$. 
Consequently, we conclude that the critical line $E_6$ is an attractor if the above condition is satisfied.
However, similarly to the case of the coupled model with a constant dissipation coefficient (see Fig.~\ref{fig-3D_phase_space} and discussion around it), phase-space trajectories that end at $E_6$ in points with an $x$-coordinate lying in the interval $0\ll |x| <1/\sqrt2$ (i.e., points corresponding to accelerated scaling solutions) do not pass close to the critical line $E_1$, which corresponds to a matter-dominated solution. 

The situation here is similar to that found in the coupled model with a constant dissipation coefficient. We indeed find a final stage of evolution corresponding to a scaling solution. Still, this stage is not preceded by a long enough matter-dominated era, as required by cosmological observations. However, despite this limitation, the appearance of scaling solutions in our coupled phantom DE cosmological model is certainly appealing.
Finding such solutions in phantom models has proven to be really challenging, with only a few examples available for specific potentials and interaction terms~\cite{Guo:2004xx,Chen:2008ft,Halder:2024aan,Paliathanasis:2024jxo}.
Therefore, the results of the present study, achieved for a generic potential, will certainly motivate future research in this direction.

The asymptotic structure of the phase space and the stability of the critical points located at infinity are investigated in~\ref{Sec-appendix3}. As for the case of a constant dissipative coefficient, we find that the interaction term $Q$, given by Eq.~(\ref{interaction}), with a variable dissipation coefficient effectively avoids Big-Rip singularities.

\section{Conclusions}
\label{Sec-conclusions}

Canonical scalar-field models have a rich historical background~\cite{Ratra:1987rm}, having produced many attractive outcomes in modern cosmology over the past three to four decades.
Specifically, after the discovery of the Universe's accelerating expansion, scalar field models came into the limelight of cosmology, the credit going to the challenges posed by the standard $\Lambda$CDM paradigm.

The dynamics of a scalar field heavily depend on its potential, thus making the selection of this potential a crucial issue in studying the evolution of the Universe.
A phantom scalar field~\cite{Caldwell:1999ew,Caldwell:2003vq} is an attractive proposal in this category, and in the literature, cosmological phantom models have been extensively studied considering different potentials.
However, as per the existing records in the literature, accelerating scaling attractors are very rare in phantom cosmology.

In this article, we studied the dynamics of a phantom DE cosmological model without specifying the potential of the phantom scalar field.
Our aim was to examine the existence of accelerating scaling attractors within this model.
For completeness, we have considered two distinct cases: when the phantom DE scalar field interacts with DM only gravitationally (uncoupled phantom scenario) and, secondly, when this field interacts directly with DM (coupled phantom scenario).
In the latter case, we have considered an interaction function $Q = \Gamma \dot{\phi}^2$ motivated by the warm inflationary paradigm, with the dissipation coefficient $\Gamma$ being either constant or variable.
Introducing dimensionless variables, we have written the evolution equations for the phantom DE cosmological model as a dynamical system, for both the uncoupled and coupled scenarios, and performed a stability analysis of the respective phase spaces by investigating the nature of the critical points.
The only condition we imposed on this analysis was the invertibility of $\lambda(\phi)$, defined by Eq.~(\ref{lambda}).
This is a minimal constraint on the potential, which allowed us to avoid any dependency of the autonomous systems on $\phi$.

The phenomenological properties of the critical points of the different dynamical systems are shown in Table~\ref{table_uncoupled} (uncoupled phantom model), Table~\ref{table_coupled_constant} (coupled phantom model with constant $\Gamma$), and Table~\ref{table_coupled_variable} (coupled phantom model with variable $\Gamma$).

The main results of our investigation are the following:
\begin{itemize}
\item In the uncoupled phantom  DE cosmological model, there are no accelerating scaling solutions, whatever the potential $V(\phi)$ we choose.

\item In the coupled phantom  DE cosmological model with a constant dissipative coefficient $\Gamma$, an accelerating scaling attractor is allowed ($C_7$ in Table~\ref{table_coupled_constant}, see also Fig.~\ref{fig-3D_phase_space}); however, this final state of the Universe's evolution is not preceded by the long enough matter-dominated era required by the observations.

\item In the coupled phantom DE cosmological model with a variable $\Gamma$ given by Eq.~(\ref{Gamma-variable}), an accelerating scaling attractor also exists ($E_6$ in Table~\ref{table_coupled_variable}), but,  similarly to the previous case, the preceding matter-dominated era is missing.
\end{itemize}

If the potential $V(\phi)$ is of the simple exponential type, given by Eq.~(\ref{expV}), $\lambda$ is not a dynamical variable, but simply a constant parameterizing the steepness of the exponential. In this case, which was investigated in Ref.~\cite{Halder:2024gag} for a constant dissipation coefficient $\Gamma$, the dynamical system has only one critical point corresponding to a scaling solution, analogous to $C_3$ of the present work. However, as here, this point is only an attractor when $\Omega_\phi=1$. When more general potentials are considered, which go beyond the simple exponential, the structure of the dynamical system becomes richer, admitting two new critical points, one of which, $C_7$, corresponds to an accelerated scaling attractor. Thus, the option for a simple exponential potential in coupled phantom cosmological models is too restrictive when it comes to the existence of accelerated scaling solutions capable of addressing the cosmic coincidence problem. Our study reveals that more complex potentials must be considered.

In a nutshell, if we assume that $\lambda(\phi)$, defined by Eq.~(\ref{lambda}), is invertible, then irrespective of the potential $V(\phi)$ chosen to the phantom DE scalar field, the uncoupled phantom model offers no accelerating scaling attractors and thus it can be argued that this model is not a viable solution for the cosmic coincidence problem.
On the contrary, the coupled phantom models are quite promising as they can provide accelerating scaling attractors and hence address the cosmic coincidence problem.
While the absence of a long enough matter-dominated era preceding the accelerated final stage of evolution is a limitation of both coupled scenarios  we have considered, the choice of a different interaction term between the dark components may allow us to overcome this limitation.
This investigation is left for future work.

\section*{Acknowledgements}
We thank the anonymous referee for useful comments and suggestions.
SH was supported by a senior research fellowship from the University Grants Commission (UGC), Govt.\ of India (NTA Ref. No: 201610019097).
SP and TS were partially supported by the Department of Science and Technology (DST), Govt.\ of India under the Scheme  “Fund for Improvement of S\&T Infrastructure (FIST)” (File No. SR/FST/MS-I/2019/41).
PS acknowledges support from Center for Research and Development in Mathematics and Applications (CIDMA) under the Portuguese Foundation for Science and Technology grants UID/04106/2025 (\url{https://doi.org/10.54499/UID/04106/2025}) and UID/PRR/04106/2025 (\url{https://doi.org/10.54499/UID/PRR/04106/2025}).

\appendix

\section{Stability of the critical surface $C_3$}
\label{Sec-appendix1}

To understand the stability features of the critical surface $C_3(0,y,1,\lambda)$, we focus on a specific point on it, namely $C_3(0,y_c,1,\lambda_c)$, with $y_c\in[0,1)$ and $\lambda_c$ being a real number.

At the chosen critical point $C_3(0,y_c,1,\lambda_c)$, we compute the Jacobian matrix of the system (\ref{DS-3}) as
\begin{equation}
 J_{C_3}=\begin{pmatrix}
     \alpha(y_c^2-1) & 0 & \frac{\sqrt6}{2}\lambda_c y_c^2 & 0
      \\
     0 & 0 & \frac32 y_c (y_c^2-1) & 0 \\
     0 & 0 & 0 & 0 \\
     0 & 0 & 0 & 0
    \end{pmatrix}.
   \label{jacA}
\end{equation}
This matrix has one non-zero eigenvalue ${\Lambda_1=\alpha(y_c^2-1)}$, while the remaining three are zero, $\Lambda_{2,3,4}=0$. The corresponding (generalized) eigenvectors are
\begin{equation}
    v_1=
    \begin{pmatrix}
    1 \\ 0 \\ 0 \\0
    \end{pmatrix},
    v_2=
   \begin{pmatrix}
    0 \\ 1 \\ 0 \\ 0
    \end{pmatrix},
     v_3=
    \begin{pmatrix}
    \frac{\sqrt6 \lambda_c y_c^2}{2\alpha(1-y_c^2)} \\ 0 \\ 1 \\0
    \end{pmatrix},
     v_4=
    \begin{pmatrix}
        0 \\ 0 \\ 0 \\ 1
    \end{pmatrix}.
    \label{gen_eig_vec}
\end{equation}

Due to the presence of three zero eigenvalues in the Jacobian matrix $J_{C_3}$, linear stability theory cannot fully characterize the stability of the point $C_3(0,y_c,1,\lambda_c)$. As a result, we turn to center manifold theory \cite{Carr:1982,Guckenheimer:1983,Bogoyavlensky:1985} for a more complete analysis.\footnote{Note that we have excluded from the analysis the value $y_c=1$, corresponding to the critical point $C_3(0,1,1,\lambda_c)$, since, for this value of $y_c$, all the eigenvalues of the Jacobian matrix~(\ref{jacA}) are zero and, consequently, we cannot resort to center manifold theory to study the stability of this critical point. However, as we will see below, the analysis of the critical points for which $0\leq y_c<1$ allows, based on continuity arguments, to draw conclusions about the stability of $C_3(0,1,1,\lambda_c)$.}

To facilitate the analysis, we translate the critical point $C_3(0,y_c,1,\lambda_c)$ to the origin via the change of variables 
\begin{equation}
    u=x, \quad v=y-y_c, \quad w=z-1, \quad r=\lambda-\lambda_c.
\end{equation}
In terms of these variables, the dynamical system (\ref{DS-3}) takes the form
\begin{subequations}
\begin{align}
 u_\tau & = -\alpha (1-y_c^2)u + \frac{\sqrt6}{2} \lambda_c y_c^2 w + f_1(u,v,w,r),
\\ 
v_\tau & = -\frac32 y_c (1-y_c^2)w + f_2(u,v,w,r),
\\
 w_\tau & = f_3(u,v,w,r), \\
 r_\tau &=f_4(u,v,w,r),
\end{align}
\end{subequations}
where $f_i=\mathcal{O}(u^2,v^2,w^2,r^2,uv,uw,ur,vw,vr,wr)$, $i=1,2,3,4$.

Performing a final change of variables
\begin{equation}
 \begin{pmatrix}
  u\\v\\w\\r
 \end{pmatrix}
  = P
 \begin{pmatrix}
 U\\V\\W\\R
 \end{pmatrix},
\end{equation}
where $P$ is a matrix whose columns are the eigenvectors given by Eq.~(\ref{gen_eig_vec}), the system simplifies to
\begin{subequations}
\begin{align}
 U_\tau & = -\alpha (1-y_c^2)U + F_1(U,V,W,R),
\\
 V_\tau & = -\frac32 y_c (1-y_c^2)W + F_2(U,V,W,R),
\\
 W_\tau & = F_3(U,V,W,R), \\
 R_\tau &=F_4(U,V,W,R),
\end{align}
\end{subequations}
with $F_i=\mathcal{O}(U^2,V^2,W^2,R^2,U\!V,U\!W,U\!R,V\!W,V\!R,W\!R)$, $i=1,2,3,4$.

Note that in the $U$ direction, trajectories approach the critical point if $\alpha>0$, and diverge if $\alpha<0$. We now proceed to determine the center manifold $U=h(V,W,R)$ and the flow on it.

To compute the center manifold, we solve the following partial differential equation
\begin{align}
 & \frac{\partial h}{\partial V} \left[
 -\frac32 y_c (1-y_c^2) W
 +F_2\Big(h(V,W,R),V,W,R\Big) \right] \nonumber
\\
 &\hspace{5mm} +\frac{\partial h}{\partial W} F_3\Big(h(V,W,R),V,W,R\Big) \nonumber \\
 &\hspace{5mm}+\frac{\partial h}{\partial R} F_4\Big(h(V,W,R),V,W,R\Big)  \nonumber
\\
 &\hspace{5mm}+\alpha (1-y_c^2)h(V,W,R) \nonumber \\ &\hspace{5mm}-F_1\Big(h(V,W,R),V,W,R\Big)=0.
 \label{CMT-hA}
\end{align}
Here, $h(V,W,R)$ is defined near the origin, and it satisfies $h(0,0,0)=0$, $\nabla h(0,0,0)=0$, ensuring tangency with the center subspace at the critical point.

We seek a solution to the above equation in the form 
\begin{equation}
    h(V,W,R) = \sum_{j=2}^{m} \sum_{i=0}^{j} \sum_{k=0}^{i}
    a_{k,i-k,j-i} V^k W^{i-k} R^{j-i}
\end{equation}
where $a_{ijk}$ are constants and $m\geq2$.
By substituting this expression into Eq.~(\ref{CMT-hA}) and keeping only the second-order terms, the center manifold $U$ is approximated by
\begin{align}
 U&=\frac{\sqrt6 \lambda_c y_c^2[3-\alpha(1-y_c^2)]}{2\alpha^2(1-y_c^2)^2} W^2+\frac{\sqrt6\lambda_c y_c}{\alpha(1-y_c^2)^2} V W \nonumber \\
 &+\frac{\sqrt{6}y_c^2}{2\alpha(1-y_c^2)} W\! R.
\end{align}

Noting that all terms in the above expression vanish for $y_c=0$, we extend the center manifold approximation to third order in $V$, $W$ and $R$, resulting in
\begin{equation}
U=\frac{\sqrt6\lambda_c}{2\alpha}V^2W,
\end{equation}
for $y_c=0$.

The dynamics restricted to the center manifold are governed by the following system of differential equations
\begin{subequations}
\label{flow-V,W,R}
\begin{align}
V_\tau & =  -\frac32 y_c (1-y_c^2 )W,
\label{flow-V}
\\
 W_\tau & = \frac32 (1+y_c^2)W^2, \label{flow-W}\\
 R_\tau&=\frac{3\lambda_c f(\lambda_c)y_c^2}{\alpha(1-y_c^2)}W^2.\label{flow-R}
 \end{align}
\end{subequations}

Due to the vanishing of the leading-order terms in Eqs.~(\ref{flow-V}) and (\ref{flow-R}) for $y_c=0$, we proceed with a higher-order approximation in powers of $V$, $W$ and $R$.
The resulting system on the center manifold takes the form 
\begin{subequations}
\label{flow-V,W_zero}
\begin{align}
V_\tau & =  -\frac32 V W,
\label{flow-V_zero}
\\
 W_\tau & = \frac32 W^2,
\label{flow-W_zero}\\
R_\tau &=\frac{3\lambda_c f(\lambda_c)}{\alpha}V^2W^2, \label{flow-R_zero}
\end{align}
\end{subequations}
for $y_c=0$.

For $0<y_c<1$, taking into account that $W<0$ near the critical point, we find that both $V_\tau$ and $W_\tau$ are positive. Moreover, $R_\tau$ is positive when $\lambda_c f(\lambda_c)/\alpha>0$, and negative for $\lambda_c f(\lambda_c)/\alpha<0$.
This implies that the trajectories approach the critical point along the $W$-direction and drift towards increasing $V$ and either increasing or decreasing $R$, depending on the sign of $\lambda_c f(\lambda_c)/\alpha$.
Similarly, for $y_c=0$, taking into account that $V>0$ in the neighborhood of the critical point, the same qualitative behavior is observed. 

When expressed in terms of the original variables $x$, $y$, $z$ and $\lambda$, these findings show that, for $\alpha>0$, the trajectories that approach the critical surface $C_3(0,y,1,\lambda)$ drift along the $y$ and $\lambda$ directions towards the critical line $C_3(0,1,1,\lambda)$, which, therefore, is an attractor.
This line coincides with the line $C_4(0,1,1,\lambda)$ belonging to the critical surface $C_4(x,\sqrt{1+x^2},1,\lambda)$.

\section{Stability of the critical surface $E_3$ and the critical line $E_6$}
\label{Sec-appendix2}

For simplicity, the stability analysis of $E_3$ and $E_6$ is illustrated in the case $n=1$. This analysis can be straightforwardly adapted to other values of $n$, specifically integer ones; however, the resulting equations are lengthy and not particularly illuminating.

To gain insight into the stability behavior of the critical surface $E_3(0,y,1,\lambda)$, we investigate a specific point on it, namely $E_3(0,y_c,1,\lambda_c)$, where $y_c\in[0,1)$ and $\lambda_c\in \mathbb{R}$.

At this point, we evaluate the Jacobian matrix corresponding to the dynamical system (\ref{DS-5}) as
\begin{equation}
 J_{E_3}=\begin{pmatrix}
     -\beta(1-y_c^2)^2 & 0 & 0 & 0
      \\
     0 & 0 & 0 & 0 \\
     0 & 0 & 0 & 0 \\
     0 & 0 & 0 & 0
    \end{pmatrix}.
   \label{jacC}
\end{equation}
 The Jacobian matrix possesses a single non-zero eigenvalue, ${\Lambda_1=-\beta(1-y_c^2)^2}$, while the remaining three eigenvalues vanish, $\Lambda_{2,3,4}=0$.

The presence of three zero eigenvalues in $J_{E_3}$, prevents a complete stability classification of $E_3(0,y_c,1,\lambda_c)$ via linear stability methods.
Consequently, we turn to the center manifold theory \cite{Carr:1982,Guckenheimer:1983,Bogoyavlensky:1985} for further analysis.

To simplify the analysis, we relocate the critical point $E_3(0,y_c,1,\lambda_c)$ to the origin by applying the coordinate transformation 
\begin{equation}
    u=x, \quad v=y-y_c, \quad w=z-1, \quad r=\lambda-\lambda_c.
\end{equation}
With this change of variables, the system (\ref{DS-5}) takes the form
\begin{subequations}
\begin{align}
 u_\sigma & = -\beta (1-y_c^2)^2u + g_1(u,v,w,r),
\\ 
v_\sigma & =  g_2(u,v,w,r),
\\
 w_\sigma & = g_3(u,v,w,r), \\
 r_\sigma &=g_4(u,v,w,r),
\end{align}
\end{subequations}
where $g_i=\mathcal{O}(u^2,v^2,w^2,r^2,uv,uw,ur,vw,vr,wr)$, $i=1,2,3,4$.

Note that along the $u$ direction, trajectories approach the critical point when $\beta>0$, and diverge when $\beta<0$.

We now proceed to determine the center manifold $u=h(v,w,r)$, which is determined by solving the partial differential equation
\begin{align}
 & \frac{\partial h}{\partial v} g_2 \left( h(v,w,r),v,w,r \right) + \frac{\partial h}{\partial w} g_3 \left( h(v,w,r),v,w,r \right) \nonumber \\ 
 &\hspace{5mm} + \frac{\partial h}{\partial r} g_4 \left( h(v,w,r),v,w,r \right) + \beta (1-y_c^2)^2 h(v,w,r) \nonumber \\ 
 &\hspace{5mm}-g_1\Big(h(v,w,r),v,w,r\Big)=0.
 \label{CMT-hC}
\end{align}
The function $h(v,w,r)$, defined near the origin, satisfies the conditions $h(0,0,0)=0$, $\nabla h(0,0,0)=0$, ensuring that the manifold is tangent to the center subspace at the critical point.

To approximate the center manifold, we assume a solution of the form
\begin{equation}
    h(v,w,r) = \sum_{j=2}^{m} \sum_{i=0}^{j} \sum_{k=0}^{i}
    a_{k,i-k,j-i} v^k w^{i-k} r^{j-i}
\end{equation}
with constants $a_{ijk}$ and $m\geq2$.
Substituting this expression into Eq.~(\ref{CMT-hC}) and collecting terms up to sixth order yields for the center manifold
\begin{widetext}
\begin{align}
 u=&
 \frac{\sqrt6 \lambda_c y_c^2}{2\beta(1-y_c^2)^2} w^3
 +\frac{\sqrt{6}y_c^2}{2\beta(1-y_c^2)^2} w^3 r
 -\frac{3\sqrt6\lambda_c y_c^2}{2\beta(1-y_c^2)^2} w^4
 +\frac{\sqrt{6}y_c \lambda_c(1+y_c^2)}{\beta(1-y_c^2)^3} v\, w^3
 -\frac{3\sqrt6 y_c^2}{2\beta(1-y_c^2)^2} w^4 r \nonumber
\\
 &+ \frac{\sqrt{6}y_c (1+y_c^2)}{\beta(1-y_c^2)^3} v \, w^3 r 
 +\frac{3\sqrt6\lambda_c y_c^2}{\beta(1-y_c^2)^2} w^5
 -\frac{3\sqrt{6}\lambda_c y_c (1+y_c^2)}{\beta(1-y_c^2)^3} v \, w^4
 +\frac{\sqrt{6}\lambda_c (1+8y_c^2+3y_c^4)}{2\beta(1-y_c^2)^4} v^2 w^3 \nonumber
\\
 &+\frac{3\sqrt6 y_c^2}{\beta(1-y_c^2)^2} w^5 r
 -\frac{3\sqrt{6} y_c (1+y_c^2)}{\beta(1-y_c^2)^3} v \, w^4 r
 +\frac{\sqrt{6} (1+8y_c^2+3y_c^4)}{2\beta(1-y_c^2)^4} v^2 w^3 r
 -\frac{5\sqrt6 \lambda_c y_c^2}{\beta(1-y_c^2)^2} w^6 \nonumber
\\
 &+\frac{6\sqrt{6} \lambda_c y_c (1+y_c^2)}{\beta(1-y_c^2)^3} v \, w^5
 -\frac{3\sqrt{6} \lambda_c (1+8y_c^2+3y_c^4)}{2\beta(1-y_c^2)^4} v^2 w^4
 +\frac{2\sqrt{6} \lambda_c y_c (2+5y_c^2+y_c^4)}{\beta (1-y_c^2)^5} v^3 w^3.
 \label{cm45}
\end{align}
\end{widetext}

The reduced system on the center manifold takes the form 
\begin{subequations}
\label{flow1-V,W,R}
\begin{align}
 v_\sigma & =  -\frac32 y_c (1-y_c^2 )w^3,
\label{flow1-V}
\\
 w_\sigma & = \frac32 (1+y_c^2)w^4, \label{flow1-W} \\
 r_\sigma&=\frac{3\lambda_c f(\lambda_c)y_c^2}{\beta(1-y_c^2)^2} w^6.\label{flow1-R}
 \end{align}
\end{subequations}
Note that the expression (\ref{cm45}) for the center manifold had to be computed up to sixth order in powers of $v$, $w$, and $r$ to obtain a non-trivial result for the flow in the $r$ direction.

For $0<y_c<1$, taking into account that $w<0$ near the critical point, we find that both $v_\sigma$ and $w_\sigma$ are positive.
Moreover, $r_\sigma$ is positive when $\lambda_c f(\lambda_c)/\beta>0$ and negative for $\lambda_c f(\lambda_c)/\beta<0$.
Thus, the trajectories approach the critical point along the $w$-direction and drift toward increasing $v$, with $r$ either increasing or decreasing depending on the sign of $\lambda_c f(\lambda_c)/\beta$.

For $y_c=0$, the leading-order terms in Eqs.~(\ref{flow1-V}) and (\ref{flow1-R}) vanish, making it necessary to extend the expression (\ref{cm45}) for the center manifold up to eighth order in powers of $v$, $w$, and $r$, yielding
\begin{align}
 u=&
 \frac{\sqrt{6}\lambda_c}{2\beta} v^2 w^3 +\frac{\sqrt{6}}{2\beta} v^2 w^3 r
 -\frac{3\sqrt{6}\lambda_c}{2\beta} v^2 w^4
 -\frac{3\sqrt{6}}{2\beta} v^2 w^4 r
 \nonumber \\
 &+\frac{3\sqrt{6}\lambda_c}{\beta} v^2 w^5
 +\frac{\sqrt{6}\lambda_c}{\beta} v^4 w^3
 +\frac{3\sqrt{6}}{\beta} v^2 w^5 r \nonumber
\\
 &+\frac{\sqrt{6}}{\beta} v^4 w^3 r
 -\frac{5\sqrt{6}\lambda_c}{\beta} v^2 w^6
 -\frac{3\sqrt{6}\lambda_c}{\beta} v^4 w^4
 \label{cm68}
\end{align}
and
\begin{subequations}
\label{flow2}
\begin{align}
 v_\sigma & =  -\frac32 v\, w^3, \label{flow2-V}
\\
 w_\sigma & = \frac32 w^4, \label{flow2-W}
\\
 r_\sigma&=\frac{3\lambda_c f(\lambda_c)}{\beta} v^2 w^6. \label{flow2-R}
\end{align}
\end{subequations}

Taking into account that $v > 0$ and $w<0$ in the vicinity of the critical point, we conclude that both $v_\sigma$ and $w_\sigma$ are positive; $r_\sigma$ is positive when $\lambda_c f(\lambda_c)/\beta>0$ and negative for $\lambda_c f(\lambda_c)/\beta<0$. Thus, the trajectories behave similarly to the case $0<y_c<1$.

In the original variables $x$, $y$, $z$, and $\lambda$, our findings show that for $\beta>0$, trajectories approaching the critical surface $E_3(0,y,1,\lambda)$ drift along the $y$ and $\lambda$ directions towards the critical line $E_3(0,1,1,\lambda)$, making it an attractor.
Note that this line coincides with the line $E_4(0,1,1,\lambda)$, which belongs to the surface $E_4(x,\sqrt{1+x^2},1,\lambda)$.

We now turn our attention to the critical line $E_6$.
Consider a representative point $E_6\left(x_c,\sqrt{1-x_c^2},\frac{3}{3-\sqrt[3]{36\beta x_c^4}},0\right)$, with $0<|x_c|< 1/\sqrt2$.
Assuming $f(\overline{\lambda}=0)=0$, the Jacobian matrix of the system~(\ref{DS-5}) yields the following eigenvalues
\begin{gather}
    \Lambda_1=0, \quad \Lambda_{2,3}=\frac{108\beta x_c^4\left( 1\pm i\sqrt{1+x_c^2} \right)}{\left(3-\sqrt[3]{36\beta x_c^4}\right)^3}, \nonumber 
\\ 
    \mbox{and} \quad 
    \Lambda_4=\frac{36\sqrt6\beta x_c^5 f'(0)}{\left(3-\sqrt[3]{36\beta x_c^4}\right)^3},
    \label{eigenvalues-E6}
\end{gather}
where a prime denotes a derivative with respect to $\lambda$.
Since only one eigenvalue vanishes, the critical line $E_6$ is normally hyperbolic.
Hence, its stability can be deduced through linear stability analysis.
Noting that $E_6$ exists only for $\beta<0$, it follows that the real parts of $\Lambda_2$ and $\Lambda_3$ are always negative.
Furthermore, $\Lambda_4<0$ if the condition $x_c f'(0)>0$ holds.
Therefore, the critical line $E_6$ behaves as an attractor for $\beta<0$, $x_c f'(0)>0$, and $0<|x_c|< 1/\sqrt2$.
Since $q=-1$ (see Table~\ref{table_coupled_variable}), this attractor represents an accelerated scaling solution.

\section{Stability of the critical points lying at infinity}
\label{Sec-appendix3}

In this Appendix, we investigate the asymptotic structure of the phase space and the stability of critical points located at infinity, both for the case of a constant dissipation coefficient and for the case of a variable coefficient.
To that end we employ the Poincar\'{e} compactification technique 
\cite{perko2013differential,dumortier2006qualitative,meiss2007differential,Bahamonde:2017ize,Halder:2024aan}.

Let us start with the case of a constant dissipation coefficient.

For convenience, we express the dynamical system (\ref{DS-3}) in terms of its associated vector field $F=(P,Q,R,S)$, where the individual components are given by the right-hand side of Eqs.~(\ref{DS-3a})--(\ref{DS-3d}), respectively.
Motivated by the dynamical requirements and without affecting the results obtained in the previous sections, we choose $f(\lambda)$ as a polynomial admitting both a zero and a non-zero root.
Accordingly, we consider $f(\lambda)=\delta\lambda(\lambda-\overline{\lambda})$, with non-zero real constants $\delta$ and $\overline{\lambda}$, a form\footnote{ We choose $f(\lambda)$ to be a second-degree polynomial for simplicity and concreteness. Nevertheless, $f(\lambda)$ itself is not restricted to this specific form and may, in principle, be any function possessing at least one zero and one non-zero root. However, the present analysis is carried out only for the polynomial case.} supported by a wide class of scalar-field potentials studied in the literature \cite{Bahamonde:2017ize}.

To analyze the asymptotic dynamics, we first map the vector field $F$ onto the upper four-dimensional sphere $\mathbb{S}^{4+}$ through a central projection.
This is achieved by introducing the projective coordinates $(x_1,x_2,x_3,x_4,x_5)$ defined as $x_1=\frac{x}{\Delta}$, $x_2=\frac{y}{\Delta}$, $x_3=\frac{z}{\Delta}$, $x_4=\frac{\lambda}{\Delta}$, $x_5=\frac{1}{\Delta}$, where $\Delta=\sqrt{1+x^2+y^2+z^2+\lambda^2}$. 
The stability properties are then examined in five local coordinate charts $(V_i,\psi_i)$, $i=1,\dots,5$, with $V_i=\left\{(x_1,x_2,x_3,x_4,x_5)\in\mathbb{S}^4:x_i>0\right\}$.
Each chart is mapped to $\mathbb{R}^4$ via $\psi_i(x_1,x_2,x_3,x_4,x_5)=\left(\frac{x_j}{x_i},\frac{x_k}{x_i},\frac{x_l}{x_i},\frac{x_m}{x_i}\right)$, where the indices $j$, $k$, $l$, and $m$ label four mutually different coordinates, ordered as $j<k<l<m$, with the additional requirement that none of them coincides with the chart index $i$.

On the upper four sphere $\mathbb{S}^{4+}$, the hypersurface $-x^2+y^2=0$ is mapped to the condition $x_1^2=x_2^2$, which, together with the spherical constraint, leads to $2x_1^2+x_3^2+x_4^2+x_5^2=1$.
Likewise, the hypersurface $-x^2+y^2=1$ transforms into $-x_1^2+x_2^2=x_5^2$, leading to $2x_2^2+x_3^2+x_4^2=1$, together with $0\leq x_3\leq x_5\leq 1$ ($0\leq z\leq 1$).
The critical points at infinity are located on the equator of the four-sphere, characterized by $x_5=0$.
Consequently, for the phase-space region under consideration, these points lie on the equator and must satisfy $\Big\{(x_1,x_2,x_3,x_4,x_5)\in\mathbb{S}^4:x_1^2=x_2^2,x_2\geq 0,x_3=0,x_5=0\Big\}$.

Focusing on the first local chart, we introduce the coordinates $(\chi,\zeta,\xi,\eta)$ through the mapping $\psi_1(x_1,x_2,x_3,x_4,x_5)=(\chi,\zeta,\xi,\eta)$.
In terms of these variables, the autonomous system (\ref{DS-3}) is correspondingly transformed into a new dynamical system 
\begin{subequations}
\label{DS-inf1}
\begin{align}
\chi_\mu&=\frac{1}{2}\chi\eta\Bigl\{6\eta^3+\sqrt{6}\eta\xi(\chi^2-1)+2(\alpha-3)\eta^2\zeta \nonumber \\
&-\zeta(\chi^2-1)(\sqrt{6}\xi+2\alpha)\Bigr\}, \label{DS-inf1a}
\\
\zeta_\mu &=\frac{1}{2}\zeta\Bigl\{6\eta^4+\eta^2\left[3(2+\zeta^2)+\chi^2(6+\sqrt{6}\xi)\right]\nonumber\\
&+\zeta^2(3+3\chi^2-4\alpha)+(2\alpha-9)\eta^3\zeta \nonumber\\
&-\eta\zeta\left[9-6\alpha+\chi^2(9+2\alpha+\sqrt{6}\xi)\right]\Bigr\}, \label{DS-inf1b}
\\
\xi_\mu &=\frac{1}{2}\eta\xi\Bigl\{3\eta^3+\eta\left[3+\chi^2(3+\sqrt{6}\xi)\right]+(2\alpha-3)\eta^2\zeta\nonumber\\
&-\zeta\left[3-2\alpha+\chi^2(3+2\alpha+\sqrt{6}\xi)\right]\nonumber\\
&+2\sqrt{6}\delta (\eta-\zeta)(\overline{\lambda}\eta-\xi)\Bigr\},
\label{DS-inf1c}
\\
\eta_\mu &=\frac{1}{2}\eta^2\Bigl\{3\eta^3+\eta\left[3+\chi^2(3+\sqrt{6}\xi)\right]+(2\alpha-3)\eta^2\zeta\nonumber\\
&-\zeta\left[3-2\alpha+\chi^2(3+2\alpha+\sqrt{6}\xi)\right]\Bigr\}, \label{DS-inf1d}
\end{align}
\end{subequations}
where the ``time'' variable $\tau$ was rescaled, $d\tau=\eta^{4}d\mu$, in order to remove the divergence arising for $\eta=0$.

To investigate the structure of the phase space at infinity, we set $\eta=0$.
The dynamical system (\ref{DS-inf1}) reduces to $\chi_\tau=\xi_\tau=\eta_\tau=0$ and $\zeta_\tau=\zeta^3(3-4\alpha+3\chi^2)/2$, yielding, for our phase space region, a continuous family of critical points $(1,0,\xi,0)$, with $\xi\in\mathbb{R}$.
These points correspond, on the compactified phase space $\mathbb{S}^4$, to 
$A^{\infty}\Bigl\{\left(x_1,x_2,x_3,x_4,x_5\right)\in\mathbb{S}^4: x_1^2=x_2^2, x_1>0, x_2>0, x_3=0, x_5=0\Bigr\}$.
As all eigenvalues vanish at $(1,0,\xi,0)$, a linear analysis fails to determine stability.
Nevertheless, restricting the flow to the invariant manifold $\zeta=0$, one finds $\chi_\tau=3\eta^4\chi+\sqrt{6}\eta^2\chi\xi(\chi^2-1)/2$, which reduces to $\chi_\tau\simeq3\eta^4$ in the vicinity of $\chi=1$.
This implies that $\chi$ always increases, leading to the conclusion that the set $A^{\infty}$ is unstable.

In the second chart, we adopt the local coordinates $(\chi,\zeta,\xi,\eta)$ defined by the map $\psi_2:(x_1,x_2,x_3,x_4,x_5)\rightarrow(\chi,\zeta,\xi,\eta)$. With this choice, the autonomous system (\ref{DS-3}) becomes
\begin{subequations}
\label{DS-inf2}
\begin{align}
\chi_\mu&=-\frac{1}{2}\eta\Bigl\{6\eta^3\chi-\sqrt{6}\eta\xi(\chi^2-1)+2(\alpha-3)\eta^2\chi\zeta \nonumber
\\
&+\zeta(\chi^2-1)(\sqrt{6}\xi+2\alpha \chi)\Bigr\},\\
\zeta_\mu &=-\frac{1}{2}\zeta(\eta-\zeta)\Bigl\{3\eta^2\zeta-\eta(6+6\chi^2+\sqrt{6}\chi\xi) \nonumber
\\
&+\zeta\left[3+(3-4\alpha) \chi^2)\right]\Bigr\},
\\
\xi_\mu &=-\frac{1}{2}\eta\xi(\eta-\zeta)\Bigl\{3\eta^2-3-3\chi^2+\sqrt{6}(2\delta-1)\chi\xi \nonumber
\\
&-2\sqrt{6}\delta\overline{\lambda}\eta\chi\Bigr\},
\\
\eta_\mu &=-\frac{1}{2}\eta^2(\eta-\zeta)(3\eta^2-3-3\chi^2-\sqrt{6}\chi\xi).
\end{align}
\end{subequations}

For $\eta=0$, this system simplifies to $\chi_\tau=\xi_\tau=\eta_\tau=0$ and $\zeta_\tau=\zeta^3\left[(3-4\alpha)\chi^2+3\right]/2$, giving rise, within our phase space region, to the critical point $(\pm1,0,\xi,0)$, with $\xi\in\mathbb{R}$. 

On the compactified phase space $\mathbb{S}^4$, these points correspond to $B^{\infty}\Bigl\{\left(x_1,x_2,x_3,x_4,x_5\right)\in\mathbb{S}^4: x_1^2=x_2^2, x_2>0, x_3=0, x_5=0\Bigr\}$.
Since the linearized system has only zero eigenvalues, stability cannot be inferred from linear theory.
However, along the invariant surface $\zeta=0$, the evolution of $\chi$ near $\chi=\pm 1$ is governed by $\chi_\tau=\mp 3\eta^4$, which confirms the unstable character of $B^{\infty}$.
In addition, one can also check that $A^{\infty}\subset B^{\infty}$.

In the third local chart, the constraint $x_3>0$ is imposed.
However, the critical points at infinity are required to satisfy $\Big\{(x_1,x_2,x_3,x_4,x_5)\in\mathbb{S}^4:x_1^2=x_2^2,x_2\geq 0,x_3=0,x_5=0\Big\}$. 
Since this set is disjoint from the region $x_3>0$, no critical points at infinity exist in the third chart.

For the fourth chart, we adopt the map $\psi_4(x_1,x_2,x_3,x_4,x_5)=(\chi,\zeta,\xi,\eta)$.
The autonomous system (\ref{DS-3}) assumes the equivalent representation 
\begin{subequations}
\label{DS-inf3}
\begin{align}
\chi_\mu &=\frac{1}{2}\eta\Bigl\{-\eta(3\chi^3+\sqrt{6}\zeta^2+3\chi \zeta^2)+(3-2\alpha)\eta^2\chi\xi \nonumber
\\
&-3\eta^3\chi+\xi\left[\sqrt{6}\zeta^2+(3-2\alpha)\chi^3+(3+2\alpha)\chi \zeta^2\right] \nonumber \\
&-2\sqrt{6}\delta \chi^2(\eta-\xi)(\overline{\lambda}\eta-1)\Bigr\},
\\
\zeta_\mu &=\frac{1}{2}\eta\zeta(\eta-\xi)\Bigl\{3\eta^2-3\chi^2-3\zeta^2+\sqrt{6}(2\delta-1)\chi \nonumber
\\
&-2\sqrt{6}\delta\overline{\lambda}\eta\chi\Bigr\},
\\
\xi_\mu &=\frac{1}{2}\xi(\eta-\xi)\Bigl\{3\eta^3+3\eta(\chi^2+\zeta^2)-3\eta^2\xi-3\zeta^2\xi \nonumber \\
&+(4\alpha-3)\chi^2\xi-2\sqrt{6}\delta \eta\chi(\overline{\lambda}\eta-1)\Bigr\},
\\
\eta_\mu &=-\sqrt{6}\delta \eta^2\chi(\eta-\xi)(\overline{\lambda}\eta-1).
\end{align}
\end{subequations}

Setting $\eta=0$, the dynamical system reduces to $\chi_\tau=\zeta_\tau=\eta_\tau=0$ and $\xi_\tau=\xi^3\left[3\zeta^2+(3-4\alpha)\chi^2\right]/2$.
This yields a family of relevant critical points at infinity for the considered phase-space region, given by $(\chi,\zeta,0,0)$ with $\chi^2=\zeta^2$.
On the compactified phase space $\mathbb{S}^4$, these points correspond to $\Bigl\{\left(x_1,x_2,x_3,x_4,x_5\right)\in\mathbb{S}^4: x_1^2=x_2^2, x_2\geq 0, x_3=0,x_4>0 ,x_5=0\Bigr\}$, which can be decomposed into $\Bigl\{\left(x_1,x_2,x_3,x_4,x_5\right)\in\mathbb{S}^4: x_1^2=x_2^2, x_2> 0, x_3=0,x_4>0 ,x_5=0\Bigr\}\cup (0,0,0,1,0)$.
It is straightforward to verify that the first subset is contained in $B^{\infty}$.
Consequently, this chart admits an additional critical point at infinity, namely $C^{\infty}(0,0,0,1,0)$. 
Since the linearized system has only zero eigenvalues, stability cannot be inferred from linear theory.
However, restricting the flow to the invariant manifold $\xi=0$, one finds $\chi_\tau=-\sqrt{6}\zeta^2\eta^2/2$ in the vicinity of $\chi=0$. Since $\chi$ always decreases, the point $C^{\infty}$ is therefore unstable. 

The analysis in the fifth chart reproduces a dynamical structure analogous to that of the system (\ref{DS-3}), without introducing any further critical points at infinity.

Overall, our analysis demonstrates that the autonomous system (\ref{DS-3}) contains only two critical points at infinity, $B^{\infty}$ and $C^{\infty}$, both displaying unstable behavior.
Therefore, we conclude that the coupled phantom cosmological model with an interaction term $Q$ given by Eq.~(\ref{interaction}), with a constant dissipation coefficient, effectively avoids Big-Rip singularities.

In the case of a variable dissipation coefficient, a similar analysis in all respects reveals the existence of the same critical points, with the same stability properties, leading, consequently, to the same conclusion.
\bibliography{biblio}

\begin{thebibliography}{110}%
\makeatletter
\providecommand \@ifxundefined [1]{%
 \@ifx{#1\undefined}
}%
\providecommand \@ifnum [1]{%
 \ifnum #1\expandafter \@firstoftwo
 \else \expandafter \@secondoftwo
 \fi
}%
\providecommand \@ifx [1]{%
 \ifx #1\expandafter \@firstoftwo
 \else \expandafter \@secondoftwo
 \fi
}%
\providecommand \natexlab [1]{#1}%
\providecommand \enquote  [1]{``#1''}%
\providecommand \bibnamefont  [1]{#1}%
\providecommand \bibfnamefont [1]{#1}%
\providecommand \citenamefont [1]{#1}%
\providecommand \href@noop [0]{\@secondoftwo}%
\providecommand \href [0]{\begingroup \@sanitize@url \@href}%
\providecommand \@href[1]{\@@startlink{#1}\@@href}%
\providecommand \@@href[1]{\endgroup#1\@@endlink}%
\providecommand \@sanitize@url [0]{\catcode `\\12\catcode `\$12\catcode
  `\&12\catcode `\#12\catcode `\^12\catcode `\_12\catcode `\%12\relax}%
\providecommand \@@startlink[1]{}%
\providecommand \@@endlink[0]{}%
\providecommand \url  [0]{\begingroup\@sanitize@url \@url }%
\providecommand \@url [1]{\endgroup\@href {#1}{\urlprefix }}%
\providecommand \urlprefix  [0]{URL }%
\providecommand \Eprint [0]{\href }%
\providecommand \doibase [0]{https://doi.org/}%
\providecommand \selectlanguage [0]{\@gobble}%
\providecommand \bibinfo  [0]{\@secondoftwo}%
\providecommand \bibfield  [0]{\@secondoftwo}%
\providecommand \translation [1]{[#1]}%
\providecommand \BibitemOpen [0]{}%
\providecommand \bibitemStop [0]{}%
\providecommand \bibitemNoStop [0]{.\EOS\space}%
\providecommand \EOS [0]{\spacefactor3000\relax}%
\providecommand \BibitemShut  [1]{\csname bibitem#1\endcsname}%
\let\auto@bib@innerbib\@empty
\bibitem [{\citenamefont {Riess}\ \emph {et~al.}(1998)\citenamefont {Riess}
  \emph {et~al.}}]{SupernovaSearchTeam:1998fmf}%
  \BibitemOpen
  \bibfield  {author} {\bibinfo {author} {\bibfnamefont {A.~G.}\ \bibnamefont
  {Riess}} \emph {et~al.} (\bibinfo {collaboration} {Supernova Search Team}),\
  }\bibfield  {title} {\bibinfo {title} {{Observational Evidence from
  Supernovae for an Accelerating Universe and a Cosmological Constant}},\
  }\href {https://doi.org/10.1086/300499} {\bibfield  {journal} {\bibinfo
  {journal} {Astron. J.}\ }\textbf {\bibinfo {volume} {116}},\ \bibinfo {pages}
  {1009} (\bibinfo {year} {1998})},\ \Eprint
  {https://arxiv.org/abs/astro-ph/9805201} {arXiv:astro-ph/9805201}
  \BibitemShut {NoStop}%
\bibitem [{\citenamefont {Perlmutter}\ \emph {et~al.}(1999)\citenamefont
  {Perlmutter} \emph {et~al.}}]{SupernovaCosmologyProject:1998vns}%
  \BibitemOpen
  \bibfield  {author} {\bibinfo {author} {\bibfnamefont {S.}~\bibnamefont
  {Perlmutter}} \emph {et~al.} (\bibinfo {collaboration} {Supernova Cosmology
  Project}),\ }\bibfield  {title} {\bibinfo {title} {{Measurements of $\Omega$
  and $\Lambda$ from 42 High-Redshift Supernovae}},\ }\href
  {https://doi.org/10.1086/307221} {\bibfield  {journal} {\bibinfo  {journal}
  {Astrophys. J.}\ }\textbf {\bibinfo {volume} {517}},\ \bibinfo {pages} {565}
  (\bibinfo {year} {1999})},\ \Eprint {https://arxiv.org/abs/astro-ph/9812133}
  {arXiv:astro-ph/9812133} \BibitemShut {NoStop}%
\bibitem [{\citenamefont {Copeland}\ \emph {et~al.}(2006)\citenamefont
  {Copeland}, \citenamefont {Sami},\ and\ \citenamefont
  {Tsujikawa}}]{Copeland:2006wr}%
  \BibitemOpen
  \bibfield  {author} {\bibinfo {author} {\bibfnamefont {E.~J.}\ \bibnamefont
  {Copeland}}, \bibinfo {author} {\bibfnamefont {M.}~\bibnamefont {Sami}},\
  and\ \bibinfo {author} {\bibfnamefont {S.}~\bibnamefont {Tsujikawa}},\
  }\bibfield  {title} {\bibinfo {title} {{Dynamics of dark energy}},\ }\href
  {https://doi.org/10.1142/S021827180600942X} {\bibfield  {journal} {\bibinfo
  {journal} {Int. J. Mod. Phys. D}\ }\textbf {\bibinfo {volume} {15}},\
  \bibinfo {pages} {1753} (\bibinfo {year} {2006})},\ \Eprint
  {https://arxiv.org/abs/hep-th/0603057} {arXiv:hep-th/0603057} \BibitemShut
  {NoStop}%
\bibitem [{\citenamefont {Bamba}\ \emph {et~al.}(2012)\citenamefont {Bamba},
  \citenamefont {Capozziello}, \citenamefont {Nojiri},\ and\ \citenamefont
  {Odintsov}}]{Bamba:2012cp}%
  \BibitemOpen
  \bibfield  {author} {\bibinfo {author} {\bibfnamefont {K.}~\bibnamefont
  {Bamba}}, \bibinfo {author} {\bibfnamefont {S.}~\bibnamefont {Capozziello}},
  \bibinfo {author} {\bibfnamefont {S.}~\bibnamefont {Nojiri}},\ and\ \bibinfo
  {author} {\bibfnamefont {S.~D.}\ \bibnamefont {Odintsov}},\ }\bibfield
  {title} {\bibinfo {title} {{Dark energy cosmology: the equivalent description
  via different theoretical models and cosmography tests}},\ }\href
  {https://doi.org/10.1007/s10509-012-1181-8} {\bibfield  {journal} {\bibinfo
  {journal} {Astrophys. Space Sci.}\ }\textbf {\bibinfo {volume} {342}},\
  \bibinfo {pages} {155} (\bibinfo {year} {2012})},\ \Eprint
  {https://arxiv.org/abs/1205.3421} {arXiv:1205.3421 [gr-qc]} \BibitemShut
  {NoStop}%
\bibitem [{\citenamefont {Nojiri}\ and\ \citenamefont
  {Odintsov}(2007)}]{Nojiri:2006ri}%
  \BibitemOpen
  \bibfield  {author} {\bibinfo {author} {\bibfnamefont {S.}~\bibnamefont
  {Nojiri}}\ and\ \bibinfo {author} {\bibfnamefont {S.~D.}\ \bibnamefont
  {Odintsov}},\ }\bibfield  {title} {\bibinfo {title} {{Introduction to
  modified gravity and gravitational alternative for dark energy}},\ }\href
  {https://doi.org/10.1142/S0219887807001928} {\bibfield  {journal} {\bibinfo
  {journal} {Int. J. Geom. Meth. Mod. Phys.}\ }\textbf {\bibinfo {volume}
  {4}},\ \bibinfo {pages} {115} (\bibinfo {year} {2007})},\ \Eprint
  {https://arxiv.org/abs/hep-th/0601213} {arXiv:hep-th/0601213} \BibitemShut
  {NoStop}%
\bibitem [{\citenamefont {De~Felice}\ and\ \citenamefont
  {Tsujikawa}(2010)}]{DeFelice:2010aj}%
  \BibitemOpen
  \bibfield  {author} {\bibinfo {author} {\bibfnamefont {A.}~\bibnamefont
  {De~Felice}}\ and\ \bibinfo {author} {\bibfnamefont {S.}~\bibnamefont
  {Tsujikawa}},\ }\bibfield  {title} {\bibinfo {title} {{$f(R)$ Theories}},\
  }\href {https://doi.org/10.12942/lrr-2010-3} {\bibfield  {journal} {\bibinfo
  {journal} {Living Rev. Rel.}\ }\textbf {\bibinfo {volume} {13}},\ \bibinfo
  {pages} {3} (\bibinfo {year} {2010})},\ \Eprint
  {https://arxiv.org/abs/1002.4928} {arXiv:1002.4928 [gr-qc]} \BibitemShut
  {NoStop}%
\bibitem [{\citenamefont {Clifton}\ \emph {et~al.}(2012)\citenamefont
  {Clifton}, \citenamefont {Ferreira}, \citenamefont {Padilla},\ and\
  \citenamefont {Skordis}}]{Clifton:2011jh}%
  \BibitemOpen
  \bibfield  {author} {\bibinfo {author} {\bibfnamefont {T.}~\bibnamefont
  {Clifton}}, \bibinfo {author} {\bibfnamefont {P.~G.}\ \bibnamefont
  {Ferreira}}, \bibinfo {author} {\bibfnamefont {A.}~\bibnamefont {Padilla}},\
  and\ \bibinfo {author} {\bibfnamefont {C.}~\bibnamefont {Skordis}},\
  }\bibfield  {title} {\bibinfo {title} {{Modified gravity and cosmology}},\
  }\href {https://doi.org/10.1016/j.physrep.2012.01.001} {\bibfield  {journal}
  {\bibinfo  {journal} {Phys. Rept.}\ }\textbf {\bibinfo {volume} {513}},\
  \bibinfo {pages} {1} (\bibinfo {year} {2012})},\ \Eprint
  {https://arxiv.org/abs/1106.2476} {arXiv:1106.2476 [astro-ph.CO]}
  \BibitemShut {NoStop}%
\bibitem [{\citenamefont {Cai}\ \emph {et~al.}(2016)\citenamefont {Cai},
  \citenamefont {Capozziello}, \citenamefont {De~Laurentis},\ and\
  \citenamefont {Saridakis}}]{Cai:2015emx}%
  \BibitemOpen
  \bibfield  {author} {\bibinfo {author} {\bibfnamefont {Y.-F.}\ \bibnamefont
  {Cai}}, \bibinfo {author} {\bibfnamefont {S.}~\bibnamefont {Capozziello}},
  \bibinfo {author} {\bibfnamefont {M.}~\bibnamefont {De~Laurentis}},\ and\
  \bibinfo {author} {\bibfnamefont {E.~N.}\ \bibnamefont {Saridakis}},\
  }\bibfield  {title} {\bibinfo {title} {{$f(T)$ teleparallel gravity and
  cosmology}},\ }\href {https://doi.org/10.1088/0034-4885/79/10/106901}
  {\bibfield  {journal} {\bibinfo  {journal} {Rept. Prog. Phys.}\ }\textbf
  {\bibinfo {volume} {79}},\ \bibinfo {pages} {106901} (\bibinfo {year}
  {2016})},\ \Eprint {https://arxiv.org/abs/1511.07586} {arXiv:1511.07586
  [gr-qc]} \BibitemShut {NoStop}%
\bibitem [{\citenamefont {Nojiri}\ \emph {et~al.}(2017)\citenamefont {Nojiri},
  \citenamefont {Odintsov},\ and\ \citenamefont {Oikonomou}}]{Nojiri:2017ncd}%
  \BibitemOpen
  \bibfield  {author} {\bibinfo {author} {\bibfnamefont {S.}~\bibnamefont
  {Nojiri}}, \bibinfo {author} {\bibfnamefont {S.~D.}\ \bibnamefont
  {Odintsov}},\ and\ \bibinfo {author} {\bibfnamefont {V.~K.}\ \bibnamefont
  {Oikonomou}},\ }\bibfield  {title} {\bibinfo {title} {{Modified gravity
  theories on a nutshell: Inflation, bounce and late-time evolution}},\ }\href
  {https://doi.org/10.1016/j.physrep.2017.06.001} {\bibfield  {journal}
  {\bibinfo  {journal} {Phys. Rept.}\ }\textbf {\bibinfo {volume} {692}},\
  \bibinfo {pages} {1} (\bibinfo {year} {2017})},\ \Eprint
  {https://arxiv.org/abs/1705.11098} {arXiv:1705.11098 [gr-qc]} \BibitemShut
  {NoStop}%
\bibitem [{\citenamefont {Bahamonde}\ \emph {et~al.}(2023)\citenamefont
  {Bahamonde}, \citenamefont {Dialektopoulos}, \citenamefont
  {Escamilla-Rivera}, \citenamefont {Farrugia}, \citenamefont {Gakis},
  \citenamefont {Hendry}, \citenamefont {Hohmann}, \citenamefont {Levi~Said},
  \citenamefont {Mifsud},\ and\ \citenamefont
  {Di~Valentino}}]{Bahamonde:2021gfp}%
  \BibitemOpen
  \bibfield  {author} {\bibinfo {author} {\bibfnamefont {S.}~\bibnamefont
  {Bahamonde}}, \bibinfo {author} {\bibfnamefont {K.~F.}\ \bibnamefont
  {Dialektopoulos}}, \bibinfo {author} {\bibfnamefont {C.}~\bibnamefont
  {Escamilla-Rivera}}, \bibinfo {author} {\bibfnamefont {G.}~\bibnamefont
  {Farrugia}}, \bibinfo {author} {\bibfnamefont {V.}~\bibnamefont {Gakis}},
  \bibinfo {author} {\bibfnamefont {M.}~\bibnamefont {Hendry}}, \bibinfo
  {author} {\bibfnamefont {M.}~\bibnamefont {Hohmann}}, \bibinfo {author}
  {\bibfnamefont {J.}~\bibnamefont {Levi~Said}}, \bibinfo {author}
  {\bibfnamefont {J.}~\bibnamefont {Mifsud}},\ and\ \bibinfo {author}
  {\bibfnamefont {E.}~\bibnamefont {Di~Valentino}},\ }\bibfield  {title}
  {\bibinfo {title} {{Teleparallel gravity: from theory to cosmology}},\ }\href
  {https://doi.org/10.1088/1361-6633/ac9cef} {\bibfield  {journal} {\bibinfo
  {journal} {Rept. Prog. Phys.}\ }\textbf {\bibinfo {volume} {86}},\ \bibinfo
  {pages} {026901} (\bibinfo {year} {2023})},\ \Eprint
  {https://arxiv.org/abs/2106.13793} {arXiv:2106.13793 [gr-qc]} \BibitemShut
  {NoStop}%
\bibitem [{\citenamefont {Weinberg}(1989)}]{Weinberg:1988cp}%
  \BibitemOpen
  \bibfield  {author} {\bibinfo {author} {\bibfnamefont {S.}~\bibnamefont
  {Weinberg}},\ }\bibfield  {title} {\bibinfo {title} {{The cosmological
  constant problem}},\ }\href {https://doi.org/10.1103/RevModPhys.61.1}
  {\bibfield  {journal} {\bibinfo  {journal} {Rev. Mod. Phys.}\ }\textbf
  {\bibinfo {volume} {61}},\ \bibinfo {pages} {1} (\bibinfo {year}
  {1989})}\BibitemShut {NoStop}%
\bibitem [{\citenamefont {Zlatev}\ \emph {et~al.}(1999)\citenamefont {Zlatev},
  \citenamefont {Wang},\ and\ \citenamefont {Steinhardt}}]{Zlatev:1998tr}%
  \BibitemOpen
  \bibfield  {author} {\bibinfo {author} {\bibfnamefont {I.}~\bibnamefont
  {Zlatev}}, \bibinfo {author} {\bibfnamefont {L.}~\bibnamefont {Wang}},\ and\
  \bibinfo {author} {\bibfnamefont {P.~J.}\ \bibnamefont {Steinhardt}},\
  }\bibfield  {title} {\bibinfo {title} {{Quintessence, Cosmic Coincidence, and
  the Cosmological Constant}},\ }\href
  {https://doi.org/10.1103/PhysRevLett.82.896} {\bibfield  {journal} {\bibinfo
  {journal} {Phys. Rev. Lett.}\ }\textbf {\bibinfo {volume} {82}},\ \bibinfo
  {pages} {896} (\bibinfo {year} {1999})},\ \Eprint
  {https://arxiv.org/abs/astro-ph/9807002} {arXiv:astro-ph/9807002}
  \BibitemShut {NoStop}%
\bibitem [{\citenamefont {Aghanim}\ \emph {et~al.}(2020)\citenamefont {Aghanim}
  \emph {et~al.}}]{Planck:2018vyg}%
  \BibitemOpen
  \bibfield  {author} {\bibinfo {author} {\bibfnamefont {N.}~\bibnamefont
  {Aghanim}} \emph {et~al.} (\bibinfo {collaboration} {Planck}),\ }\bibfield
  {title} {\bibinfo {title} {{Planck 2018 results. VI. Cosmological
  parameters}},\ }\href {https://doi.org/10.1051/0004-6361/201833910}
  {\bibfield  {journal} {\bibinfo  {journal} {Astron. Astrophys.}\ }\textbf
  {\bibinfo {volume} {641}},\ \bibinfo {pages} {A6} (\bibinfo {year} {2020})},\
  \bibinfo {note} {[Erratum: Astron.Astrophys. 652, C4 (2021)]},\ \Eprint
  {https://arxiv.org/abs/1807.06209} {arXiv:1807.06209 [astro-ph.CO]}
  \BibitemShut {NoStop}%
\bibitem [{\citenamefont {Riess}\ \emph {et~al.}(2022)\citenamefont {Riess}
  \emph {et~al.}}]{Riess:2021jrx}%
  \BibitemOpen
  \bibfield  {author} {\bibinfo {author} {\bibfnamefont {A.~G.}\ \bibnamefont
  {Riess}} \emph {et~al.},\ }\bibfield  {title} {\bibinfo {title} {{A
  Comprehensive Measurement of the Local Value of the Hubble Constant with 1 km
  s$^{-1}$ Mpc$^{-1}$ Uncertainty from the Hubble Space Telescope and the SH0ES
  Team}},\ }\href {https://doi.org/10.3847/2041-8213/ac5c5b} {\bibfield
  {journal} {\bibinfo  {journal} {Astrophys. J. Lett.}\ }\textbf {\bibinfo
  {volume} {934}},\ \bibinfo {pages} {L7} (\bibinfo {year} {2022})},\ \Eprint
  {https://arxiv.org/abs/2112.04510} {arXiv:2112.04510 [astro-ph.CO]}
  \BibitemShut {NoStop}%
\bibitem [{\citenamefont {Caldwell}(2002)}]{Caldwell:1999ew}%
  \BibitemOpen
  \bibfield  {author} {\bibinfo {author} {\bibfnamefont {R.~R.}\ \bibnamefont
  {Caldwell}},\ }\bibfield  {title} {\bibinfo {title} {{A phantom menace?
  Cosmological consequences of a dark energy component with super-negative
  equation of state}},\ }\href {https://doi.org/10.1016/S0370-2693(02)02589-3}
  {\bibfield  {journal} {\bibinfo  {journal} {Phys. Lett. B}\ }\textbf
  {\bibinfo {volume} {545}},\ \bibinfo {pages} {23} (\bibinfo {year} {2002})},\
  \Eprint {https://arxiv.org/abs/astro-ph/9908168} {arXiv:astro-ph/9908168}
  \BibitemShut {NoStop}%
\bibitem [{\citenamefont {Caldwell}\ \emph {et~al.}(2003)\citenamefont
  {Caldwell}, \citenamefont {Kamionkowski},\ and\ \citenamefont
  {Weinberg}}]{Caldwell:2003vq}%
  \BibitemOpen
  \bibfield  {author} {\bibinfo {author} {\bibfnamefont {R.~R.}\ \bibnamefont
  {Caldwell}}, \bibinfo {author} {\bibfnamefont {M.}~\bibnamefont
  {Kamionkowski}},\ and\ \bibinfo {author} {\bibfnamefont {N.~N.}\ \bibnamefont
  {Weinberg}},\ }\bibfield  {title} {\bibinfo {title} {{Phantom energy and
  cosmic doomsday}},\ }\href {https://doi.org/10.1103/PhysRevLett.91.071301}
  {\bibfield  {journal} {\bibinfo  {journal} {Phys. Rev. Lett.}\ }\textbf
  {\bibinfo {volume} {91}},\ \bibinfo {pages} {071301} (\bibinfo {year}
  {2003})},\ \Eprint {https://arxiv.org/abs/astro-ph/0302506}
  {arXiv:astro-ph/0302506} \BibitemShut {NoStop}%
\bibitem [{\citenamefont {Elizalde}\ \emph {et~al.}(2004)\citenamefont
  {Elizalde}, \citenamefont {Nojiri},\ and\ \citenamefont
  {Odintsov}}]{Elizalde:2004mq}%
  \BibitemOpen
  \bibfield  {author} {\bibinfo {author} {\bibfnamefont {E.}~\bibnamefont
  {Elizalde}}, \bibinfo {author} {\bibfnamefont {S.}~\bibnamefont {Nojiri}},\
  and\ \bibinfo {author} {\bibfnamefont {S.~D.}\ \bibnamefont {Odintsov}},\
  }\bibfield  {title} {\bibinfo {title} {{Late-time cosmology in (phantom)
  scalar-tensor theory: Dark energy and the cosmic speed-up}},\ }\href
  {https://doi.org/10.1103/PhysRevD.70.043539} {\bibfield  {journal} {\bibinfo
  {journal} {Phys. Rev. D}\ }\textbf {\bibinfo {volume} {70}},\ \bibinfo
  {pages} {043539} (\bibinfo {year} {2004})},\ \Eprint
  {https://arxiv.org/abs/hep-th/0405034} {arXiv:hep-th/0405034} \BibitemShut
  {NoStop}%
\bibitem [{\citenamefont {Amendola}(2000)}]{Amendola:1999er}%
  \BibitemOpen
  \bibfield  {author} {\bibinfo {author} {\bibfnamefont {L.}~\bibnamefont
  {Amendola}},\ }\bibfield  {title} {\bibinfo {title} {{Coupled
  quintessence}},\ }\href {https://doi.org/10.1103/PhysRevD.62.043511}
  {\bibfield  {journal} {\bibinfo  {journal} {Phys. Rev. D}\ }\textbf {\bibinfo
  {volume} {62}},\ \bibinfo {pages} {043511} (\bibinfo {year} {2000})},\
  \Eprint {https://arxiv.org/abs/astro-ph/9908023} {arXiv:astro-ph/9908023}
  \BibitemShut {NoStop}%
\bibitem [{\citenamefont {Cai}\ and\ \citenamefont {Wang}(2005)}]{Cai:2004dk}%
  \BibitemOpen
  \bibfield  {author} {\bibinfo {author} {\bibfnamefont {R.-G.}\ \bibnamefont
  {Cai}}\ and\ \bibinfo {author} {\bibfnamefont {A.}~\bibnamefont {Wang}},\
  }\bibfield  {title} {\bibinfo {title} {{Cosmology with interaction between
  phantom dark energy and dark matter and the coincidence problem}},\ }\href
  {https://doi.org/10.1088/1475-7516/2005/03/002} {\bibfield  {journal}
  {\bibinfo  {journal} {{{JCAP}}}\ }\textbf {\bibinfo {volume} {03}},\ \bibinfo
  {pages} {002} (\bibinfo {year} {{2005}})},\ \Eprint
  {https://arxiv.org/abs/hep-th/0411025} {arXiv:hep-th/0411025} \BibitemShut
  {NoStop}%
\bibitem [{\citenamefont {Huey}\ and\ \citenamefont
  {Wandelt}(2006)}]{Huey:2004qv}%
  \BibitemOpen
  \bibfield  {author} {\bibinfo {author} {\bibfnamefont {G.}~\bibnamefont
  {Huey}}\ and\ \bibinfo {author} {\bibfnamefont {B.~D.}\ \bibnamefont
  {Wandelt}},\ }\bibfield  {title} {\bibinfo {title} {{Interacting
  quintessence, the coincidence problem, and cosmic acceleration}},\ }\href
  {https://doi.org/10.1103/PhysRevD.74.023519} {\bibfield  {journal} {\bibinfo
  {journal} {Phys. Rev. D}\ }\textbf {\bibinfo {volume} {74}},\ \bibinfo
  {pages} {023519} (\bibinfo {year} {2006})},\ \Eprint
  {https://arxiv.org/abs/astro-ph/0407196} {arXiv:astro-ph/0407196}
  \BibitemShut {NoStop}%
\bibitem [{\citenamefont {Curbelo}\ \emph {et~al.}(2006)\citenamefont
  {Curbelo}, \citenamefont {Gonzalez}, \citenamefont {Leon},\ and\
  \citenamefont {Quiros}}]{Curbelo:2005dh}%
  \BibitemOpen
  \bibfield  {author} {\bibinfo {author} {\bibfnamefont {R.}~\bibnamefont
  {Curbelo}}, \bibinfo {author} {\bibfnamefont {T.}~\bibnamefont {Gonzalez}},
  \bibinfo {author} {\bibfnamefont {G.}~\bibnamefont {Leon}},\ and\ \bibinfo
  {author} {\bibfnamefont {I.}~\bibnamefont {Quiros}},\ }\bibfield  {title}
  {\bibinfo {title} {{Interacting phantom energy and avoidance of the big rip
  singularity}},\ }\href {https://doi.org/10.1088/0264-9381/23/5/010}
  {\bibfield  {journal} {\bibinfo  {journal} {Class. Quant. Grav.}\ }\textbf
  {\bibinfo {volume} {23}},\ \bibinfo {pages} {1585} (\bibinfo {year}
  {2006})},\ \Eprint {https://arxiv.org/abs/astro-ph/0502141}
  {arXiv:astro-ph/0502141} \BibitemShut {NoStop}%
\bibitem [{\citenamefont {Barrow}\ and\ \citenamefont
  {Clifton}(2006)}]{Barrow:2006hia}%
  \BibitemOpen
  \bibfield  {author} {\bibinfo {author} {\bibfnamefont {J.~D.}\ \bibnamefont
  {Barrow}}\ and\ \bibinfo {author} {\bibfnamefont {T.}~\bibnamefont
  {Clifton}},\ }\bibfield  {title} {\bibinfo {title} {{Cosmologies with energy
  exchange}},\ }\href {https://doi.org/10.1103/PhysRevD.73.103520} {\bibfield
  {journal} {\bibinfo  {journal} {Phys. Rev. D}\ }\textbf {\bibinfo {volume}
  {73}},\ \bibinfo {pages} {103520} (\bibinfo {year} {2006})},\ \Eprint
  {https://arxiv.org/abs/gr-qc/0604063} {arXiv:gr-qc/0604063} \BibitemShut
  {NoStop}%
\bibitem [{\citenamefont {del Campo}\ \emph {et~al.}(2008)\citenamefont {del
  Campo}, \citenamefont {Herrera},\ and\ \citenamefont
  {Pavon}}]{delCampo:2008sr}%
  \BibitemOpen
  \bibfield  {author} {\bibinfo {author} {\bibfnamefont {S.}~\bibnamefont {del
  Campo}}, \bibinfo {author} {\bibfnamefont {R.}~\bibnamefont {Herrera}},\ and\
  \bibinfo {author} {\bibfnamefont {D.}~\bibnamefont {Pavon}},\ }\bibfield
  {title} {\bibinfo {title} {{Toward a solution of the coincidence problem}},\
  }\href {https://doi.org/10.1103/PhysRevD.78.021302} {\bibfield  {journal}
  {\bibinfo  {journal} {Phys. Rev. D}\ }\textbf {\bibinfo {volume} {78}},\
  \bibinfo {pages} {021302} (\bibinfo {year} {2008})},\ \Eprint
  {https://arxiv.org/abs/0806.2116} {arXiv:0806.2116 [astro-ph]} \BibitemShut
  {NoStop}%
\bibitem [{\citenamefont {Valiviita}\ \emph {et~al.}(2008)\citenamefont
  {Valiviita}, \citenamefont {Majerotto},\ and\ \citenamefont
  {Maartens}}]{Valiviita:2008iv}%
  \BibitemOpen
  \bibfield  {author} {\bibinfo {author} {\bibfnamefont {J.}~\bibnamefont
  {Valiviita}}, \bibinfo {author} {\bibfnamefont {E.}~\bibnamefont
  {Majerotto}},\ and\ \bibinfo {author} {\bibfnamefont {R.}~\bibnamefont
  {Maartens}},\ }\bibfield  {title} {\bibinfo {title} {{Large-scale instability
  in interacting dark energy and dark matter fluids}},\ }\href
  {https://doi.org/10.1088/1475-7516/2008/07/020} {\bibfield  {journal}
  {\bibinfo  {journal} {{{JCAP}}}\ }\textbf {\bibinfo {volume} {07}},\ \bibinfo
  {pages} {020} (\bibinfo {year} {2008})},\ \Eprint
  {https://arxiv.org/abs/0804.0232} {arXiv:0804.0232 [astro-ph]} \BibitemShut
  {NoStop}%
\bibitem [{\citenamefont {del Campo}\ \emph {et~al.}(2009)\citenamefont {del
  Campo}, \citenamefont {Herrera},\ and\ \citenamefont
  {Pavon}}]{delCampo:2008jx}%
  \BibitemOpen
  \bibfield  {author} {\bibinfo {author} {\bibfnamefont {S.}~\bibnamefont {del
  Campo}}, \bibinfo {author} {\bibfnamefont {R.}~\bibnamefont {Herrera}},\ and\
  \bibinfo {author} {\bibfnamefont {D.}~\bibnamefont {Pavon}},\ }\bibfield
  {title} {\bibinfo {title} {{Interacting models may be key to solve the cosmic
  coincidence problem}},\ }\href
  {https://doi.org/10.1088/1475-7516/2009/01/020} {\bibfield  {journal}
  {\bibinfo  {journal} {{{JCAP}}}\ }\textbf {\bibinfo {volume} {01}},\ \bibinfo
  {pages} {020} (\bibinfo {year} {2009})},\ \Eprint
  {https://arxiv.org/abs/0812.2210} {arXiv:0812.2210 [gr-qc]} \BibitemShut
  {NoStop}%
\bibitem [{\citenamefont {Leon}\ and\ \citenamefont
  {Saridakis}(2010)}]{Leon:2009dt}%
  \BibitemOpen
  \bibfield  {author} {\bibinfo {author} {\bibfnamefont {G.}~\bibnamefont
  {Leon}}\ and\ \bibinfo {author} {\bibfnamefont {E.~N.}\ \bibnamefont
  {Saridakis}},\ }\bibfield  {title} {\bibinfo {title} {{Phantom dark energy
  with varying-mass dark matter particles: acceleration and cosmic coincidence
  problem}},\ }\href {https://doi.org/10.1016/j.physletb.2010.08.016}
  {\bibfield  {journal} {\bibinfo  {journal} {Phys. Lett. B}\ }\textbf
  {\bibinfo {volume} {693}},\ \bibinfo {pages} {1} (\bibinfo {year} {2010})},\
  \Eprint {https://arxiv.org/abs/0904.1577} {arXiv:0904.1577 [gr-qc]}
  \BibitemShut {NoStop}%
\bibitem [{\citenamefont {Clemson}\ \emph {et~al.}(2012)\citenamefont
  {Clemson}, \citenamefont {Koyama}, \citenamefont {Zhao}, \citenamefont
  {Maartens},\ and\ \citenamefont {Valiviita}}]{Clemson:2011an}%
  \BibitemOpen
  \bibfield  {author} {\bibinfo {author} {\bibfnamefont {T.}~\bibnamefont
  {Clemson}}, \bibinfo {author} {\bibfnamefont {K.}~\bibnamefont {Koyama}},
  \bibinfo {author} {\bibfnamefont {G.-B.}\ \bibnamefont {Zhao}}, \bibinfo
  {author} {\bibfnamefont {R.}~\bibnamefont {Maartens}},\ and\ \bibinfo
  {author} {\bibfnamefont {J.}~\bibnamefont {Valiviita}},\ }\bibfield  {title}
  {\bibinfo {title} {{Interacting Dark Energy -- constraints and
  degeneracies}},\ }\href {https://doi.org/10.1103/PhysRevD.85.043007}
  {\bibfield  {journal} {\bibinfo  {journal} {Phys. Rev. D}\ }\textbf {\bibinfo
  {volume} {85}},\ \bibinfo {pages} {043007} (\bibinfo {year} {2012})},\
  \Eprint {https://arxiv.org/abs/1109.6234} {arXiv:1109.6234 [astro-ph.CO]}
  \BibitemShut {NoStop}%
\bibitem [{\citenamefont {Li}\ and\ \citenamefont {Zhang}(2014)}]{Li:2013bya}%
  \BibitemOpen
  \bibfield  {author} {\bibinfo {author} {\bibfnamefont {Y.-H.}\ \bibnamefont
  {Li}}\ and\ \bibinfo {author} {\bibfnamefont {X.}~\bibnamefont {Zhang}},\
  }\bibfield  {title} {\bibinfo {title} {{Large-scale stable interacting dark
  energy model: Cosmological perturbations and observational constraints}},\
  }\href {https://doi.org/10.1103/PhysRevD.89.083009} {\bibfield  {journal}
  {\bibinfo  {journal} {Phys. Rev. D}\ }\textbf {\bibinfo {volume} {89}},\
  \bibinfo {pages} {083009} (\bibinfo {year} {2014})},\ \Eprint
  {https://arxiv.org/abs/1312.6328} {arXiv:1312.6328 [astro-ph.CO]}
  \BibitemShut {NoStop}%
\bibitem [{\citenamefont {Yang}\ and\ \citenamefont
  {Xu}(2014{\natexlab{a}})}]{Yang:2014gza}%
  \BibitemOpen
  \bibfield  {author} {\bibinfo {author} {\bibfnamefont {W.}~\bibnamefont
  {Yang}}\ and\ \bibinfo {author} {\bibfnamefont {L.}~\bibnamefont {Xu}},\
  }\bibfield  {title} {\bibinfo {title} {{Cosmological constraints on
  interacting dark energy with redshift-space distortion after Planck data}},\
  }\href {https://doi.org/10.1103/PhysRevD.89.083517} {\bibfield  {journal}
  {\bibinfo  {journal} {Phys. Rev. D}\ }\textbf {\bibinfo {volume} {89}},\
  \bibinfo {pages} {083517} (\bibinfo {year} {2014}{\natexlab{a}})},\ \Eprint
  {https://arxiv.org/abs/1401.1286} {arXiv:1401.1286 [astro-ph.CO]}
  \BibitemShut {NoStop}%
\bibitem [{\citenamefont {Yang}\ and\ \citenamefont
  {Xu}(2014{\natexlab{b}})}]{Yang:2014okp}%
  \BibitemOpen
  \bibfield  {author} {\bibinfo {author} {\bibfnamefont {W.}~\bibnamefont
  {Yang}}\ and\ \bibinfo {author} {\bibfnamefont {L.}~\bibnamefont {Xu}},\
  }\bibfield  {title} {\bibinfo {title} {{Testing coupled dark energy with
  large scale structure observation}},\ }\href
  {https://doi.org/10.1088/1475-7516/2014/08/034} {\bibfield  {journal}
  {\bibinfo  {journal} {{{JCAP}}}\ }\textbf {\bibinfo {volume} {08}},\ \bibinfo
  {pages} {034} (\bibinfo {year} {2014}{\natexlab{b}})},\ \Eprint
  {https://arxiv.org/abs/1401.5177} {arXiv:1401.5177 [astro-ph.CO]}
  \BibitemShut {NoStop}%
\bibitem [{\citenamefont {Salvatelli}\ \emph {et~al.}(2014)\citenamefont
  {Salvatelli}, \citenamefont {Said}, \citenamefont {Bruni}, \citenamefont
  {Melchiorri},\ and\ \citenamefont {Wands}}]{Salvatelli:2014zta}%
  \BibitemOpen
  \bibfield  {author} {\bibinfo {author} {\bibfnamefont {V.}~\bibnamefont
  {Salvatelli}}, \bibinfo {author} {\bibfnamefont {N.}~\bibnamefont {Said}},
  \bibinfo {author} {\bibfnamefont {M.}~\bibnamefont {Bruni}}, \bibinfo
  {author} {\bibfnamefont {A.}~\bibnamefont {Melchiorri}},\ and\ \bibinfo
  {author} {\bibfnamefont {D.}~\bibnamefont {Wands}},\ }\bibfield  {title}
  {\bibinfo {title} {{Indications of a late-time interaction in the dark
  sector}},\ }\href {https://doi.org/10.1103/PhysRevLett.113.181301} {\bibfield
   {journal} {\bibinfo  {journal} {Phys. Rev. Lett.}\ }\textbf {\bibinfo
  {volume} {113}},\ \bibinfo {pages} {181301} (\bibinfo {year} {2014})},\
  \Eprint {https://arxiv.org/abs/1406.7297} {arXiv:1406.7297 [astro-ph.CO]}
  \BibitemShut {NoStop}%
\bibitem [{\citenamefont {Duniya}\ \emph {et~al.}(2015)\citenamefont {Duniya},
  \citenamefont {Bertacca},\ and\ \citenamefont {Maartens}}]{Duniya:2015nva}%
  \BibitemOpen
  \bibfield  {author} {\bibinfo {author} {\bibfnamefont {D.~G.~A.}\
  \bibnamefont {Duniya}}, \bibinfo {author} {\bibfnamefont {D.}~\bibnamefont
  {Bertacca}},\ and\ \bibinfo {author} {\bibfnamefont {R.}~\bibnamefont
  {Maartens}},\ }\bibfield  {title} {\bibinfo {title} {{Probing the imprint of
  interacting dark energy on very large scales}},\ }\href
  {https://doi.org/10.1103/PhysRevD.91.063530} {\bibfield  {journal} {\bibinfo
  {journal} {Phys. Rev. D}\ }\textbf {\bibinfo {volume} {91}},\ \bibinfo
  {pages} {063530} (\bibinfo {year} {2015})},\ \Eprint
  {https://arxiv.org/abs/1502.06424} {arXiv:1502.06424 [astro-ph.CO]}
  \BibitemShut {NoStop}%
\bibitem [{\citenamefont {Nunes}\ \emph {et~al.}(2016)\citenamefont {Nunes},
  \citenamefont {Pan},\ and\ \citenamefont {Saridakis}}]{Nunes:2016dlj}%
  \BibitemOpen
  \bibfield  {author} {\bibinfo {author} {\bibfnamefont {R.~C.}\ \bibnamefont
  {Nunes}}, \bibinfo {author} {\bibfnamefont {S.}~\bibnamefont {Pan}},\ and\
  \bibinfo {author} {\bibfnamefont {E.~N.}\ \bibnamefont {Saridakis}},\
  }\bibfield  {title} {\bibinfo {title} {{New constraints on interacting dark
  energy from cosmic chronometers}},\ }\href
  {https://doi.org/10.1103/PhysRevD.94.023508} {\bibfield  {journal} {\bibinfo
  {journal} {Phys. Rev. D}\ }\textbf {\bibinfo {volume} {94}},\ \bibinfo
  {pages} {023508} (\bibinfo {year} {2016})},\ \Eprint
  {https://arxiv.org/abs/1605.01712} {arXiv:1605.01712 [astro-ph.CO]}
  \BibitemShut {NoStop}%
\bibitem [{\citenamefont {Yang}\ \emph {et~al.}(2016)\citenamefont {Yang},
  \citenamefont {Li}, \citenamefont {Wu},\ and\ \citenamefont
  {Lu}}]{Yang:2016evp}%
  \BibitemOpen
  \bibfield  {author} {\bibinfo {author} {\bibfnamefont {W.}~\bibnamefont
  {Yang}}, \bibinfo {author} {\bibfnamefont {H.}~\bibnamefont {Li}}, \bibinfo
  {author} {\bibfnamefont {Y.}~\bibnamefont {Wu}},\ and\ \bibinfo {author}
  {\bibfnamefont {J.}~\bibnamefont {Lu}},\ }\bibfield  {title} {\bibinfo
  {title} {{Cosmological constraints on coupled dark energy}},\ }\href
  {https://doi.org/10.1088/1475-7516/2016/10/007} {\bibfield  {journal}
  {\bibinfo  {journal} {{{JCAP}}}\ }\textbf {\bibinfo {volume} {10}},\ \bibinfo
  {pages} {007} (\bibinfo {year} {2016})},\ \Eprint
  {https://arxiv.org/abs/1608.07039} {arXiv:1608.07039 [astro-ph.CO]}
  \BibitemShut {NoStop}%
\bibitem [{\citenamefont {Sharov}\ \emph {et~al.}(2017)\citenamefont {Sharov},
  \citenamefont {Bhattacharya}, \citenamefont {Pan}, \citenamefont {Nunes},\
  and\ \citenamefont {Chakraborty}}]{Sharov:2017iue}%
  \BibitemOpen
  \bibfield  {author} {\bibinfo {author} {\bibfnamefont {G.~S.}\ \bibnamefont
  {Sharov}}, \bibinfo {author} {\bibfnamefont {S.}~\bibnamefont
  {Bhattacharya}}, \bibinfo {author} {\bibfnamefont {S.}~\bibnamefont {Pan}},
  \bibinfo {author} {\bibfnamefont {R.~C.}\ \bibnamefont {Nunes}},\ and\
  \bibinfo {author} {\bibfnamefont {S.}~\bibnamefont {Chakraborty}},\
  }\bibfield  {title} {\bibinfo {title} {{A new interacting two fluid model and
  its consequences}},\ }\href {https://doi.org/10.1093/mnras/stw3358}
  {\bibfield  {journal} {\bibinfo  {journal} {Mon. Not. Roy. Astron. Soc.}\
  }\textbf {\bibinfo {volume} {466}},\ \bibinfo {pages} {3497} (\bibinfo {year}
  {2017})},\ \Eprint {https://arxiv.org/abs/1701.00780} {arXiv:1701.00780
  [gr-qc]} \BibitemShut {NoStop}%
\bibitem [{\citenamefont {Di~Valentino}\ \emph {et~al.}(2017)\citenamefont
  {Di~Valentino}, \citenamefont {Melchiorri},\ and\ \citenamefont
  {Mena}}]{DiValentino:2017iww}%
  \BibitemOpen
  \bibfield  {author} {\bibinfo {author} {\bibfnamefont {E.}~\bibnamefont
  {Di~Valentino}}, \bibinfo {author} {\bibfnamefont {A.}~\bibnamefont
  {Melchiorri}},\ and\ \bibinfo {author} {\bibfnamefont {O.}~\bibnamefont
  {Mena}},\ }\bibfield  {title} {\bibinfo {title} {{Can interacting dark energy
  solve the $H_0$ tension?}},\ }\href
  {https://doi.org/10.1103/PhysRevD.96.043503} {\bibfield  {journal} {\bibinfo
  {journal} {Phys. Rev. D}\ }\textbf {\bibinfo {volume} {96}},\ \bibinfo
  {pages} {043503} (\bibinfo {year} {2017})},\ \Eprint
  {https://arxiv.org/abs/1704.08342} {arXiv:1704.08342 [astro-ph.CO]}
  \BibitemShut {NoStop}%
\bibitem [{\citenamefont {Mifsud}\ and\ \citenamefont {Van
  De~Bruck}(2017)}]{Mifsud:2017fsy}%
  \BibitemOpen
  \bibfield  {author} {\bibinfo {author} {\bibfnamefont {J.}~\bibnamefont
  {Mifsud}}\ and\ \bibinfo {author} {\bibfnamefont {C.}~\bibnamefont {Van
  De~Bruck}},\ }\bibfield  {title} {\bibinfo {title} {{Probing the imprints of
  generalized interacting dark energy on the growth of perturbations}},\ }\href
  {https://doi.org/10.1088/1475-7516/2017/11/001} {\bibfield  {journal}
  {\bibinfo  {journal} {{{JCAP}}}\ }\textbf {\bibinfo {volume} {11}},\ \bibinfo
  {pages} {001} (\bibinfo {year} {2017})},\ \Eprint
  {https://arxiv.org/abs/1707.07667} {arXiv:1707.07667 [astro-ph.CO]}
  \BibitemShut {NoStop}%
\bibitem [{\citenamefont {Yang}\ \emph {et~al.}(2017)\citenamefont {Yang},
  \citenamefont {Xu}, \citenamefont {Li}, \citenamefont {Wu},\ and\
  \citenamefont {Lu}}]{Yang:2017iew}%
  \BibitemOpen
  \bibfield  {author} {\bibinfo {author} {\bibfnamefont {W.}~\bibnamefont
  {Yang}}, \bibinfo {author} {\bibfnamefont {L.}~\bibnamefont {Xu}}, \bibinfo
  {author} {\bibfnamefont {H.}~\bibnamefont {Li}}, \bibinfo {author}
  {\bibfnamefont {Y.}~\bibnamefont {Wu}},\ and\ \bibinfo {author}
  {\bibfnamefont {J.}~\bibnamefont {Lu}},\ }\bibfield  {title} {\bibinfo
  {title} {{Testing the Interacting Dark Energy Model with Cosmic Microwave
  Background Anisotropy and Observational Hubble Data}},\ }\href
  {https://doi.org/10.3390/e19070327} {\bibfield  {journal} {\bibinfo
  {journal} {Entropy}\ }\textbf {\bibinfo {volume} {19}},\ \bibinfo {pages}
  {327} (\bibinfo {year} {2017})}\BibitemShut {NoStop}%
\bibitem [{\citenamefont {Yang}\ \emph
  {et~al.}(2018{\natexlab{a}})\citenamefont {Yang}, \citenamefont {Pan},\ and\
  \citenamefont {Barrow}}]{Yang:2017zjs}%
  \BibitemOpen
  \bibfield  {author} {\bibinfo {author} {\bibfnamefont {W.}~\bibnamefont
  {Yang}}, \bibinfo {author} {\bibfnamefont {S.}~\bibnamefont {Pan}},\ and\
  \bibinfo {author} {\bibfnamefont {J.~D.}\ \bibnamefont {Barrow}},\ }\bibfield
   {title} {\bibinfo {title} {{Large-scale stability and astronomical
  constraints for coupled dark-energy models}},\ }\href
  {https://doi.org/10.1103/PhysRevD.97.043529} {\bibfield  {journal} {\bibinfo
  {journal} {Phys. Rev. D}\ }\textbf {\bibinfo {volume} {97}},\ \bibinfo
  {pages} {043529} (\bibinfo {year} {2018}{\natexlab{a}})},\ \Eprint
  {https://arxiv.org/abs/1706.04953} {arXiv:1706.04953 [astro-ph.CO]}
  \BibitemShut {NoStop}%
\bibitem [{\citenamefont {Linton}\ \emph {et~al.}(2018)\citenamefont {Linton},
  \citenamefont {Pourtsidou}, \citenamefont {Crittenden},\ and\ \citenamefont
  {Maartens}}]{Linton:2017ged}%
  \BibitemOpen
  \bibfield  {author} {\bibinfo {author} {\bibfnamefont {M.~S.}\ \bibnamefont
  {Linton}}, \bibinfo {author} {\bibfnamefont {A.}~\bibnamefont {Pourtsidou}},
  \bibinfo {author} {\bibfnamefont {R.}~\bibnamefont {Crittenden}},\ and\
  \bibinfo {author} {\bibfnamefont {R.}~\bibnamefont {Maartens}},\ }\bibfield
  {title} {\bibinfo {title} {{Variable sound speed in interacting dark energy
  models}},\ }\href {https://doi.org/10.1088/1475-7516/2018/04/043} {\bibfield
  {journal} {\bibinfo  {journal} {{{JCAP}}}\ }\textbf {\bibinfo {volume}
  {04}},\ \bibinfo {pages} {043} (\bibinfo {year} {2018})},\ \Eprint
  {https://arxiv.org/abs/1711.05196} {arXiv:1711.05196 [astro-ph.CO]}
  \BibitemShut {NoStop}%
\bibitem [{\citenamefont {Yang}\ \emph {et~al.}(2019)\citenamefont {Yang},
  \citenamefont {Banerjee}, \citenamefont {Paliathanasis},\ and\ \citenamefont
  {Pan}}]{Yang:2018qec}%
  \BibitemOpen
  \bibfield  {author} {\bibinfo {author} {\bibfnamefont {W.}~\bibnamefont
  {Yang}}, \bibinfo {author} {\bibfnamefont {N.}~\bibnamefont {Banerjee}},
  \bibinfo {author} {\bibfnamefont {A.}~\bibnamefont {Paliathanasis}},\ and\
  \bibinfo {author} {\bibfnamefont {S.}~\bibnamefont {Pan}},\ }\bibfield
  {title} {\bibinfo {title} {{Reconstructing the dark matter and dark energy
  interaction scenarios from observations}},\ }\href
  {https://doi.org/10.1016/j.dark.2019.100383} {\bibfield  {journal} {\bibinfo
  {journal} {Phys. Dark Univ.}\ }\textbf {\bibinfo {volume} {26}},\ \bibinfo
  {pages} {100383} (\bibinfo {year} {2019})},\ \Eprint
  {https://arxiv.org/abs/1812.06854} {arXiv:1812.06854 [astro-ph.CO]}
  \BibitemShut {NoStop}%
\bibitem [{\citenamefont {Martinelli}\ \emph {et~al.}(2019)\citenamefont
  {Martinelli}, \citenamefont {Hogg}, \citenamefont {Peirone}, \citenamefont
  {Bruni},\ and\ \citenamefont {Wands}}]{Martinelli:2019dau}%
  \BibitemOpen
  \bibfield  {author} {\bibinfo {author} {\bibfnamefont {M.}~\bibnamefont
  {Martinelli}}, \bibinfo {author} {\bibfnamefont {N.~B.}\ \bibnamefont
  {Hogg}}, \bibinfo {author} {\bibfnamefont {S.}~\bibnamefont {Peirone}},
  \bibinfo {author} {\bibfnamefont {M.}~\bibnamefont {Bruni}},\ and\ \bibinfo
  {author} {\bibfnamefont {D.}~\bibnamefont {Wands}},\ }\bibfield  {title}
  {\bibinfo {title} {{Constraints on the interacting
  vacuum{\textendash}geodesic CDM scenario}},\ }\href
  {https://doi.org/10.1093/mnras/stz1915} {\bibfield  {journal} {\bibinfo
  {journal} {Mon. Not. Roy. Astron. Soc.}\ }\textbf {\bibinfo {volume} {488}},\
  \bibinfo {pages} {3423} (\bibinfo {year} {2019})},\ \Eprint
  {https://arxiv.org/abs/1902.10694} {arXiv:1902.10694 [astro-ph.CO]}
  \BibitemShut {NoStop}%
\bibitem [{\citenamefont {Paliathanasis}\ \emph {et~al.}(2019)\citenamefont
  {Paliathanasis}, \citenamefont {Pan},\ and\ \citenamefont
  {Yang}}]{Paliathanasis:2019hbi}%
  \BibitemOpen
  \bibfield  {author} {\bibinfo {author} {\bibfnamefont {A.}~\bibnamefont
  {Paliathanasis}}, \bibinfo {author} {\bibfnamefont {S.}~\bibnamefont {Pan}},\
  and\ \bibinfo {author} {\bibfnamefont {W.}~\bibnamefont {Yang}},\ }\bibfield
  {title} {\bibinfo {title} {{Dynamics of nonlinear interacting dark energy
  models}},\ }\href {https://doi.org/10.1142/S021827181950161X} {\bibfield
  {journal} {\bibinfo  {journal} {Int. J. Mod. Phys. D}\ }\textbf {\bibinfo
  {volume} {28}},\ \bibinfo {pages} {1950161} (\bibinfo {year} {2019})},\
  \Eprint {https://arxiv.org/abs/1903.02370} {arXiv:1903.02370 [gr-qc]}
  \BibitemShut {NoStop}%
\bibitem [{\citenamefont {Di~Valentino}\ and\ \citenamefont
  {Mena}(2020)}]{DiValentino:2020leo}%
  \BibitemOpen
  \bibfield  {author} {\bibinfo {author} {\bibfnamefont {E.}~\bibnamefont
  {Di~Valentino}}\ and\ \bibinfo {author} {\bibfnamefont {O.}~\bibnamefont
  {Mena}},\ }\bibfield  {title} {\bibinfo {title} {{A fake Interacting Dark
  Energy detection?}},\ }\href {https://doi.org/10.1093/mnrasl/slaa175}
  {\bibfield  {journal} {\bibinfo  {journal} {Mon. Not. Roy. Astron. Soc.}\
  }\textbf {\bibinfo {volume} {500}},\ \bibinfo {pages} {L22} (\bibinfo {year}
  {2020})},\ \Eprint {https://arxiv.org/abs/2009.12620} {arXiv:2009.12620
  [astro-ph.CO]} \BibitemShut {NoStop}%
\bibitem [{\citenamefont {Pan}\ \emph {et~al.}(2020{\natexlab{a}})\citenamefont
  {Pan}, \citenamefont {Yang},\ and\ \citenamefont
  {Paliathanasis}}]{Pan:2020bur}%
  \BibitemOpen
  \bibfield  {author} {\bibinfo {author} {\bibfnamefont {S.}~\bibnamefont
  {Pan}}, \bibinfo {author} {\bibfnamefont {W.}~\bibnamefont {Yang}},\ and\
  \bibinfo {author} {\bibfnamefont {A.}~\bibnamefont {Paliathanasis}},\
  }\bibfield  {title} {\bibinfo {title} {{Non-linear interacting cosmological
  models after Planck 2018 legacy release and the $H_0$ tension}},\ }\href
  {https://doi.org/10.1093/mnras/staa213} {\bibfield  {journal} {\bibinfo
  {journal} {Mon. Not. Roy. Astron. Soc.}\ }\textbf {\bibinfo {volume} {493}},\
  \bibinfo {pages} {3114} (\bibinfo {year} {2020}{\natexlab{a}})},\ \Eprint
  {https://arxiv.org/abs/2002.03408} {arXiv:2002.03408 [astro-ph.CO]}
  \BibitemShut {NoStop}%
\bibitem [{\citenamefont {Yang}\ \emph {et~al.}(2020)\citenamefont {Yang},
  \citenamefont {Di~Valentino}, \citenamefont {Pan}, \citenamefont
  {Basilakos},\ and\ \citenamefont {Paliathanasis}}]{Yang:2020zuk}%
  \BibitemOpen
  \bibfield  {author} {\bibinfo {author} {\bibfnamefont {W.}~\bibnamefont
  {Yang}}, \bibinfo {author} {\bibfnamefont {E.}~\bibnamefont {Di~Valentino}},
  \bibinfo {author} {\bibfnamefont {S.}~\bibnamefont {Pan}}, \bibinfo {author}
  {\bibfnamefont {S.}~\bibnamefont {Basilakos}},\ and\ \bibinfo {author}
  {\bibfnamefont {A.}~\bibnamefont {Paliathanasis}},\ }\bibfield  {title}
  {\bibinfo {title} {{Metastable dark energy models in light of $Planck$ 2018
  data: Alleviating the $H_0$ tension}},\ }\href
  {https://doi.org/10.1103/PhysRevD.102.063503} {\bibfield  {journal} {\bibinfo
   {journal} {Phys. Rev. D}\ }\textbf {\bibinfo {volume} {102}},\ \bibinfo
  {pages} {063503} (\bibinfo {year} {2020})},\ \Eprint
  {https://arxiv.org/abs/2001.04307} {arXiv:2001.04307 [astro-ph.CO]}
  \BibitemShut {NoStop}%
\bibitem [{\citenamefont {Pan}\ \emph {et~al.}(2020{\natexlab{b}})\citenamefont
  {Pan}, \citenamefont {Sharov},\ and\ \citenamefont {Yang}}]{Pan:2020zza}%
  \BibitemOpen
  \bibfield  {author} {\bibinfo {author} {\bibfnamefont {S.}~\bibnamefont
  {Pan}}, \bibinfo {author} {\bibfnamefont {G.~S.}\ \bibnamefont {Sharov}},\
  and\ \bibinfo {author} {\bibfnamefont {W.}~\bibnamefont {Yang}},\ }\bibfield
  {title} {\bibinfo {title} {{Field theoretic interpretations of interacting
  dark energy scenarios and recent observations}},\ }\href
  {https://doi.org/10.1103/PhysRevD.101.103533} {\bibfield  {journal} {\bibinfo
   {journal} {Phys. Rev. D}\ }\textbf {\bibinfo {volume} {101}},\ \bibinfo
  {pages} {103533} (\bibinfo {year} {2020}{\natexlab{b}})},\ \Eprint
  {https://arxiv.org/abs/2001.03120} {arXiv:2001.03120 [astro-ph.CO]}
  \BibitemShut {NoStop}%
\bibitem [{\citenamefont {S\'a}(2020{\natexlab{a}})}]{Sa:2020a}%
  \BibitemOpen
  \bibfield  {author} {\bibinfo {author} {\bibfnamefont {P.~M.}\ \bibnamefont
  {S\'a}},\ }\bibfield  {title} {\bibinfo {title} {{Unified Description of Dark
  Energy and Dark Matter within the Generalized Hybrid Metric-Palatini Theory
  of Gravity}},\ }\href {https://doi.org/10.3390/universe6060078} {\bibfield
  {journal} {\bibinfo  {journal} {Universe}\ }\textbf {\bibinfo {volume} {6}},\
  \bibinfo {pages} {78} (\bibinfo {year} {2020}{\natexlab{a}})},\ \Eprint
  {https://arxiv.org/abs/2002.09446} {arXiv:2002.09446 [gr-qc]} \BibitemShut
  {NoStop}%
\bibitem [{\citenamefont {Pan}\ \emph {et~al.}(2020{\natexlab{c}})\citenamefont
  {Pan}, \citenamefont {de~Haro}, \citenamefont {Yang},\ and\ \citenamefont
  {Amor\'os}}]{Pan:2020mst}%
  \BibitemOpen
  \bibfield  {author} {\bibinfo {author} {\bibfnamefont {S.}~\bibnamefont
  {Pan}}, \bibinfo {author} {\bibfnamefont {J.}~\bibnamefont {de~Haro}},
  \bibinfo {author} {\bibfnamefont {W.}~\bibnamefont {Yang}},\ and\ \bibinfo
  {author} {\bibfnamefont {J.}~\bibnamefont {Amor\'os}},\ }\bibfield  {title}
  {\bibinfo {title} {{Understanding the phenomenology of interacting dark
  energy scenarios and their theoretical bounds}},\ }\href
  {https://doi.org/10.1103/PhysRevD.101.123506} {\bibfield  {journal} {\bibinfo
   {journal} {Phys. Rev. D}\ }\textbf {\bibinfo {volume} {101}},\ \bibinfo
  {pages} {123506} (\bibinfo {year} {2020}{\natexlab{c}})},\ \Eprint
  {https://arxiv.org/abs/2001.09885} {arXiv:2001.09885 [gr-qc]} \BibitemShut
  {NoStop}%
\bibitem [{\citenamefont {S\'a}(2020{\natexlab{b}})}]{Sa:2020b}%
  \BibitemOpen
  \bibfield  {author} {\bibinfo {author} {\bibfnamefont {P.~M.}\ \bibnamefont
  {S\'a}},\ }\bibfield  {title} {\bibinfo {title} {{Triple unification of
  inflation, dark energy, and dark matter in two-scalar-field cosmology}},\
  }\href {https://doi.org/10.1103/PhysRevD.102.103519} {\bibfield  {journal}
  {\bibinfo  {journal} {Phys. Rev. D}\ }\textbf {\bibinfo {volume} {102}},\
  \bibinfo {pages} {103519} (\bibinfo {year} {2020}{\natexlab{b}})},\ \Eprint
  {https://arxiv.org/abs/2007.07109} {arXiv:2007.07109 [gr-qc]} \BibitemShut
  {NoStop}%
\bibitem [{\citenamefont {Di~Valentino}\ \emph
  {et~al.}(2020{\natexlab{a}})\citenamefont {Di~Valentino}, \citenamefont
  {Melchiorri}, \citenamefont {Mena},\ and\ \citenamefont
  {Vagnozzi}}]{DiValentino:2019ffd}%
  \BibitemOpen
  \bibfield  {author} {\bibinfo {author} {\bibfnamefont {E.}~\bibnamefont
  {Di~Valentino}}, \bibinfo {author} {\bibfnamefont {A.}~\bibnamefont
  {Melchiorri}}, \bibinfo {author} {\bibfnamefont {O.}~\bibnamefont {Mena}},\
  and\ \bibinfo {author} {\bibfnamefont {S.}~\bibnamefont {Vagnozzi}},\
  }\bibfield  {title} {\bibinfo {title} {{Interacting dark energy in the early
  2020s: A promising solution to the $H_0$ and cosmic shear tensions}},\ }\href
  {https://doi.org/10.1016/j.dark.2020.100666} {\bibfield  {journal} {\bibinfo
  {journal} {Phys. Dark Univ.}\ }\textbf {\bibinfo {volume} {30}},\ \bibinfo
  {pages} {100666} (\bibinfo {year} {2020}{\natexlab{a}})},\ \Eprint
  {https://arxiv.org/abs/1908.04281} {arXiv:1908.04281 [astro-ph.CO]}
  \BibitemShut {NoStop}%
\bibitem [{\citenamefont {Di~Valentino}\ \emph
  {et~al.}(2020{\natexlab{b}})\citenamefont {Di~Valentino}, \citenamefont
  {Melchiorri}, \citenamefont {Mena},\ and\ \citenamefont
  {Vagnozzi}}]{DiValentino:2019jae}%
  \BibitemOpen
  \bibfield  {author} {\bibinfo {author} {\bibfnamefont {E.}~\bibnamefont
  {Di~Valentino}}, \bibinfo {author} {\bibfnamefont {A.}~\bibnamefont
  {Melchiorri}}, \bibinfo {author} {\bibfnamefont {O.}~\bibnamefont {Mena}},\
  and\ \bibinfo {author} {\bibfnamefont {S.}~\bibnamefont {Vagnozzi}},\
  }\bibfield  {title} {\bibinfo {title} {{Nonminimal dark sector physics and
  cosmological tensions}},\ }\href
  {https://doi.org/10.1103/PhysRevD.101.063502} {\bibfield  {journal} {\bibinfo
   {journal} {Phys. Rev. D}\ }\textbf {\bibinfo {volume} {101}},\ \bibinfo
  {pages} {063502} (\bibinfo {year} {2020}{\natexlab{b}})},\ \Eprint
  {https://arxiv.org/abs/1910.09853} {arXiv:1910.09853 [astro-ph.CO]}
  \BibitemShut {NoStop}%
\bibitem [{\citenamefont {Di~Valentino}\ \emph {et~al.}(2021)\citenamefont
  {Di~Valentino}, \citenamefont {Melchiorri}, \citenamefont {Mena},
  \citenamefont {Pan},\ and\ \citenamefont {Yang}}]{DiValentino:2020kpf}%
  \BibitemOpen
  \bibfield  {author} {\bibinfo {author} {\bibfnamefont {E.}~\bibnamefont
  {Di~Valentino}}, \bibinfo {author} {\bibfnamefont {A.}~\bibnamefont
  {Melchiorri}}, \bibinfo {author} {\bibfnamefont {O.}~\bibnamefont {Mena}},
  \bibinfo {author} {\bibfnamefont {S.}~\bibnamefont {Pan}},\ and\ \bibinfo
  {author} {\bibfnamefont {W.}~\bibnamefont {Yang}},\ }\bibfield  {title}
  {\bibinfo {title} {{Interacting dark energy in a closed universe}},\ }\href
  {https://doi.org/10.1093/mnrasl/slaa207} {\bibfield  {journal} {\bibinfo
  {journal} {Mon. Not. Roy. Astron. Soc.}\ }\textbf {\bibinfo {volume} {502}},\
  \bibinfo {pages} {L23} (\bibinfo {year} {2021})},\ \Eprint
  {https://arxiv.org/abs/2011.00283} {arXiv:2011.00283 [astro-ph.CO]}
  \BibitemShut {NoStop}%
\bibitem [{\citenamefont {Paliathanasis}\ and\ \citenamefont
  {Leon}(2021)}]{Paliathanasis:2020sfe}%
  \BibitemOpen
  \bibfield  {author} {\bibinfo {author} {\bibfnamefont {A.}~\bibnamefont
  {Paliathanasis}}\ and\ \bibinfo {author} {\bibfnamefont {G.}~\bibnamefont
  {Leon}},\ }\bibfield  {title} {\bibinfo {title} {{Dynamics of a two scalar
  field cosmological model with phantom terms}},\ }\href
  {https://doi.org/10.1088/1361-6382/abe2d7} {\bibfield  {journal} {\bibinfo
  {journal} {Class. Quant. Grav.}\ }\textbf {\bibinfo {volume} {38}},\ \bibinfo
  {pages} {075013} (\bibinfo {year} {2021})},\ \Eprint
  {https://arxiv.org/abs/2009.12874} {arXiv:2009.12874 [gr-qc]} \BibitemShut
  {NoStop}%
\bibitem [{\citenamefont {Gao}\ \emph {et~al.}(2021)\citenamefont {Gao},
  \citenamefont {Zhao}, \citenamefont {Xue},\ and\ \citenamefont
  {Zhang}}]{Gao:2021xnk}%
  \BibitemOpen
  \bibfield  {author} {\bibinfo {author} {\bibfnamefont {L.-Y.}\ \bibnamefont
  {Gao}}, \bibinfo {author} {\bibfnamefont {Z.-W.}\ \bibnamefont {Zhao}},
  \bibinfo {author} {\bibfnamefont {S.-S.}\ \bibnamefont {Xue}},\ and\ \bibinfo
  {author} {\bibfnamefont {X.}~\bibnamefont {Zhang}},\ }\bibfield  {title}
  {\bibinfo {title} {{Relieving the $H_0$ tension with a new interacting dark
  energy model}},\ }\href {https://doi.org/10.1088/1475-7516/2021/07/005}
  {\bibfield  {journal} {\bibinfo  {journal} {{JCAP}}\ }\textbf {\bibinfo
  {volume} {07}},\ \bibinfo {pages} {005} (\bibinfo {year} {{2021}})},\ \Eprint
  {https://arxiv.org/abs/2101.10714} {arXiv:2101.10714 [astro-ph.CO]}
  \BibitemShut {NoStop}%
\bibitem [{\citenamefont {Lucca}(2021)}]{Lucca:2021eqy}%
  \BibitemOpen
  \bibfield  {author} {\bibinfo {author} {\bibfnamefont {M.}~\bibnamefont
  {Lucca}},\ }\bibfield  {title} {\bibinfo {title} {{Multi-interacting dark
  energy and its cosmological implications}},\ }\href
  {https://doi.org/10.1103/PhysRevD.104.083510} {\bibfield  {journal} {\bibinfo
   {journal} {Phys. Rev. D}\ }\textbf {\bibinfo {volume} {104}},\ \bibinfo
  {pages} {083510} (\bibinfo {year} {2021})},\ \Eprint
  {https://arxiv.org/abs/2106.15196} {arXiv:2106.15196 [astro-ph.CO]}
  \BibitemShut {NoStop}%
\bibitem [{\citenamefont {S\'a}(2021)}]{Sa:2021}%
  \BibitemOpen
  \bibfield  {author} {\bibinfo {author} {\bibfnamefont {P.~M.}\ \bibnamefont
  {S\'a}},\ }\bibfield  {title} {\bibinfo {title} {{Late-time evolution of the
  Universe within a two-scalar-field cosmological model}},\ }\href
  {https://doi.org/10.1103/PhysRevD.103.123517} {\bibfield  {journal} {\bibinfo
   {journal} {Phys. Rev. D}\ }\textbf {\bibinfo {volume} {103}},\ \bibinfo
  {pages} {123517} (\bibinfo {year} {2021})},\ \Eprint
  {https://arxiv.org/abs/2103.01693} {arXiv:2103.01693 [gr-qc]} \BibitemShut
  {NoStop}%
\bibitem [{\citenamefont {Gomez}\ and\ \citenamefont
  {Rodriguez}(2021)}]{Gomez:2020sfz}%
  \BibitemOpen
  \bibfield  {author} {\bibinfo {author} {\bibfnamefont {L.~G.}\ \bibnamefont
  {Gomez}}\ and\ \bibinfo {author} {\bibfnamefont {Y.}~\bibnamefont
  {Rodriguez}},\ }\bibfield  {title} {\bibinfo {title} {{Coupled multi-Proca
  vector dark energy}},\ }\href {https://doi.org/10.1016/j.dark.2020.100759}
  {\bibfield  {journal} {\bibinfo  {journal} {Phys. Dark Univ.}\ }\textbf
  {\bibinfo {volume} {31}},\ \bibinfo {pages} {100759} (\bibinfo {year}
  {2021})},\ \Eprint {https://arxiv.org/abs/2004.06466} {arXiv:2004.06466
  [gr-qc]} \BibitemShut {NoStop}%
\bibitem [{\citenamefont {Nunes}\ \emph {et~al.}(2022)\citenamefont {Nunes},
  \citenamefont {Vagnozzi}, \citenamefont {Kumar}, \citenamefont
  {Di~Valentino},\ and\ \citenamefont {Mena}}]{Nunes:2022bhn}%
  \BibitemOpen
  \bibfield  {author} {\bibinfo {author} {\bibfnamefont {R.~C.}\ \bibnamefont
  {Nunes}}, \bibinfo {author} {\bibfnamefont {S.}~\bibnamefont {Vagnozzi}},
  \bibinfo {author} {\bibfnamefont {S.}~\bibnamefont {Kumar}}, \bibinfo
  {author} {\bibfnamefont {E.}~\bibnamefont {Di~Valentino}},\ and\ \bibinfo
  {author} {\bibfnamefont {O.}~\bibnamefont {Mena}},\ }\bibfield  {title}
  {\bibinfo {title} {{New tests of dark sector interactions from the full-shape
  galaxy power spectrum}},\ }\href
  {https://doi.org/10.1103/PhysRevD.105.123506} {\bibfield  {journal} {\bibinfo
   {journal} {Phys. Rev. D}\ }\textbf {\bibinfo {volume} {105}},\ \bibinfo
  {pages} {123506} (\bibinfo {year} {2022})},\ \Eprint
  {https://arxiv.org/abs/2203.08093} {arXiv:2203.08093 [astro-ph.CO]}
  \BibitemShut {NoStop}%
\bibitem [{\citenamefont {Chatzidakis}\ \emph {et~al.}(2022)\citenamefont
  {Chatzidakis}, \citenamefont {Giacomini}, \citenamefont {Leach},
  \citenamefont {Leon}, \citenamefont {Paliathanasis},\ and\ \citenamefont
  {Pan}}]{Chatzidakis:2022mpf}%
  \BibitemOpen
  \bibfield  {author} {\bibinfo {author} {\bibfnamefont {S.}~\bibnamefont
  {Chatzidakis}}, \bibinfo {author} {\bibfnamefont {A.}~\bibnamefont
  {Giacomini}}, \bibinfo {author} {\bibfnamefont {P.~G.~L.}\ \bibnamefont
  {Leach}}, \bibinfo {author} {\bibfnamefont {G.}~\bibnamefont {Leon}},
  \bibinfo {author} {\bibfnamefont {A.}~\bibnamefont {Paliathanasis}},\ and\
  \bibinfo {author} {\bibfnamefont {S.}~\bibnamefont {Pan}},\ }\bibfield
  {title} {\bibinfo {title} {{Interacting dark energy in curved FLRW spacetime
  from Weyl Integrable Spacetime}},\ }\href
  {https://doi.org/10.1016/j.jheap.2022.10.001} {\bibfield  {journal} {\bibinfo
   {journal} {JHEAp}\ }\textbf {\bibinfo {volume} {36}},\ \bibinfo {pages}
  {141} (\bibinfo {year} {2022})},\ \Eprint {https://arxiv.org/abs/2206.06639}
  {arXiv:2206.06639 [gr-qc]} \BibitemShut {NoStop}%
\bibitem [{\citenamefont {Potting}\ and\ \citenamefont {S\'a}(2022)}]{Sa:2022}%
  \BibitemOpen
  \bibfield  {author} {\bibinfo {author} {\bibfnamefont {R.}~\bibnamefont
  {Potting}}\ and\ \bibinfo {author} {\bibfnamefont {P.~M.}\ \bibnamefont
  {S\'a}},\ }\bibfield  {title} {\bibinfo {title} {{Coupled quintessence with a
  generalized interaction term}},\ }\href
  {https://doi.org/10.1142/S0218271822501115} {\bibfield  {journal} {\bibinfo
  {journal} {Int. J. Mod. Phys. D}\ }\textbf {\bibinfo {volume} {31}},\
  \bibinfo {pages} {2250111} (\bibinfo {year} {2022})},\ \Eprint
  {https://arxiv.org/abs/2112.07608} {arXiv:2112.07608 [gr-qc]} \BibitemShut
  {NoStop}%
\bibitem [{\citenamefont {Teixeira}\ \emph {et~al.}(2023)\citenamefont
  {Teixeira}, \citenamefont {Daniel}, \citenamefont {Frusciante},\ and\
  \citenamefont {van~de Bruck}}]{Teixeira:2023zjt}%
  \BibitemOpen
  \bibfield  {author} {\bibinfo {author} {\bibfnamefont {E.~M.}\ \bibnamefont
  {Teixeira}}, \bibinfo {author} {\bibfnamefont {R.}~\bibnamefont {Daniel}},
  \bibinfo {author} {\bibfnamefont {N.}~\bibnamefont {Frusciante}},\ and\
  \bibinfo {author} {\bibfnamefont {C.}~\bibnamefont {van~de Bruck}},\
  }\bibfield  {title} {\bibinfo {title} {{Forecasts on interacting dark energy
  with standard sirens}},\ }\href {https://doi.org/10.1103/PhysRevD.108.084070}
  {\bibfield  {journal} {\bibinfo  {journal} {Phys. Rev. D}\ }\textbf {\bibinfo
  {volume} {108}},\ \bibinfo {pages} {084070} (\bibinfo {year} {2023})},\
  \Eprint {https://arxiv.org/abs/2309.06544} {arXiv:2309.06544 [astro-ph.CO]}
  \BibitemShut {NoStop}%
\bibitem [{\citenamefont {Zhai}\ \emph {et~al.}(2023)\citenamefont {Zhai},
  \citenamefont {Giar\`e}, \citenamefont {van~de Bruck}, \citenamefont
  {Di~Valentino}, \citenamefont {Mena},\ and\ \citenamefont
  {Nunes}}]{Zhai:2023yny}%
  \BibitemOpen
  \bibfield  {author} {\bibinfo {author} {\bibfnamefont {Y.}~\bibnamefont
  {Zhai}}, \bibinfo {author} {\bibfnamefont {W.}~\bibnamefont {Giar\`e}},
  \bibinfo {author} {\bibfnamefont {C.}~\bibnamefont {van~de Bruck}}, \bibinfo
  {author} {\bibfnamefont {E.}~\bibnamefont {Di~Valentino}}, \bibinfo {author}
  {\bibfnamefont {O.}~\bibnamefont {Mena}},\ and\ \bibinfo {author}
  {\bibfnamefont {R.~C.}\ \bibnamefont {Nunes}},\ }\bibfield  {title} {\bibinfo
  {title} {{A consistent view of interacting dark energy from multiple CMB
  probes}},\ }\href {https://doi.org/10.1088/1475-7516/2023/07/032} {\bibfield
  {journal} {\bibinfo  {journal} {{JCAP}}\ }\textbf {\bibinfo {volume} {07}},\
  \bibinfo {pages} {032} (\bibinfo {year} {2023})},\ \Eprint
  {https://arxiv.org/abs/2303.08201} {arXiv:2303.08201 [astro-ph.CO]}
  \BibitemShut {NoStop}%
\bibitem [{\citenamefont {Paliathanasis}(2023)}]{Paliathanasis:2023moe}%
  \BibitemOpen
  \bibfield  {author} {\bibinfo {author} {\bibfnamefont {A.}~\bibnamefont
  {Paliathanasis}},\ }\bibfield  {title} {\bibinfo {title} {{Unified dark
  energy from Chiral-Quintom model with a mixed potential in
  Friedmann{\textendash}Lema{\^\i}tre{\textendash}Robertson{\textendash}Walker
  cosmology}},\ }\href {https://doi.org/10.1140/epjc/s10052-023-11946-5}
  {\bibfield  {journal} {\bibinfo  {journal} {Eur. Phys. J. C}\ }\textbf
  {\bibinfo {volume} {83}},\ \bibinfo {pages} {756} (\bibinfo {year} {2023})},\
  \Eprint {https://arxiv.org/abs/2308.10460} {arXiv:2308.10460 [gr-qc]}
  \BibitemShut {NoStop}%
\bibitem [{\citenamefont {G{\'o}mez}(2023)}]{Gomez:2022okq}%
  \BibitemOpen
  \bibfield  {author} {\bibinfo {author} {\bibfnamefont {G.}~\bibnamefont
  {G{\'o}mez}},\ }\bibfield  {title} {\bibinfo {title} {{Conformally and
  disformally coupled vector field models of dark energy}},\ }\href
  {https://doi.org/10.1103/PhysRevD.107.123535} {\bibfield  {journal} {\bibinfo
   {journal} {Phys. Rev. D}\ }\textbf {\bibinfo {volume} {107}},\ \bibinfo
  {pages} {123535} (\bibinfo {year} {2023})},\ \Eprint
  {https://arxiv.org/abs/2202.07027} {arXiv:2202.07027 [gr-qc]} \BibitemShut
  {NoStop}%
\bibitem [{\citenamefont {Li}\ \emph {et~al.}(2024{\natexlab{a}})\citenamefont
  {Li}, \citenamefont {Jin}, \citenamefont {Li}, \citenamefont {Zhang},\ and\
  \citenamefont {Zhang}}]{Li:2023gtu}%
  \BibitemOpen
  \bibfield  {author} {\bibinfo {author} {\bibfnamefont {T.-N.}\ \bibnamefont
  {Li}}, \bibinfo {author} {\bibfnamefont {S.-J.}\ \bibnamefont {Jin}},
  \bibinfo {author} {\bibfnamefont {H.-L.}\ \bibnamefont {Li}}, \bibinfo
  {author} {\bibfnamefont {J.-F.}\ \bibnamefont {Zhang}},\ and\ \bibinfo
  {author} {\bibfnamefont {X.}~\bibnamefont {Zhang}},\ }\bibfield  {title}
  {\bibinfo {title} {{Prospects for Probing the Interaction between Dark Energy
  and Dark Matter Using Gravitational-wave Dark Sirens with Neutron Star Tidal
  Deformation}},\ }\href {https://doi.org/10.3847/1538-4357/ad1bc9} {\bibfield
  {journal} {\bibinfo  {journal} {Astrophys. J.}\ }\textbf {\bibinfo {volume}
  {963}},\ \bibinfo {pages} {52} (\bibinfo {year} {2024}{\natexlab{a}})},\
  \Eprint {https://arxiv.org/abs/2310.15879} {arXiv:2310.15879 [astro-ph.CO]}
  \BibitemShut {NoStop}%
\bibitem [{\citenamefont {Giar{\`e}}\ \emph
  {et~al.}(2024{\natexlab{a}})\citenamefont {Giar{\`e}}, \citenamefont
  {Sabogal}, \citenamefont {Nunes},\ and\ \citenamefont
  {Di~Valentino}}]{Giare:2024smz}%
  \BibitemOpen
  \bibfield  {author} {\bibinfo {author} {\bibfnamefont {W.}~\bibnamefont
  {Giar{\`e}}}, \bibinfo {author} {\bibfnamefont {M.~A.}\ \bibnamefont
  {Sabogal}}, \bibinfo {author} {\bibfnamefont {R.~C.}\ \bibnamefont {Nunes}},\
  and\ \bibinfo {author} {\bibfnamefont {E.}~\bibnamefont {Di~Valentino}},\
  }\bibfield  {title} {\bibinfo {title} {{Interacting Dark Energy after DESI
  Baryon Acoustic Oscillation Measurements}},\ }\href
  {https://doi.org/10.1103/PhysRevLett.133.251003} {\bibfield  {journal}
  {\bibinfo  {journal} {Phys. Rev. Lett.}\ }\textbf {\bibinfo {volume} {133}},\
  \bibinfo {pages} {251003} (\bibinfo {year} {2024}{\natexlab{a}})},\ \Eprint
  {https://arxiv.org/abs/2404.15232} {arXiv:2404.15232 [astro-ph.CO]}
  \BibitemShut {NoStop}%
\bibitem [{\citenamefont {Giar{\`e}}\ \emph
  {et~al.}(2024{\natexlab{b}})\citenamefont {Giar{\`e}}, \citenamefont {Zhai},
  \citenamefont {Pan}, \citenamefont {Di~Valentino}, \citenamefont {Nunes},\
  and\ \citenamefont {van~de Bruck}}]{Giare:2024ytc}%
  \BibitemOpen
  \bibfield  {author} {\bibinfo {author} {\bibfnamefont {W.}~\bibnamefont
  {Giar{\`e}}}, \bibinfo {author} {\bibfnamefont {Y.}~\bibnamefont {Zhai}},
  \bibinfo {author} {\bibfnamefont {S.}~\bibnamefont {Pan}}, \bibinfo {author}
  {\bibfnamefont {E.}~\bibnamefont {Di~Valentino}}, \bibinfo {author}
  {\bibfnamefont {R.~C.}\ \bibnamefont {Nunes}},\ and\ \bibinfo {author}
  {\bibfnamefont {C.}~\bibnamefont {van~de Bruck}},\ }\bibfield  {title}
  {\bibinfo {title} {{Tightening the reins on nonminimal dark sector physics:
  Interacting dark energy with dynamical and nondynamical equation of state}},\
  }\href {https://doi.org/10.1103/PhysRevD.110.063527} {\bibfield  {journal}
  {\bibinfo  {journal} {Phys. Rev. D}\ }\textbf {\bibinfo {volume} {110}},\
  \bibinfo {pages} {063527} (\bibinfo {year} {2024}{\natexlab{b}})},\ \Eprint
  {https://arxiv.org/abs/2404.02110} {arXiv:2404.02110 [astro-ph.CO]}
  \BibitemShut {NoStop}%
\bibitem [{\citenamefont {Li}\ \emph {et~al.}(2024{\natexlab{b}})\citenamefont
  {Li}, \citenamefont {Wu}, \citenamefont {Du}, \citenamefont {Jin},
  \citenamefont {Li}, \citenamefont {Zhang},\ and\ \citenamefont
  {Zhang}}]{Li:2024qso}%
  \BibitemOpen
  \bibfield  {author} {\bibinfo {author} {\bibfnamefont {T.-N.}\ \bibnamefont
  {Li}}, \bibinfo {author} {\bibfnamefont {P.-J.}\ \bibnamefont {Wu}}, \bibinfo
  {author} {\bibfnamefont {G.-H.}\ \bibnamefont {Du}}, \bibinfo {author}
  {\bibfnamefont {S.-J.}\ \bibnamefont {Jin}}, \bibinfo {author} {\bibfnamefont
  {H.-L.}\ \bibnamefont {Li}}, \bibinfo {author} {\bibfnamefont {J.-F.}\
  \bibnamefont {Zhang}},\ and\ \bibinfo {author} {\bibfnamefont
  {X.}~\bibnamefont {Zhang}},\ }\bibfield  {title} {\bibinfo {title}
  {{Constraints on Interacting Dark Energy Models from the DESI Baryon Acoustic
  Oscillation and DES Supernovae Data}},\ }\href
  {https://doi.org/10.3847/1538-4357/ad87f0} {\bibfield  {journal} {\bibinfo
  {journal} {Astrophys. J.}\ }\textbf {\bibinfo {volume} {976}},\ \bibinfo
  {pages} {1} (\bibinfo {year} {2024}{\natexlab{b}})},\ \Eprint
  {https://arxiv.org/abs/2407.14934} {arXiv:2407.14934 [astro-ph.CO]}
  \BibitemShut {NoStop}%
\bibitem [{\citenamefont {Paliathanasis}\ \emph
  {et~al.}(2025{\natexlab{a}})\citenamefont {Paliathanasis}, \citenamefont
  {Duffy}, \citenamefont {Halder},\ and\ \citenamefont
  {Abebe}}]{Paliathanasis:2024abl}%
  \BibitemOpen
  \bibfield  {author} {\bibinfo {author} {\bibfnamefont {A.}~\bibnamefont
  {Paliathanasis}}, \bibinfo {author} {\bibfnamefont {K.}~\bibnamefont
  {Duffy}}, \bibinfo {author} {\bibfnamefont {A.}~\bibnamefont {Halder}},\ and\
  \bibinfo {author} {\bibfnamefont {A.}~\bibnamefont {Abebe}},\ }\bibfield
  {title} {\bibinfo {title} {{Compartmentalization and coexistence in the dark
  sector of the universe}},\ }\href
  {https://doi.org/10.1016/j.dark.2024.101750} {\bibfield  {journal} {\bibinfo
  {journal} {Phys. Dark Univ.}\ }\textbf {\bibinfo {volume} {47}},\ \bibinfo
  {pages} {101750} (\bibinfo {year} {2025}{\natexlab{a}})},\ \Eprint
  {https://arxiv.org/abs/2409.05348} {arXiv:2409.05348 [gr-qc]} \BibitemShut
  {NoStop}%
\bibitem [{\citenamefont {Li}\ \emph {et~al.}(2025{\natexlab{a}})\citenamefont
  {Li}, \citenamefont {Du}, \citenamefont {Li}, \citenamefont {Wu},
  \citenamefont {Jin}, \citenamefont {Zhang},\ and\ \citenamefont
  {Zhang}}]{Li:2025owk}%
  \BibitemOpen
  \bibfield  {author} {\bibinfo {author} {\bibfnamefont {T.-N.}\ \bibnamefont
  {Li}}, \bibinfo {author} {\bibfnamefont {G.-H.}\ \bibnamefont {Du}}, \bibinfo
  {author} {\bibfnamefont {Y.-H.}\ \bibnamefont {Li}}, \bibinfo {author}
  {\bibfnamefont {P.-J.}\ \bibnamefont {Wu}}, \bibinfo {author} {\bibfnamefont
  {S.-J.}\ \bibnamefont {Jin}}, \bibinfo {author} {\bibfnamefont {J.-F.}\
  \bibnamefont {Zhang}},\ and\ \bibinfo {author} {\bibfnamefont
  {X.}~\bibnamefont {Zhang}},\ }\href@noop {} {\bibinfo {title} {{Probing the
  sign-changeable interaction between dark energy and dark matter with DESI
  baryon acoustic oscillations and DES supernovae data}}} (\bibinfo {year}
  {2025}{\natexlab{a}}),\ \Eprint {https://arxiv.org/abs/2501.07361}
  {arXiv:2501.07361 [astro-ph.CO]} \BibitemShut {NoStop}%
\bibitem [{\citenamefont {Leon}\ \emph {et~al.}(2025)\citenamefont {Leon},
  \citenamefont {Shankar}, \citenamefont {Halder},\ and\ \citenamefont
  {Paliathanasis}}]{Leon:2025sfd}%
  \BibitemOpen
  \bibfield  {author} {\bibinfo {author} {\bibfnamefont {G.}~\bibnamefont
  {Leon}}, \bibinfo {author} {\bibfnamefont {D.}~\bibnamefont {Shankar}},
  \bibinfo {author} {\bibfnamefont {A.}~\bibnamefont {Halder}},\ and\ \bibinfo
  {author} {\bibfnamefont {A.}~\bibnamefont {Paliathanasis}},\ }\bibfield
  {title} {\bibinfo {title} {{Cosmological interactions with phantom scalar
  field: Revisiting background phase-space analysis with compactified
  variables}},\ }\href {https://doi.org/10.1016/j.chaos.2025.117170} {\bibfield
   {journal} {\bibinfo  {journal} {Chaos Solitons Fractals}\ }\textbf {\bibinfo
  {volume} {200}},\ \bibinfo {pages} {117170} (\bibinfo {year} {2025})},\
  \Eprint {https://arxiv.org/abs/2501.09177} {arXiv:2501.09177 [gr-qc]}
  \BibitemShut {NoStop}%
\bibitem [{\citenamefont {Tsedrik}\ \emph {et~al.}(2025)\citenamefont {Tsedrik}
  \emph {et~al.}}]{Tsedrik:2025cwc}%
  \BibitemOpen
  \bibfield  {author} {\bibinfo {author} {\bibfnamefont {M.}~\bibnamefont
  {Tsedrik}} \emph {et~al.},\ }\href {https://doi.org/10.1093/mnrasl/slaf055}
  {\bibinfo {title} {{Interacting dark energy constraints from the full-shape
  analyses of BOSS DR12 and DES Year 3 measurements}}} (\bibinfo {year}
  {2025}),\ \Eprint {https://arxiv.org/abs/2502.03390} {arXiv:2502.03390
  [astro-ph.CO]} \BibitemShut {NoStop}%
\bibitem [{\citenamefont {Liu}\ \emph {et~al.}(2025)\citenamefont {Liu},
  \citenamefont {Fu}, \citenamefont {Xu}, \citenamefont {Ding}, \citenamefont
  {Huang},\ and\ \citenamefont {Qing}}]{Liu:2025pxy}%
  \BibitemOpen
  \bibfield  {author} {\bibinfo {author} {\bibfnamefont {K.}~\bibnamefont
  {Liu}}, \bibinfo {author} {\bibfnamefont {X.}~\bibnamefont {Fu}}, \bibinfo
  {author} {\bibfnamefont {B.}~\bibnamefont {Xu}}, \bibinfo {author}
  {\bibfnamefont {C.}~\bibnamefont {Ding}}, \bibinfo {author} {\bibfnamefont
  {Y.}~\bibnamefont {Huang}},\ and\ \bibinfo {author} {\bibfnamefont
  {X.}~\bibnamefont {Qing}},\ }\href@noop {} {\bibinfo {title} {{The growth of
  linear perturbations in the interacting dark energy models and observational
  constraints}}} (\bibinfo {year} {2025}),\ \Eprint
  {https://arxiv.org/abs/2503.05208} {arXiv:2503.05208 [astro-ph.CO]}
  \BibitemShut {NoStop}%
\bibitem [{\citenamefont {Feng}\ \emph {et~al.}(2025)\citenamefont {Feng},
  \citenamefont {Li}, \citenamefont {Du}, \citenamefont {Zhang},\ and\
  \citenamefont {Zhang}}]{Feng:2025mlo}%
  \BibitemOpen
  \bibfield  {author} {\bibinfo {author} {\bibfnamefont {L.}~\bibnamefont
  {Feng}}, \bibinfo {author} {\bibfnamefont {T.-N.}\ \bibnamefont {Li}},
  \bibinfo {author} {\bibfnamefont {G.-H.}\ \bibnamefont {Du}}, \bibinfo
  {author} {\bibfnamefont {J.-F.}\ \bibnamefont {Zhang}},\ and\ \bibinfo
  {author} {\bibfnamefont {X.}~\bibnamefont {Zhang}},\ }\bibfield  {title}
  {\bibinfo {title} {{A search for sterile neutrinos in interacting dark energy
  models using DESI baryon acoustic oscillations and DES supernovae data}},\
  }\href {https://doi.org/10.1016/j.dark.2025.101935} {\bibfield  {journal}
  {\bibinfo  {journal} {Phys. Dark Univ.}\ }\textbf {\bibinfo {volume} {48}},\
  \bibinfo {pages} {101935} (\bibinfo {year} {2025})},\ \Eprint
  {https://arxiv.org/abs/2503.10423} {arXiv:2503.10423 [astro-ph.CO]}
  \BibitemShut {NoStop}%
\bibitem [{\citenamefont {Zhai}\ \emph {et~al.}(2025)\citenamefont {Zhai},
  \citenamefont {de~Cesare}, \citenamefont {van~de Bruck}, \citenamefont
  {Di~Valentino},\ and\ \citenamefont {Wilson-Ewing}}]{Zhai:2025hfi}%
  \BibitemOpen
  \bibfield  {author} {\bibinfo {author} {\bibfnamefont {Y.}~\bibnamefont
  {Zhai}}, \bibinfo {author} {\bibfnamefont {M.}~\bibnamefont {de~Cesare}},
  \bibinfo {author} {\bibfnamefont {C.}~\bibnamefont {van~de Bruck}}, \bibinfo
  {author} {\bibfnamefont {E.}~\bibnamefont {Di~Valentino}},\ and\ \bibinfo
  {author} {\bibfnamefont {E.}~\bibnamefont {Wilson-Ewing}},\ }\bibfield
  {title} {\bibinfo {title} {{A low-redshift preference for an interacting dark
  energy model}},\ }\href {https://doi.org/10.1088/1475-7516/2025/11/010}
  {\bibfield  {journal} {\bibinfo  {journal} {JCAP}\ }\textbf {\bibinfo
  {volume} {11}},\ \bibinfo {pages} {010}},\ \Eprint
  {https://arxiv.org/abs/2503.15659} {arXiv:2503.15659 [astro-ph.CO]}
  \BibitemShut {NoStop}%
\bibitem [{\citenamefont {Silva}\ \emph {et~al.}(2025)\citenamefont {Silva},
  \citenamefont {Sabogal}, \citenamefont {Scherer}, \citenamefont {Nunes},
  \citenamefont {Di~Valentino},\ and\ \citenamefont {Kumar}}]{Silva:2025hxw}%
  \BibitemOpen
  \bibfield  {author} {\bibinfo {author} {\bibfnamefont {E.}~\bibnamefont
  {Silva}}, \bibinfo {author} {\bibfnamefont {M.~A.}\ \bibnamefont {Sabogal}},
  \bibinfo {author} {\bibfnamefont {M.}~\bibnamefont {Scherer}}, \bibinfo
  {author} {\bibfnamefont {R.~C.}\ \bibnamefont {Nunes}}, \bibinfo {author}
  {\bibfnamefont {E.}~\bibnamefont {Di~Valentino}},\ and\ \bibinfo {author}
  {\bibfnamefont {S.}~\bibnamefont {Kumar}},\ }\bibfield  {title} {\bibinfo
  {title} {{New constraints on interacting dark energy from DESI DR2 BAO
  observations}},\ }\href {https://doi.org/10.1103/qqc6-76z4} {\bibfield
  {journal} {\bibinfo  {journal} {Phys. Rev. D}\ }\textbf {\bibinfo {volume}
  {111}},\ \bibinfo {pages} {123511} (\bibinfo {year} {2025})},\ \Eprint
  {https://arxiv.org/abs/2503.23225} {arXiv:2503.23225 [astro-ph.CO]}
  \BibitemShut {NoStop}%
\bibitem [{\citenamefont {Li}\ and\ \citenamefont {Zhang}(2025)}]{Li:2025ula}%
  \BibitemOpen
  \bibfield  {author} {\bibinfo {author} {\bibfnamefont {Y.-H.}\ \bibnamefont
  {Li}}\ and\ \bibinfo {author} {\bibfnamefont {X.}~\bibnamefont {Zhang}},\
  }\bibfield  {title} {\bibinfo {title} {{Cosmic sign-reversal: non-parametric
  reconstruction of interacting dark energy with DESI DR2}},\ }\href
  {https://doi.org/10.1088/1475-7516/2025/12/018} {\bibfield  {journal}
  {\bibinfo  {journal} {JCAP}\ }\textbf {\bibinfo {volume} {12}},\ \bibinfo
  {pages} {018}},\ \Eprint {https://arxiv.org/abs/2506.18477} {arXiv:2506.18477
  [astro-ph.CO]} \BibitemShut {NoStop}%
\bibitem [{\citenamefont {van~der Westhuizen}\ \emph
  {et~al.}(2025{\natexlab{a}})\citenamefont {van~der Westhuizen}, \citenamefont
  {Figueruelo}, \citenamefont {Thubisi}, \citenamefont {Sahlu}, \citenamefont
  {Abebe},\ and\ \citenamefont {Paliathanasis}}]{vanderWesthuizen:2025iam}%
  \BibitemOpen
  \bibfield  {author} {\bibinfo {author} {\bibfnamefont {M.}~\bibnamefont
  {van~der Westhuizen}}, \bibinfo {author} {\bibfnamefont {D.}~\bibnamefont
  {Figueruelo}}, \bibinfo {author} {\bibfnamefont {R.}~\bibnamefont {Thubisi}},
  \bibinfo {author} {\bibfnamefont {S.}~\bibnamefont {Sahlu}}, \bibinfo
  {author} {\bibfnamefont {A.}~\bibnamefont {Abebe}},\ and\ \bibinfo {author}
  {\bibfnamefont {A.}~\bibnamefont {Paliathanasis}},\ }\bibfield  {title}
  {\bibinfo {title} {{Compartmentalization in the dark sector of the universe
  after DESI DR2 BAO data}},\ }\href
  {https://doi.org/10.1016/j.dark.2025.102107} {\bibfield  {journal} {\bibinfo
  {journal} {Phys. Dark Univ.}\ }\textbf {\bibinfo {volume} {50}},\ \bibinfo
  {pages} {102107} (\bibinfo {year} {2025}{\natexlab{a}})},\ \Eprint
  {https://arxiv.org/abs/2505.23306} {arXiv:2505.23306 [astro-ph.CO]}
  \BibitemShut {NoStop}%
\bibitem [{\citenamefont {Paliathanasis}(2025)}]{Paliathanasis:2025xxm}%
  \BibitemOpen
  \bibfield  {author} {\bibinfo {author} {\bibfnamefont {A.}~\bibnamefont
  {Paliathanasis}},\ }\href@noop {} {\bibinfo {title} {{Observational
  Constraints on Scalar Field--Matter Interaction in Weyl Integrable
  Spacetime}}} (\bibinfo {year} {2025}),\ \Eprint
  {https://arxiv.org/abs/2506.16223} {arXiv:2506.16223 [gr-qc]} \BibitemShut
  {NoStop}%
\bibitem [{\citenamefont {Yan}\ \emph {et~al.}(2025)\citenamefont {Yan},
  \citenamefont {Pan}, \citenamefont {Wang}, \citenamefont {Xu},\ and\
  \citenamefont {Peng}}]{Yan:2025iga}%
  \BibitemOpen
  \bibfield  {author} {\bibinfo {author} {\bibfnamefont {H.}~\bibnamefont
  {Yan}}, \bibinfo {author} {\bibfnamefont {Y.}~\bibnamefont {Pan}}, \bibinfo
  {author} {\bibfnamefont {J.-X.}\ \bibnamefont {Wang}}, \bibinfo {author}
  {\bibfnamefont {W.-X.}\ \bibnamefont {Xu}},\ and\ \bibinfo {author}
  {\bibfnamefont {Z.-H.}\ \bibnamefont {Peng}},\ }\href@noop {} {\bibinfo
  {title} {{Investigating Interacting Dark Energy Models Using Fast Radio Burst
  Observations}}} (\bibinfo {year} {2025}),\ \Eprint
  {https://arxiv.org/abs/2507.16308} {arXiv:2507.16308 [astro-ph.CO]}
  \BibitemShut {NoStop}%
\bibitem [{\citenamefont {Wang}\ \emph {et~al.}(2025)\citenamefont {Wang},
  \citenamefont {Cai}, \citenamefont {Guo},\ and\ \citenamefont
  {Wang}}]{Wang:2025znm}%
  \BibitemOpen
  \bibfield  {author} {\bibinfo {author} {\bibfnamefont {J.-Q.}\ \bibnamefont
  {Wang}}, \bibinfo {author} {\bibfnamefont {R.-G.}\ \bibnamefont {Cai}},
  \bibinfo {author} {\bibfnamefont {Z.-K.}\ \bibnamefont {Guo}},\ and\ \bibinfo
  {author} {\bibfnamefont {S.-J.}\ \bibnamefont {Wang}},\ }\href@noop {}
  {\bibinfo {title} {{Resolving the Planck-DESI tension by non-minimally
  coupled quintessence}}} (\bibinfo {year} {2025}),\ \Eprint
  {https://arxiv.org/abs/2508.01759} {arXiv:2508.01759 [astro-ph.CO]}
  \BibitemShut {NoStop}%
\bibitem [{\citenamefont {van~der Westhuizen}\ \emph
  {et~al.}(2025{\natexlab{b}})\citenamefont {van~der Westhuizen}, \citenamefont
  {Abebe},\ and\ \citenamefont {Di~Valentino}}]{vanderWesthuizen:2025vcb}%
  \BibitemOpen
  \bibfield  {author} {\bibinfo {author} {\bibfnamefont {M.}~\bibnamefont
  {van~der Westhuizen}}, \bibinfo {author} {\bibfnamefont {A.}~\bibnamefont
  {Abebe}},\ and\ \bibinfo {author} {\bibfnamefont {E.}~\bibnamefont
  {Di~Valentino}},\ }\bibfield  {title} {\bibinfo {title} {{I. Linear
  interacting dark energy: Analytical solutions and theoretical pathologies}},\
  }\href {https://doi.org/10.1016/j.dark.2025.102119} {\bibfield  {journal}
  {\bibinfo  {journal} {Phys. Dark Univ.}\ }\textbf {\bibinfo {volume} {50}},\
  \bibinfo {pages} {102119} (\bibinfo {year} {2025}{\natexlab{b}})},\ \Eprint
  {https://arxiv.org/abs/2509.04495} {arXiv:2509.04495 [gr-qc]} \BibitemShut
  {NoStop}%
\bibitem [{\citenamefont {van~der Westhuizen}\ \emph
  {et~al.}(2025{\natexlab{c}})\citenamefont {van~der Westhuizen}, \citenamefont
  {Abebe},\ and\ \citenamefont {Di~Valentino}}]{vanderWesthuizen:2025mnw}%
  \BibitemOpen
  \bibfield  {author} {\bibinfo {author} {\bibfnamefont {M.}~\bibnamefont
  {van~der Westhuizen}}, \bibinfo {author} {\bibfnamefont {A.}~\bibnamefont
  {Abebe}},\ and\ \bibinfo {author} {\bibfnamefont {E.}~\bibnamefont
  {Di~Valentino}},\ }\href@noop {} {\bibinfo {title} {{II. Non-Linear
  Interacting Dark Energy: Analytical Solutions and Theoretical Pathologies}}}
  (\bibinfo {year} {2025}{\natexlab{c}}),\ \Eprint
  {https://arxiv.org/abs/2509.04494} {arXiv:2509.04494 [gr-qc]} \BibitemShut
  {NoStop}%
\bibitem [{\citenamefont {Li}\ \emph {et~al.}(2025{\natexlab{b}})\citenamefont
  {Li}, \citenamefont {Du}, \citenamefont {Li}, \citenamefont {Li},
  \citenamefont {Ling}, \citenamefont {Zhang},\ and\ \citenamefont
  {Zhang}}]{Li:2025muv}%
  \BibitemOpen
  \bibfield  {author} {\bibinfo {author} {\bibfnamefont {T.-N.}\ \bibnamefont
  {Li}}, \bibinfo {author} {\bibfnamefont {G.-H.}\ \bibnamefont {Du}}, \bibinfo
  {author} {\bibfnamefont {Y.-H.}\ \bibnamefont {Li}}, \bibinfo {author}
  {\bibfnamefont {Y.}~\bibnamefont {Li}}, \bibinfo {author} {\bibfnamefont
  {J.-L.}\ \bibnamefont {Ling}}, \bibinfo {author} {\bibfnamefont {J.-F.}\
  \bibnamefont {Zhang}},\ and\ \bibinfo {author} {\bibfnamefont
  {X.}~\bibnamefont {Zhang}},\ }\href@noop {} {\bibinfo {title} {{Updated
  constraints on interacting dark energy: A comprehensive analysis using
  multiple CMB probes, DESI DR2, and supernovae observations}}} (\bibinfo
  {year} {2025}{\natexlab{b}}),\ \Eprint {https://arxiv.org/abs/2510.11363}
  {arXiv:2510.11363 [astro-ph.CO]} \BibitemShut {NoStop}%
\bibitem [{\citenamefont {Zhang}\ \emph {et~al.}(2025)\citenamefont {Zhang},
  \citenamefont {Li}, \citenamefont {Du}, \citenamefont {Zhou}, \citenamefont
  {Gao}, \citenamefont {Zhang},\ and\ \citenamefont {Zhang}}]{Zhang:2025dwu}%
  \BibitemOpen
  \bibfield  {author} {\bibinfo {author} {\bibfnamefont {Y.-M.}\ \bibnamefont
  {Zhang}}, \bibinfo {author} {\bibfnamefont {T.-N.}\ \bibnamefont {Li}},
  \bibinfo {author} {\bibfnamefont {G.-H.}\ \bibnamefont {Du}}, \bibinfo
  {author} {\bibfnamefont {S.-H.}\ \bibnamefont {Zhou}}, \bibinfo {author}
  {\bibfnamefont {L.-Y.}\ \bibnamefont {Gao}}, \bibinfo {author} {\bibfnamefont
  {J.-F.}\ \bibnamefont {Zhang}},\ and\ \bibinfo {author} {\bibfnamefont
  {X.}~\bibnamefont {Zhang}},\ }\href@noop {} {\bibinfo {title} {{Alleviating
  the $H_0$ tension through new interacting dark energy model in light of DESI
  DR2}}} (\bibinfo {year} {2025}),\ \Eprint {https://arxiv.org/abs/2510.12627}
  {arXiv:2510.12627 [astro-ph.CO]} \BibitemShut {NoStop}%
\bibitem [{\citenamefont {Kumar}\ and\ \citenamefont
  {Nunes}(2017)}]{Kumar:2017dnp}%
  \BibitemOpen
  \bibfield  {author} {\bibinfo {author} {\bibfnamefont {S.}~\bibnamefont
  {Kumar}}\ and\ \bibinfo {author} {\bibfnamefont {R.~C.}\ \bibnamefont
  {Nunes}},\ }\bibfield  {title} {\bibinfo {title} {{Echo of interactions in
  the dark sector}},\ }\href {https://doi.org/10.1103/PhysRevD.96.103511}
  {\bibfield  {journal} {\bibinfo  {journal} {Phys. Rev. D}\ }\textbf {\bibinfo
  {volume} {96}},\ \bibinfo {pages} {103511} (\bibinfo {year} {2017})},\
  \Eprint {https://arxiv.org/abs/1702.02143} {arXiv:1702.02143 [astro-ph.CO]}
  \BibitemShut {NoStop}%
\bibitem [{\citenamefont {Yang}\ \emph
  {et~al.}(2018{\natexlab{b}})\citenamefont {Yang}, \citenamefont {Pan},
  \citenamefont {Di~Valentino}, \citenamefont {Nunes}, \citenamefont
  {Vagnozzi},\ and\ \citenamefont {Mota}}]{Yang:2018euj}%
  \BibitemOpen
  \bibfield  {author} {\bibinfo {author} {\bibfnamefont {W.}~\bibnamefont
  {Yang}}, \bibinfo {author} {\bibfnamefont {S.}~\bibnamefont {Pan}}, \bibinfo
  {author} {\bibfnamefont {E.}~\bibnamefont {Di~Valentino}}, \bibinfo {author}
  {\bibfnamefont {R.~C.}\ \bibnamefont {Nunes}}, \bibinfo {author}
  {\bibfnamefont {S.}~\bibnamefont {Vagnozzi}},\ and\ \bibinfo {author}
  {\bibfnamefont {D.~F.}\ \bibnamefont {Mota}},\ }\bibfield  {title} {\bibinfo
  {title} {{Tale of stable interacting dark energy, observational signatures,
  and the $H_0$ tension}},\ }\href
  {https://doi.org/10.1088/1475-7516/2018/09/019} {\bibfield  {journal}
  {\bibinfo  {journal} {{{JCAP}}}\ }\textbf {\bibinfo {volume} {09}},\ \bibinfo
  {pages} {019} (\bibinfo {year} {2018}{\natexlab{b}})},\ \Eprint
  {https://arxiv.org/abs/1805.08252} {arXiv:1805.08252 [astro-ph.CO]}
  \BibitemShut {NoStop}%
\bibitem [{\citenamefont {Bahamonde}\ \emph {et~al.}(2018)\citenamefont
  {Bahamonde}, \citenamefont {B\"ohmer}, \citenamefont {Carloni}, \citenamefont
  {Copeland}, \citenamefont {Fang},\ and\ \citenamefont
  {Tamanini}}]{Bahamonde:2017ize}%
  \BibitemOpen
  \bibfield  {author} {\bibinfo {author} {\bibfnamefont {S.}~\bibnamefont
  {Bahamonde}}, \bibinfo {author} {\bibfnamefont {C.~G.}\ \bibnamefont
  {B\"ohmer}}, \bibinfo {author} {\bibfnamefont {S.}~\bibnamefont {Carloni}},
  \bibinfo {author} {\bibfnamefont {E.~J.}\ \bibnamefont {Copeland}}, \bibinfo
  {author} {\bibfnamefont {W.}~\bibnamefont {Fang}},\ and\ \bibinfo {author}
  {\bibfnamefont {N.}~\bibnamefont {Tamanini}},\ }\bibfield  {title} {\bibinfo
  {title} {{Dynamical systems applied to cosmology: Dark energy and modified
  gravity}},\ }\href {https://doi.org/10.1016/j.physrep.2018.09.001} {\bibfield
   {journal} {\bibinfo  {journal} {Phys. Rept.}\ }\textbf {\bibinfo {volume}
  {775-777}},\ \bibinfo {pages} {1} (\bibinfo {year} {2018})},\ \Eprint
  {https://arxiv.org/abs/1712.03107} {arXiv:1712.03107 [gr-qc]} \BibitemShut
  {NoStop}%
\bibitem [{\citenamefont {Guo}\ \emph {et~al.}(2005)\citenamefont {Guo},
  \citenamefont {Cai},\ and\ \citenamefont {Zhang}}]{Guo:2004xx}%
  \BibitemOpen
  \bibfield  {author} {\bibinfo {author} {\bibfnamefont {Z.-K.}\ \bibnamefont
  {Guo}}, \bibinfo {author} {\bibfnamefont {R.-G.}\ \bibnamefont {Cai}},\ and\
  \bibinfo {author} {\bibfnamefont {Y.-Z.}\ \bibnamefont {Zhang}},\ }\bibfield
  {title} {\bibinfo {title} {{Cosmological evolution of interacting phantom
  energy with dark matter}},\ }\href
  {https://doi.org/10.1088/1475-7516/2005/05/002} {\bibfield  {journal}
  {\bibinfo  {journal} {{{JCAP}}}\ }\textbf {\bibinfo {volume} {05}},\ \bibinfo
  {pages} {002} (\bibinfo {year} {2005})},\ \Eprint
  {https://arxiv.org/abs/astro-ph/0412624} {arXiv:astro-ph/0412624}
  \BibitemShut {NoStop}%
\bibitem [{\citenamefont {Chen}\ \emph {et~al.}(2009)\citenamefont {Chen},
  \citenamefont {Gong},\ and\ \citenamefont {Saridakis}}]{Chen:2008ft}%
  \BibitemOpen
  \bibfield  {author} {\bibinfo {author} {\bibfnamefont {X.-m.}\ \bibnamefont
  {Chen}}, \bibinfo {author} {\bibfnamefont {Y.-g.}\ \bibnamefont {Gong}},\
  and\ \bibinfo {author} {\bibfnamefont {E.~N.}\ \bibnamefont {Saridakis}},\
  }\bibfield  {title} {\bibinfo {title} {{Phase-space analysis of interacting
  phantom cosmology}},\ }\href {https://doi.org/10.1088/1475-7516/2009/04/001}
  {\bibfield  {journal} {\bibinfo  {journal} {{{JCAP}}}\ }\textbf {\bibinfo
  {volume} {04}},\ \bibinfo {pages} {001} (\bibinfo {year} {2009})},\ \Eprint
  {https://arxiv.org/abs/0812.1117} {arXiv:0812.1117 [gr-qc]} \BibitemShut
  {NoStop}%
\bibitem [{\citenamefont {Halder}\ \emph {et~al.}(2025)\citenamefont {Halder},
  \citenamefont {Odintsov}, \citenamefont {Pan}, \citenamefont {Saha},\ and\
  \citenamefont {Saridakis}}]{Halder:2024aan}%
  \BibitemOpen
  \bibfield  {author} {\bibinfo {author} {\bibfnamefont {S.}~\bibnamefont
  {Halder}}, \bibinfo {author} {\bibfnamefont {S.~D.}\ \bibnamefont
  {Odintsov}}, \bibinfo {author} {\bibfnamefont {S.}~\bibnamefont {Pan}},
  \bibinfo {author} {\bibfnamefont {T.}~\bibnamefont {Saha}},\ and\ \bibinfo
  {author} {\bibfnamefont {E.~N.}\ \bibnamefont {Saridakis}},\ }\bibfield
  {title} {\bibinfo {title} {{Interacting phantom dark energy: New accelerating
  scaling attractors}},\ }\href {https://doi.org/10.1103/7jrj-sn1z} {\bibfield
  {journal} {\bibinfo  {journal} {Phys. Rev. D}\ }\textbf {\bibinfo {volume}
  {112}},\ \bibinfo {pages} {023519} (\bibinfo {year} {2025})},\ \Eprint
  {https://arxiv.org/abs/2411.18300} {arXiv:2411.18300 [gr-qc]} \BibitemShut
  {NoStop}%
\bibitem [{\citenamefont {Paliathanasis}\ \emph
  {et~al.}(2025{\natexlab{b}})\citenamefont {Paliathanasis}, \citenamefont
  {Halder},\ and\ \citenamefont {Leon}}]{Paliathanasis:2024jxo}%
  \BibitemOpen
  \bibfield  {author} {\bibinfo {author} {\bibfnamefont {A.}~\bibnamefont
  {Paliathanasis}}, \bibinfo {author} {\bibfnamefont {A.}~\bibnamefont
  {Halder}},\ and\ \bibinfo {author} {\bibfnamefont {G.}~\bibnamefont {Leon}},\
  }\bibfield  {title} {\bibinfo {title} {{Revise the dark matter-phantom scalar
  field interaction}},\ }\href {https://doi.org/10.1088/1402-4896/adf14e}
  {\bibfield  {journal} {\bibinfo  {journal} {Phys. Scripta}\ }\textbf
  {\bibinfo {volume} {100}},\ \bibinfo {pages} {085007} (\bibinfo {year}
  {2025}{\natexlab{b}})},\ \Eprint {https://arxiv.org/abs/2412.06501}
  {arXiv:2412.06501 [gr-qc]} \BibitemShut {NoStop}%
\bibitem [{\citenamefont {Halder}\ \emph {et~al.}(2024)\citenamefont {Halder},
  \citenamefont {Pan}, \citenamefont {S\'a},\ and\ \citenamefont
  {Saha}}]{Halder:2024gag}%
  \BibitemOpen
  \bibfield  {author} {\bibinfo {author} {\bibfnamefont {S.}~\bibnamefont
  {Halder}}, \bibinfo {author} {\bibfnamefont {S.}~\bibnamefont {Pan}},
  \bibinfo {author} {\bibfnamefont {P.~M.}\ \bibnamefont {S\'a}},\ and\
  \bibinfo {author} {\bibfnamefont {T.}~\bibnamefont {Saha}},\ }\bibfield
  {title} {\bibinfo {title} {{Coupled phantom cosmological model motivated by
  the warm inflationary paradigm}},\ }\href
  {https://doi.org/10.1103/PhysRevD.110.063529} {\bibfield  {journal} {\bibinfo
   {journal} {Phys. Rev. D}\ }\textbf {\bibinfo {volume} {110}},\ \bibinfo
  {pages} {063529} (\bibinfo {year} {2024})},\ \Eprint
  {https://arxiv.org/abs/2407.15804} {arXiv:2407.15804 [gr-qc]} \BibitemShut
  {NoStop}%
\bibitem [{\citenamefont {Steinhardt}\ \emph {et~al.}(1999)\citenamefont
  {Steinhardt}, \citenamefont {Wang},\ and\ \citenamefont
  {Zlatev}}]{Steinhardt:1999nw}%
  \BibitemOpen
  \bibfield  {author} {\bibinfo {author} {\bibfnamefont {P.~J.}\ \bibnamefont
  {Steinhardt}}, \bibinfo {author} {\bibfnamefont {L.-M.}\ \bibnamefont
  {Wang}},\ and\ \bibinfo {author} {\bibfnamefont {I.}~\bibnamefont {Zlatev}},\
  }\bibfield  {title} {\bibinfo {title} {{Cosmological tracking solutions}},\
  }\href {https://doi.org/10.1103/PhysRevD.59.123504} {\bibfield  {journal}
  {\bibinfo  {journal} {Phys. Rev. D}\ }\textbf {\bibinfo {volume} {59}},\
  \bibinfo {pages} {123504} (\bibinfo {year} {1999})},\ \Eprint
  {https://arxiv.org/abs/astro-ph/9812313} {arXiv:astro-ph/9812313}
  \BibitemShut {NoStop}%
\bibitem [{\citenamefont {de~la Macorra}\ and\ \citenamefont
  {Piccinelli}(2000)}]{delaMacorra:1999ff}%
  \BibitemOpen
  \bibfield  {author} {\bibinfo {author} {\bibfnamefont {A.}~\bibnamefont
  {de~la Macorra}}\ and\ \bibinfo {author} {\bibfnamefont {G.}~\bibnamefont
  {Piccinelli}},\ }\bibfield  {title} {\bibinfo {title} {{General scalar fields
  as quintessence}},\ }\href {https://doi.org/10.1103/PhysRevD.61.123503}
  {\bibfield  {journal} {\bibinfo  {journal} {Phys. Rev. D}\ }\textbf {\bibinfo
  {volume} {61}},\ \bibinfo {pages} {123503} (\bibinfo {year} {2000})},\
  \Eprint {https://arxiv.org/abs/hep-ph/9909459} {arXiv:hep-ph/9909459}
  \BibitemShut {NoStop}%
\bibitem [{\citenamefont {Ng}\ \emph {et~al.}(2001)\citenamefont {Ng},
  \citenamefont {Nunes},\ and\ \citenamefont {Rosati}}]{Ng:2001hs}%
  \BibitemOpen
  \bibfield  {author} {\bibinfo {author} {\bibfnamefont {S.~C.~C.}\
  \bibnamefont {Ng}}, \bibinfo {author} {\bibfnamefont {N.~J.}\ \bibnamefont
  {Nunes}},\ and\ \bibinfo {author} {\bibfnamefont {F.}~\bibnamefont
  {Rosati}},\ }\bibfield  {title} {\bibinfo {title} {{Applications of scalar
  attractor solutions to cosmology}},\ }\href
  {https://doi.org/10.1103/PhysRevD.64.083510} {\bibfield  {journal} {\bibinfo
  {journal} {Phys. Rev. D}\ }\textbf {\bibinfo {volume} {64}},\ \bibinfo
  {pages} {083510} (\bibinfo {year} {2001})},\ \Eprint
  {https://arxiv.org/abs/astro-ph/0107321} {arXiv:astro-ph/0107321}
  \BibitemShut {NoStop}%
\bibitem [{\citenamefont {S\'a}(2024)}]{Sa:2023coi}%
  \BibitemOpen
  \bibfield  {author} {\bibinfo {author} {\bibfnamefont {P.~M.}\ \bibnamefont
  {S\'a}},\ }\bibfield  {title} {\bibinfo {title} {{Coupled quintessence
  inspired by warm inflation}},\ }\href
  {https://doi.org/10.3390/universe10080324} {\bibfield  {journal} {\bibinfo
  {journal} {Universe}\ }\textbf {\bibinfo {volume} {10}},\ \bibinfo {pages}
  {324} (\bibinfo {year} {2024})},\ \Eprint {https://arxiv.org/abs/2312.09171}
  {arXiv:2312.09171 [gr-qc]} \BibitemShut {NoStop}%
\bibitem [{\citenamefont {Berera}(1995)}]{Berera:1995ie}%
  \BibitemOpen
  \bibfield  {author} {\bibinfo {author} {\bibfnamefont {A.}~\bibnamefont
  {Berera}},\ }\bibfield  {title} {\bibinfo {title} {{Warm inflation}},\ }\href
  {https://doi.org/10.1103/PhysRevLett.75.3218} {\bibfield  {journal} {\bibinfo
   {journal} {Phys. Rev. Lett.}\ }\textbf {\bibinfo {volume} {75}},\ \bibinfo
  {pages} {3218} (\bibinfo {year} {1995})},\ \Eprint
  {https://arxiv.org/abs/astro-ph/9509049} {arXiv:astro-ph/9509049}
  \BibitemShut {NoStop}%
\bibitem [{\citenamefont {Berera}(2023)}]{Berera:2023liv}%
  \BibitemOpen
  \bibfield  {author} {\bibinfo {author} {\bibfnamefont {A.}~\bibnamefont
  {Berera}},\ }\bibfield  {title} {\bibinfo {title} {{The Warm Inflation
  Story}},\ }\href {https://doi.org/10.3390/universe9060272} {\bibfield
  {journal} {\bibinfo  {journal} {Universe}\ }\textbf {\bibinfo {volume} {9}},\
  \bibinfo {pages} {272} (\bibinfo {year} {2023})},\ \Eprint
  {https://arxiv.org/abs/2305.10879} {arXiv:2305.10879 [hep-ph]} \BibitemShut
  {NoStop}%
\bibitem [{\citenamefont {Kamali}\ \emph {et~al.}(2023)\citenamefont {Kamali},
  \citenamefont {Motaharfar},\ and\ \citenamefont {Ramos}}]{Kamali:2023}%
  \BibitemOpen
  \bibfield  {author} {\bibinfo {author} {\bibfnamefont {V.}~\bibnamefont
  {Kamali}}, \bibinfo {author} {\bibfnamefont {M.}~\bibnamefont {Motaharfar}},\
  and\ \bibinfo {author} {\bibfnamefont {R.~O.}\ \bibnamefont {Ramos}},\
  }\bibfield  {title} {\bibinfo {title} {{Recent Developments in Warm
  Inflation.}},\ }\href {https://doi.org/doi.org/10.3390/universe9030124}
  {\bibfield  {journal} {\bibinfo  {journal} {Universe}\ }\textbf {\bibinfo
  {volume} {9}},\ \bibinfo {pages} {124} (\bibinfo {year} {2023})},\ \Eprint
  {https://arxiv.org/abs/2302.02827} {arXiv:2302.02827 [hep-ph]} \BibitemShut
  {NoStop}%
\bibitem [{\citenamefont {Lima}\ and\ \citenamefont {Ramos}(2019)}]{Lima:2019}%
  \BibitemOpen
  \bibfield  {author} {\bibinfo {author} {\bibfnamefont {G.~B.~F.}\
  \bibnamefont {Lima}}\ and\ \bibinfo {author} {\bibfnamefont {R.~O.}\
  \bibnamefont {Ramos}},\ }\bibfield  {title} {\bibinfo {title} {{Unified early
  and late Universe cosmology through dissipative effects in steep
  quintessential inflation potential models}},\ }\href
  {https://doi.org/doi.org/10.1103/PhysRevD.100.123529} {\bibfield  {journal}
  {\bibinfo  {journal} {Phys. Rev. D}\ }\textbf {\bibinfo {volume} {100}},\
  \bibinfo {pages} {123529} (\bibinfo {year} {2019})},\ \Eprint
  {https://arxiv.org/abs/1910.05185} {arXiv:1910.05185 [astro-ph]} \BibitemShut
  {NoStop}%
\bibitem [{\citenamefont {Das}\ \emph {et~al.}(2023)\citenamefont {Das},
  \citenamefont {Hussain}, \citenamefont {Nandi}, \citenamefont {Ramos},\ and\
  \citenamefont {Silva}}]{Das:2023}%
  \BibitemOpen
  \bibfield  {author} {\bibinfo {author} {\bibfnamefont {S.}~\bibnamefont
  {Das}}, \bibinfo {author} {\bibfnamefont {S.}~\bibnamefont {Hussain}},
  \bibinfo {author} {\bibfnamefont {D.}~\bibnamefont {Nandi}}, \bibinfo
  {author} {\bibfnamefont {R.}~\bibnamefont {Ramos}},\ and\ \bibinfo {author}
  {\bibfnamefont {R.}~\bibnamefont {Silva}},\ }\bibfield  {title} {\bibinfo
  {title} {{Stability analysis of warm quintessential dark energy model}},\
  }\href {https://doi.org/doi.org/10.1103/PhysRevD.108.083517} {\bibfield
  {journal} {\bibinfo  {journal} {Phys. Rev. D}\ }\textbf {\bibinfo {volume}
  {108}},\ \bibinfo {pages} {083517} (\bibinfo {year} {2023})},\ \Eprint
  {https://arxiv.org/abs/2306.09369} {arXiv:2306.09369 [gr-qc]} \BibitemShut
  {NoStop}%
\bibitem [{\citenamefont {Ratra}\ and\ \citenamefont
  {Peebles}(1988)}]{Ratra:1987rm}%
  \BibitemOpen
  \bibfield  {author} {\bibinfo {author} {\bibfnamefont {B.}~\bibnamefont
  {Ratra}}\ and\ \bibinfo {author} {\bibfnamefont {P.~J.~E.}\ \bibnamefont
  {Peebles}},\ }\bibfield  {title} {\bibinfo {title} {{Cosmological
  Consequences of a Rolling Homogeneous Scalar Field}},\ }\href
  {https://doi.org/10.1103/PhysRevD.37.3406} {\bibfield  {journal} {\bibinfo
  {journal} {Phys. Rev. D}\ }\textbf {\bibinfo {volume} {37}},\ \bibinfo
  {pages} {3406} (\bibinfo {year} {1988})}\BibitemShut {NoStop}%
\bibitem [{\citenamefont {Carr}(1982)}]{Carr:1982}%
  \BibitemOpen
  \bibfield  {author} {\bibinfo {author} {\bibfnamefont {J.}~\bibnamefont
  {Carr}},\ }\href {https://doi.org/10.1007/978-1-4612-5929-9} {\emph {\bibinfo
  {title} {{Applications of Centre Manifold Theory}}}}\ (\bibinfo  {publisher}
  {Sprin\-ger},\ \bibinfo {address} {New York},\ \bibinfo {year}
  {1982})\BibitemShut {NoStop}%
\bibitem [{\citenamefont {Guckenheimer}\ and\ \citenamefont
  {Holmes}(1983)}]{Guckenheimer:1983}%
  \BibitemOpen
  \bibfield  {author} {\bibinfo {author} {\bibfnamefont {J.}~\bibnamefont
  {Guckenheimer}}\ and\ \bibinfo {author} {\bibfnamefont {P.}~\bibnamefont
  {Holmes}},\ }\href {https://doi.org/10.1007/978-1-4612-1140-2} {\emph
  {\bibinfo {title} {{Nonlinear Oscillations, Dynamical Systems, and
  Bifurcations of Vector Fields}}}}\ (\bibinfo  {publisher} {Springer},\
  \bibinfo {address} {New York},\ \bibinfo {year} {1983})\BibitemShut {NoStop}%
\bibitem [{\citenamefont {Bogoyavlensky}(1985)}]{Bogoyavlensky:1985}%
  \BibitemOpen
  \bibfield  {author} {\bibinfo {author} {\bibfnamefont {O.~I.}\ \bibnamefont
  {Bogoyavlensky}},\ }\href
  {https://doi.org/https://link.springer.com/book/9783642649028} {\emph
  {\bibinfo {title} {{Methods in the Qualitative Theory of Dynamical Systems in
  Astrophysics and Gas Dynamics}}}}\ (\bibinfo  {publisher} {Springer-Verlag},\
  \bibinfo {address} {Berlin, Heidelberg},\ \bibinfo {year} {1985})\BibitemShut
  {NoStop}%
\bibitem [{\citenamefont {Perko}(2013)}]{perko2013differential}%
  \BibitemOpen
  \bibfield  {author} {\bibinfo {author} {\bibfnamefont {L.}~\bibnamefont
  {Perko}},\ }\href@noop {} {\emph {\bibinfo {title} {Differential equations
  and dynamical systems}}},\ Vol.~\bibinfo {volume} {7}\ (\bibinfo  {publisher}
  {Springer Science \& Business Media},\ \bibinfo {year} {2013})\BibitemShut
  {NoStop}%
\bibitem [{\citenamefont {Dumortier}\ \emph {et~al.}(2006)\citenamefont
  {Dumortier}, \citenamefont {Llibre},\ and\ \citenamefont
  {Art{\'e}s}}]{dumortier2006qualitative}%
  \BibitemOpen
  \bibfield  {author} {\bibinfo {author} {\bibfnamefont {F.}~\bibnamefont
  {Dumortier}}, \bibinfo {author} {\bibfnamefont {J.}~\bibnamefont {Llibre}},\
  and\ \bibinfo {author} {\bibfnamefont {J.~C.}\ \bibnamefont {Art{\'e}s}},\
  }\href@noop {} {\emph {\bibinfo {title} {Qualitative theory of planar
  differential systems}}},\ Vol.~\bibinfo {volume} {2}\ (\bibinfo  {publisher}
  {Springer},\ \bibinfo {year} {2006})\BibitemShut {NoStop}%
\bibitem [{\citenamefont {Meiss}(2007)}]{meiss2007differential}%
  \BibitemOpen
  \bibfield  {author} {\bibinfo {author} {\bibfnamefont {J.~D.}\ \bibnamefont
  {Meiss}},\ }\href@noop {} {\emph {\bibinfo {title} {Differential dynamical
  systems}}}\ (\bibinfo  {publisher} {SIAM, Philadelphia},\ \bibinfo {year}
  {2007})\BibitemShut {NoStop}%
\end{thebibliography}%

\end{document}